\documentclass[a4paper,11pt]{article}
\pdfoutput=1 

\usepackage{jheppub} 

\usepackage[T1]{fontenc} 
\usepackage{parskip}
\usepackage{bm}         
\usepackage{array}
\usepackage{multirow}   
\usepackage{booktabs}   
\usepackage{mathtools} 
\usepackage{multicol}
\usepackage{mleftright}
\usepackage[utf8x]{inputenc}
\usepackage{bbold}
\usepackage{color}
\usepackage[dvipsnames]{xcolor}
\usepackage{amsmath,amscd}

\renewcommand{\d}{\mathrm d}
\renewcommand{\i}{\mathrm i}
\renewcommand{\(}{\left(}
\renewcommand{\)}{\right)}
\renewcommand{\[}{\left[}
\renewcommand{\]}{\right]}

\newcommand{\U}[1]{\mathrm{U}(1)_{\mathrm{#1}}}			
\newcommand{\SU}[2]{\mathrm{SU}(#1)_{\mathrm{#2}}}		

\DeclareMathOperator{\Tr}{Tr}

\title{Electroweak phase transitions in multi-Higgs models: the case of Trinification-inspired THDSM}

\author[a,b]{Thibault~Vieu}
\author[a]{Ant\'onio~P.~Morais}
\author[c]{Roman~Pasechnik}

\affiliation[a]{Departamento de F\'\i sica, Universidade de Aveiro and CIDMA,\\ 
Campus de Santiago, 3810-183 Aveiro, Portugal}
\affiliation[b]{Magist\`ere de Physique Fondamentale, Universit\'e Paris-Saclay, \\ 
B\^at. 470, F-91405 Orsay, France}
\affiliation[c]{Department of Astronomy and Theoretical Physics, Lund University,\\ 
221 00 Lund, Sweden}

\emailAdd{thibault.vieu@u-psud.fr}
\emailAdd{aapmorais@ua.pt}
\emailAdd{roman.pasechnik@thep.lu.se}

\abstract{
The rich vacuum structure of multi-Higgs extensions of the Standard Model (SM) may have interesting cosmological implications for the electroweak phase transition (EWPT). 
As an important example of such class of models, we consider a particularly simple low-energy SM-like limit of a recently proposed Grand-Unified Trinification model with the scalar sector 
composed of two Higgs doublets and a complex singlet and with a global $\U{}$ family symmetry. The fermion sector of this model is extended with 
a family of vector-like quarks which enhances CP violation. With the current study, we aim at exploring the generic vacuum structure and uncovering the features of the EWPT in this model relevant for cosmology. We show the existence of different phase transition patterns providing strong departure from thermal equilibrium. 
Most of these observations are not specific to the considered model and may generically be expected in other multi-Higgs extensions of the SM.}

\begin{document} 

\maketitle
\flushbottom

\section{Introduction}

According to Sakharov \cite{Sakharov:1967dj}, among the important prerequisites for efficient generation of the observed 
baryon asymmetry in the early universe are (i) the baryon number ($B$) violating processes, (ii) sufficient $C$ and $CP$ violation, 
and (iii) the presence of a strong first-order phase transition. The search for an adequate model that is capable to accommodate 
all three conditions is one of the most active research directions in physics beyond the Standard Model (SM).

In fact, the first condition, $B$-number violation, is already satisfied within the SM framework due to the so-called sphaleron boundary 
crossing\footnote{The baryon number current $j_B^\mu$ is not exactly conserved in the SM. Instead, it satisfies $\partial_\mu j_B^\mu 
= 3/(16 \pi^2) F_{\mu \nu}^a \tilde{F}_{\mu \nu}^a$ \citep{Dine:2003ax}.} \citep{Klinkhamer1984,Arnold1987}: thanks to the anomalous 
fermion number, a jump between two physically equivalent but topologically different vacua leads to the creation of three baryons. 
This is the main motivation behind the electroweak (EW) baryogenesis (EWBG) mechanism, one of the most theoretically attractive and 
phenomenologically verifiable scenarios capable of explaining the baryon asymmetry in the universe. For a detailed overview of the EWBG 
concepts and existing approaches, see e.g.~Refs.~\cite{Morrissey:2012db, White-book} and references therein.

While the Sakharov conditions (i) and (ii) are connected to the microscopic properties of the interactions in the underlying theory
and are responsible for creation of a local excess of baryons over antibaryons in the cosmological plasma, due to the CPT-theorem 
such an excess would be quickly washed out unless the plasma was away from thermal equilibrium at the corresponding energy scale.
A required departure from thermal equilibrium is realized e.g.~by means of the first-order cosmological phase transitions through the nucleation of vacuum bubbles of a new EW-broken phase. So, for the efficient EWBG one expects that the first-order EW 
phase transition (EWPT) happens at the characteristic energy scale of the sphaleron ($B$-violating) processes in the early universe.
These processes are active only outside the bubbles while inside them the Higgs field(s) get(s) 
non-vanishing vacuum expectation values (VEVs), thus, exponentially suppressing the $B$-violating reactions there. The criterion 
for washing out the sphaleron processes inside the bubbles in order for the created asymmetry not to be canceled by thermal 
equilibration at the surface reads
\begin{equation}
v_c/T_c \gtrsim 1 \,,
\label{sphaleron_washout}
\end{equation}
where $v_c$ is the Higgs VEV inside the bubbles (or square root of the sum of the Higgs VEVs squares in models with several Higgs doublet fields), 
and $T_c$ is the critical temperature of the transition. One refers to transitions where Eq.~(\ref{sphaleron_washout}) is satisfied as strong first-order phase transition, whereas others are denoted as weak transitions closed to a crossover.

The EWBG does not work in the conventional SM framework containing single Higgs doublet since it fails to satisfy the Sakharov conditions (ii) and (iii). 
Indeed, while the SM has some $C$ and $CP$ violation via the non-vanishing physical phase of the Cabibbo–Kobayashi– Maskawa (CKM) matrix, the latter 
predicts way too small baryon-to-entropy ratio compared to its observed value \cite{Xiao:2015tja,Gavela:1993ts,Konstandin:2003dx}. Moreover, in the SM, 
the Higgs mass of 125 GeV measured at the LHC \cite{Aad:2012tfa,Chatrchyan:2012xdj} implies that there is no first-order phase transition, only 
a crossover \cite{Rummukainen:1998as}, thus, a sufficient departure from thermal equilibrium is not realised. These are the standard arguments that 
motivate an extension beyond the SM having an enhanced $C$ and $CP$ violation and enabling a strong first-order phase transition at the EW scale.

The observed baryon asymmetry has inspired a lot of models beyond the SM being actively developed during the last three decades. Already some of the simplest 
SM extensions have successfully brought the missing ingredients. On general grounds, the only way to reach a first-order EW phase transition is to extend 
the scalar sector of the SM by means of additional Higgs (EW doublet and/or singlet) fields which may also introduce additional sources of $CP$ violation 
due to non-trivial phases between different Higgs VEVs. Also, already in such simplest SM extensions as the real-singlet \citep{Cline:2012hg,Li:2014wia,
Vaskonen:2016yiu,Beniwal:2017eik,Kurup:2017dzf} and complex-singlet \citep{Barger:2008jx,Jiang:2015cwa,Chiang:2017nmu,Chen:2017qcz} extended SM the strength of the EW phase transition 
can become sufficient for the EWBG. Thus, the possibility for an efficient EWBG with a strong first-order phase transition is among the most attractive features in multi-Higgs extensions of the SM.

Particularly important examples of such extensions is the Two-Higgs Doublet Model (THDM) \cite{Branco:2011iw} (and, consequently, NHDM for $N$ doublets) and 
Singlet-Extended SM \cite{Barger:2007im,Barger:2008jx,Costa:2014qga,Costa:2017gky}. There has been much work done on baryogenesis in THDM so far (see e.g. Refs.~\cite{Turok:1990zg,Turok:1991uc,
Funakubo:1993jg,Davies:1994id,Cline:1995dg,Laine:2000rm,Fromme:2006cm}) with the simplest source of dynamical $CP$ violation being the spatially varying phase of 
the top quark mass. An enhancement of $C$ and $CP$ violation in this type of models can be achieved via either complex Higgs VEVs \cite{Lee:1973iz,Lee:1974jb,
Weinberg:1976hu}, higher dimensional operators \cite{Bodeker:2004ws,Fromme:2006wx,Huang:2015izx,Balazs:2016yvi}, or by introducing  vector-like quarks 
\cite{Uesugi:1996rd,McDonald:1996uz,Chen:2015uza,Xiao:2015tja}.

In general, the vacuum structure of multi-Higgs extensions of the SM \citep{Ivanov:2017dad} is very rich and typically is very difficult to explore. For example, the most general scalar potential 
of the THDM contains 14 parameters and exhibits various $CP$-conserving, $CP$-violating and charge-violating minima (see e.g. Refs.~\cite{Ginzburg:2010wa,
Branco:2011iw,Dorsch:2013wja,Basler:2016obg,Basler:2017uxn}). Also, it is often a rather non-trivial task to find the key set of physical parameters of the potential by identifying 
and eliminating those parameters that emerge due to a freedom in choice of the basis. In practice, however, one often employs various simplifying assumptions 
such as by considering scenarios without a spontaneous $CP$ breaking in the Higgs sector and by imposing certain global (continuous or discrete) symmetries 
either in the Higgs sector only or in both Higgs and Yukawa sectors. One particularly simple scenario is the so-called Inert Doublet Model (IDM) \cite{Ma:2006km,
Barbieri:2006dq,Majumdar:2006nt,LopezHonorez:2006gr} where an additional Higgs doublet $H_2$ is $\mathbb{Z}_2$-symmetric $H_2\leftrightarrow -H_2$. 
In this simple model, one acomplishes both the strong first-order EWPT and a suitable scalar Dark Matter candidate in one of the $H_2$ components 
simultaneously (see e.g.~Refs.~\cite{Chowdhury:2011ga,Borah:2012pu} and references therein).

The additional symmetries not only simplify the study of multi-Higgs models but also enable to avoid potentially dangerous interactions 
leading e.g.~to large (tree-level) FCNCs that are present in a generic NHDM \citep{Ivanov:2017dad}. Indeed, one of the ways to suppress FCNCs in a multi-scalar doublet 
model is to impose an {\it ad hoc} discrete symmetry on the couplings of quarks and leptons to the two doublets following Glashow 
and Weinberg \cite{Glashow:1976nt}. Another sufficiently generic way to avoid large FCNCs is the Minimal Flavor Violation (MFV) hypothesis 
\cite{Chivukula:1987py,Hall:1990ac,DAmbrosio:2002vsn} (see also Ref.~\cite{Isidori:2012ts}) providing a consistency with the flavor physics constraints 
without imposing any discrete symmetries. The EWBG in the MFV THDM has been thoroughly discussed in Ref.~\cite{Cline:2011mm} where it was shown 
to be a challenge to get a large enough baryon asymmetry in this model.

Most of the existing models beyond the SM which aim at accommodating the EWBG mechanism typically predict new states at the EW scale, while the simplest 
models suffer from a lack of experimental evidence, thus, further limiting their relevance \cite{Cline2017}. This is why, since the pioneering works of the 80's, 
more and more complicated models have been considered that may not only accommodate the EWBG but also explain other observed features such as strong hierarchies in the SM fermion spectra etc. A large majority of these models are built in a bottom-up approach, i.e.~inspired by the SM, often extended 
in a minimal way. However, the existing phenomenological information which would have point out a correct pathway towards a consistent SM extension is rather 
scarse and insufficient, with no significant (or a few yet uncertain) observed deviations from the SM in collider measurements still leaving us with a landscape of 
possibilities.

The aim of our work is to show that Grand Unified Theories (GUTs), besides making a big step towards unification of fundamental interactions, may also be able 
to accommodate an efficient EWBG mechanism. While a direct access to physics of the GUTs is beyond the reach of laboratory measurements, they can be indirectly probed 
through their low-energy (e.g.~collider and cosmological) implications. Our goal in this paper is to explore the EW phase transition in a SM-like effective 
field theory (EFT) inspired by such an attractive high-scale GUT as the Higgs-unified supersymmetric (SUSY) trinification (SHUT) with the global $\SU{3}{F}$ family symmetry 
proposed recently in Refs.~\citep{Camargo-Molina:2016yqm,Camargo-Molina:2017kxd}. Here, we perform a case study of the minimal low-energy EFT of the SHUT 
model given by an effective THDM plus a singlet scalar (THDSM) augmented with a sector of light vector-like quarks (VLQ) and possessing a family-non-universal global $\U{}$ 
symmetry in Higgs and fermion sectors (for other realisations of the THDSM scenario, see e.g.~Refs.~\citep{Alanne:2016wtx,Banik:2014cfa,Blinov:2015sna,Chao:2017vrq}). 
The basic features of the EW phase transition in this model are generic and can be attributed to a variety of other multi-Higgs SM extensions.

The paper is organised as follows. In Sect.~\ref{sec:MD}, the basic elements and features of the considering THDSM are presented. In Sect.~\ref{sec:structure},
we discuss the vacuum structure and classify the phase transitions in this model at tree level considering both the limit of vanishing singlet scalar and the complete 
$\mathbb{Z}_2$-symmetric version of the THDSM. Sect.~\ref{sec:EWPT} is devoted to a discussion of the EWPT accounting for the complete one-loop (Coleman-Weinberg 
plus thermal) correction in the effective potential. In Sect.~\ref{Sec:numan}, a thorough numerical analysis of distinct features of the EWPT in the considering model
has been performed. Final remarks and a summary of the main results are given in Sect.~\ref{sec:conclusions}.

\section{SHUT-inspired THDSM scenario} 
\label{sec:MD}

A novel set of trinification-based GUTs (or T-GUTs, in what follows) was proposed by some of the authors in both SUSY 
\citep{Camargo-Molina:2016yqm,Camargo-Molina:2017kxd} and non-SUSY realisations \cite{Camargo-Molina:2016bwm}. The minimal field content of each of these realisations 
transforms according to the gauge $\SU{3}{C} \times \SU{3}{L} \times \SU{3}{R}$ and global-family\footnote{It is well known that global continuous symmetries 
are problematic. As already discussed in Refs.~\citep{Camargo-Molina:2016yqm,Camargo-Molina:2017kxd} we take this as a simplistic approach of 
a complete anomaly-free model where the family symmetry is also local. We argue that the low-scale properties of such a complete model and its simplified 
version are identical in the limit of small $\SU{3}{F}$ gauge couplings. A full analysis is however mandatory and is subject of a forthcoming work.} $\SU{3}{F}$ 
symmetries, which are sequentially broken down to the SM gauge group supplemented with an additional family symmetry. Such a sequential multi-stage breaking may 
be in principle achieved e.g.~by means of the renormalization group (RG) evolution of the theory parameters as was demonstrated in the case of the spontaneous 
Left-Right (LR) symmetry breaking in the non-SUSY T-GUT in Ref.~\cite{Camargo-Molina:2016bwm}.

The non-SUSY model, with limited possibilities for realistic VEV settings, yields one generation of SM quarks and leptons massless\footnote{This is a result of 
accidental symmetries that emerge due to the interplay between the gauge and family ones.} after the EW symmetry breaking (EWSB) and contains a large amount 
of fine-tuning. Notably, the low-scale limit of the SHUT model does not suffer from any of such issues. This is a direct consequence of a larger field content and of the fact 
that a soft-SUSY breaking sector is naturally decoupled from the GUT scale determined by the mass terms in the superpotential. Indeed, a richer scalar field content 
of the SHUT model offers a possibility for an appropriate rank reduction, and the existence of a soft SUSY breaking scale well below the GUT scale stabilizes the hierarchy. 
Note, however, that the soft-SUSY breaking mass terms in the SHUT model do not necessarily need to be as small as the EW or even a few $\rm TeV$ scale as does happen e.g.~in the MSSM. Instead, they sit at an intermediate scale sufficiently decoupled from the EW one such that the low-energy (SM-like) theory contains only a subset of the original heavy\footnote{We denote heavy states as those that are above or at the same order as the intermediate symmetry breaking scales.} states. While a complete 
RG analysis of the T-GUT EFT scenarios accounting for their matching to the high-scale theory is beyond the scope of the present article, here we will focus on a simple minimal 
low-scale scenario inspired by the SHUT model and study its relevance for the EWPT.
\begin{table}[h!]
\centering
\begin{tabular}{|c|c|c|c|c|c|} \hline
& Particle		    & $\SU{3}{C}$ 
& $\SU{2}{L}$ & $\U{Y}$ & $\U{T}$ \\\hline
\multirow{3}{*}{Scalars}
& $H_1$    			&  \bf{1}   &  \bf{2}  &   1	 & 1  \\
& $H_2$    			&  \bf{1}   &  \bf{2}  &   1	 & 5  \\
& $\Phi$    	    	&  \bf{1}   &  \bf{1}  &   0	 & 4 \\
\hline
\multirow{2}{*}{Left-handed quarks}
& $Q_L^1$    		&   \bf{3}  &  \bf{2}  &   1/3   & 3   \\
& $Q_L^{2,3}$		&   \bf{3}  &  \bf{2}  &   1/3   & $-1$  \\
\hline
\multirow{5}{*}{Right-handed quarks}
& $u_R^1$    		&   \bf{3}$^*$  &  \bf{1}  &   $-4/3$  & 0   \\
& $u_R^{2,3}$		&   \bf{3}$^*$  &  \bf{1}  &   $-4/3$  & $-4$  \\
& $d_R^1$			&   \bf{3}$^*$  &  \bf{1}  &   2/3   & 6   \\
& $d_R^2$ 		&   \bf{3}$^*$  &  \bf{1}  &   2/3   & $-2$  \\
& $d_R^3$ 		&   \bf{3}$^*$  &  \bf{1}  &   2/3   & 2   \\
\hline
\multirow{2}{*}{Vector-like quarks}
& $D_L$    		&   \bf{3}  &  \bf{1}  &   $-2/3$  & $-2$  \\
& $D_R$			&   \bf{3}$^*$  &  \bf{1}  &   2/3   & 2   \\
\hline
\end{tabular}
\caption{\label{tab:ingredients} 
\it Quantum numbers for the scalar and quark fields in the SHUT-inspired THDSM.}
\end{table}

As shown in Ref.~\citep{Camargo-Molina:2017kxd}, the field content of the low-scale EFT transforms according to the symmetry
\begin{equation}
\label{eq:group}
\SU{3}{C} \times \SU{2}{L} \times \U{Y} \times \U{T}\,,
\end{equation}
where the generator of the $\U{T}$ symmetry is a remnant of the original $\SU{3}{F}$ family symmetry. In this article, we consider the minimal low-scale EFT limit of the SHUT model where the scalar sector is composed of two Higgs doublets and a complex singlet. As discussed in \citep{Camargo-Molina:2017kxd}, with a minimum of three-Higgs-doublets the Cabibbo mixing is naturally emergent at tree-level in the CKM matrix with small deviations from unitarity resulting from quantum effects. While a THDM scenario can still provide a realistic mass spectrum, the CKM mixing can only be realized upon tuning of theory parameters. On the other hand, the scalar potential of a model with two doublets is rather simpler than its 3HDM extension and both share common features with respect to the EWPT. 

In addition to the SM fermions, the model also contains one generation of vector-like quarks. Note, in the SHUT model there are three families of vector-like quarks. 
However, on the T-GUT symmetry breaking path down to the SM gauge symmetry, two of them become as heavy as the intermediate scale and are absent in the low-scale EFT spectrum. In what follows, 
in practical computations of the effective potential we will only account for the largest top and vector-like quark contributions such that the details and a discussion 
of the lepton sector in the considering model are omitted, for simplicity. The relevant field content and quantum numbers of the scalar and quark sectors in 
the considering SHUT-inspired THDSM are specified in Tab.~\ref{tab:ingredients}.

The most general scalar potential allowed by the symmetry \eqref{eq:group} in our THDSM scenario reads
\begin{align} 
\label{eq:VH}
\begin{aligned}
V\left[H_1,H_2,\Phi \right] =& 
m_1^2 H_1^{\dagger} H_1 
+ m_2^2 H_2^{\dagger} H_2 
+ m_s^2 \Phi \Phi ^*
+ \frac{\lambda_1}{2} \(H_1^{\dagger} H_1 \)^2
+ \frac{\lambda_2}{2} \(H_2^{\dagger} H_2 \)^2 
\\
+& \frac{\lambda_s}{2} (\Phi \Phi ^*)^2
+ \lambda_{3} (H_1^{\dagger} H_1) (H_2^{\dagger} H_2) 
+ \lambda_{s1} (H_1^{\dagger} H_1) (\Phi \Phi^*)
\\
+& \lambda_{s2} (H_2^{\dagger} H_2) (\Phi \Phi^*)
+ \lambda_{3}' (H_1^{\dagger} H_2) (H_2^{\dagger} H_1) \\
+& A_{HH\Phi} \( e^{i\theta} H_1^{\dagger} H_2 \Phi^* + e^{-i\theta} H_2^{\dagger} H_1 \Phi \)
\end{aligned}
\end{align}
with $H_1$, $H_2$ and $\Phi$ defined as
\begin{align}
\begin{aligned}
H_1 &= \frac{1}{\sqrt{2}} 
\begin{pmatrix} 
\chi_1 + i \chi_1' \\ 
\phi_1 + h_1 + i \eta_1 
\end{pmatrix}\,,
\end{aligned} 
\qquad
\begin{aligned}
H_2 = \frac{1}{\sqrt{2}} 
\begin{pmatrix} 
\chi_2 + i \chi_2' \\ 
\phi_2 + h_2 + i \eta_2 
\end{pmatrix}
\end{aligned} 
\end{align} 
and
\begin{equation}
\Phi = \tilde{\phi}_s +  \dfrac{1}{\sqrt{2}} \( S_R + i S_I \)\,,
\end{equation}
with $h_1$, $\eta_1$, $\chi_1$, $\chi_1'$, $h_2$, $\eta_2$, $\chi_2$, $\chi_2'$, $S_R$, $S_I$ real scalars. As usual, $h_1$ and $h_2$ parametrize quantum fluctuations about the $\phi_1$ and $\phi_2$ real classical field configurations, whereas the $S_R$ and $S_I$ components represent the fluctuations about the complex classical field $\tilde{\phi}_s$.
Note that, in the SHUT model \cite{Camargo-Molina:2017kxd}, the $H_1^{\dagger} H_2 \Phi^* + {\rm h.c.}$ interactions do not originate from the initial
tree-level SHUT Lagrangian and thus can only be generated radiatively. It is then instructive to imply a regime where the coupling $A_{HH\Phi}$ is naturally 
suppressed in comparison to the remaining quartic couplings. Motivated by this observation, if one adopts a vanishing $A_{HH\Phi} \sim 0$ one recovers 
three approximate accidental $\mathbb{Z}_2$ symmetries w.r.t. $H_1 \leftrightarrow -H_1$, $H_2 \leftrightarrow -H_2$ and $\Phi \leftrightarrow -\Phi$ 
separate transformations in the scalar potential \eqref{eq:VH}, respectively. 

The first two $\mathbb{Z}_2$ symmetries w.r.t. switching the sign of each Higgs doublet are analogous to $\mathbb{Z}_2$ in the IDM \cite{Ma:2006km,Barbieri:2006dq,
Majumdar:2006nt,LopezHonorez:2006gr}, but with an additional singlet $\Phi$ and an extra $\U{T}$ symmetry in the scalar potential (for analysis of the EW phase 
transitions in the IDM, see Ref.~\citep{Blinov:2015vma}). In principle, a small but non-vanishing $A_{HH\Phi}$ could be responsible for transmitting the $\U{T}$ 
breaking effect (by means of a VEV in $\Phi$) to the rest of the scalar sector. For the purposes of this work and for simplicity, below we discuss the exact 
$\mathbb{Z}_2$ limit of this model unless noted otherwise.

The third $\mathbb{Z}_2$ symmetry w.r.t.~$\Phi \leftrightarrow -\Phi$ gives rise to the existence of a scalar Dark Matter (DM) candidate which could, in principle,
be destabilized by a non-vanishing $A_{HH\Phi}$ coupling. Such a metastable DM candidate is a characteristic prediction of the considering THDSM model that could
be studied in more detail elsewhere.

The physical Goldstone state, the familon, emerging due to the global family $\U{T}$ symmetry breaking can become a pseudo-Goldstone one acquiring a very small 
mass due to nonperturbative interactions with the QCD vacuum, namely, with the gluon condensate. Indeed, the latter provides an additional contribution to the masses
of scalar bosons through its interactions with the Higgs condensate. This contribution is negligibly small for massive scalars but is significant for the massless ones
such as the Goldstone familon inducing its tiny pseudo-Goldstone mass, $m_{\rm F}$. An order-of-magnitude estimate for the upper bound on the familon mass 
generated by such nonperturbative interactions reads
\begin{eqnarray}
m_{\rm F}^2 < \frac{\langle 0| \frac{\alpha_s}{\pi} G^a_{\mu\nu}G_a^{\mu\nu} |0 \rangle}{\Lambda_{\rm F}^2} \sim 1\,{\rm MeV}^2
\end{eqnarray}
for the global family $\U{T}$ symmetry breaking scale, $\Lambda_{\rm F}$, taken to be at the EW scale, 
i.e.~$\Lambda_{\rm F}\sim 100$ GeV, for simplicity. $\alpha_s$ is the strong coupling constant in QCD, and $G^a_{\mu\nu}$ is the standard 
gluon stress-tensor. Provided such small typical values for the familon mass, in practical calculations it has been neglected compared 
to the masses of all other particles.

The Yukawa interactions and fermion bilinear terms in this model read
\begin{align} 
\label{eq:LY}
\begin{aligned}
\mathcal{L}_\mathrm{Y} =\; & 
\sum_{i=2}^3 Y_{1\,i}^{u} Q_\mathrm{L}^1 H_1 u_\mathrm{R}^{i} 
+ Y_{1\,2}^{d} Q_\mathrm{L}^1 H_1^{\dagger} d_\mathrm{R}^2 
+ \sum_{i=2}^3 Y_{i\,1}^{u} Q_\mathrm{L}^{i} H_1 u_\mathrm{R}^1 
+ \sum_{i=2}^3 Y_{i\,3}^{d} Q_\mathrm{L}^{i} H_1^{\dagger} d_\mathrm{R}^3
\\
&
+ Y_{1\,3}^{d}Q_\mathrm{L}^1 H_2^{\dagger} d_\mathrm{R}^3
+ \sum_{i,j=2}^3Y_{i\,j}^{u} Q_\mathrm{L}^{i} H_2 u_\mathrm{R}^{j} 
+ \sum_{i=2}^3 Y_{i\,1}^{d} Q_\mathrm{L}^{i} H_2^{\dagger} d_R^1 
+ Y_{1\,4}^{d} Q_\mathrm{L}^1 H_2^{\dagger} D_\mathrm{R} 
\\
&
+ \sum_{i=2}^3 Y_{i\,4}^{d} Q_\mathrm{L}^{i} H_1^{\dagger} D_\mathrm{R} 
+ m_D D_\mathrm{L} D_\mathrm{R} + m_{dD} D_\mathrm{L} d_R^3 
\\
&
+ Y_{\Phi D}^{1}\Phi^* D_\mathrm{L} d_R^1 + Y_{\Phi D}^{2} \Phi D_\mathrm{L} d_R^2
+ {\rm h.c.} \,.
\end{aligned}
\end{align}

Given the choice of light Higgs doublets and fixing the mixing among the original components of $\SU{3}{R}$ down-type quarks in the original 
SHUT model \citep{Camargo-Molina:2017kxd}, the only Yukawa interactions already present at tree level are $Q_\mathrm{L}^2 H_2 u_\mathrm{R}^{3}$ and 
$Q_\mathrm{L}^3 H_2 u_\mathrm{R}^{2}$ for the up sector, and $Q_\mathrm{L}^2 H_1^\dagger d_\mathrm{R}^{3}$ for the down sector. This can be seen 
by comparing our Yukawa interactions in Eq.~\eqref{eq:LY} with the leading ones given in Eq.~(C.20) of Ref.~\citep{Camargo-Molina:2017kxd}. As a consequence 
of this choice, the $Y^u_{2\,3}$, $Y^u_{3\,2}$ and $Y^d_{2\,3}$ are expected to be larger than the remaining loop-generated Yukawa couplings 
(just as $m_D \gg m_{dD}$ due to the same reason). We will then consider the leading tree-level Yukawa couplings at the EW scale to be of the order 
\begin{align}
\label{eq:leadingY}
\mathcal{O}(10^{-2}) \leq Y^u_{2\,3},\,Y^u_{3\,2},\,Y^d_{2\,3} \leq \mathcal{O}(1)\,.
\end{align}
Besides, in practical calculations we consider the VLQ decoupling limit, $|m_D| \gg v_{1,2}$. Keeping only the dominant Yukawa couplings in the mass forms,
the physical quark masses simplify to
\begin{align}
\label{eq:Mq-simp}
\begin{aligned}
m_u^2 &= m_d^2 = m_s^2 = 0\,,  \\
m_c &\simeq \frac{v_2}{\sqrt{2}} |Y_{2\,3}^u|\,, \\
m_t &\simeq \frac{v_2}{\sqrt{2}} |Y_{3\,2}^u| \,,\\
\end{aligned} \qquad
\begin{aligned}
m_b &\simeq \frac{v_1}{\sqrt{2}} |Y_{2\,3}^d|\,, \\
m_D' &\simeq \frac{1}{\sqrt{2}} |m_D|\,.
\end{aligned}
\end{align}
With such simple relations we can obtain approximate values for the leading Yukawa couplings and for the Higgs VEVs. In particular, using the PDG data for quark masses \citep{Patrignani:2016xqp}, $m_c = 1.27~\mathrm{GeV}$, $m_b = 4.18~\mathrm{GeV}$ and $m_t = 173.21~\mathrm{GeV}$, we can choose an example point $|Y_{2\,3}^d| = 0.035$ from where we fix the values the VEVs and of the $|Y_{3\,2}^u|$ and $|Y_{2\,3}^u|$ Yukawa couplings as
\begin{align}
\label{eq:benchmark}
\begin{aligned}
|Y_{2\,3}^d| = 0.035 \Rightarrow
\begin{cases}
v_1 = 168.90~\mathrm{GeV} \,, \\
v_2 = 178.86~\mathrm{GeV}
\end{cases} \Rightarrow
\begin{cases}
|Y_{3\,2}^u| = 1.37 \,, \\
|Y_{2\,3}^u| = 0.01 \,,
\end{cases}
\end{aligned}
\end{align}
consistent with Eq.~\eqref{eq:leadingY}. Note that if we increase the value of $|Y_{2\,3}^d|$ it results in a decrease of $|Y_{2\,3}^u|$ that rapidly becomes too small for the estimated values of the leading Yukawa couplings in Eq.~\eqref{eq:leadingY}. As the freedom in Eq.~\eqref{eq:Mq-simp} is limited, we consider that Eq.~\eqref{eq:benchmark} is representative, but by no means unique, of the type of structure that we expect to obtain in the current low-scale limit of the SHUT model. In our numerical analysis of the EWPTs below, we allow the VEVs $\{v_1,v_2\}$ to vary such that they satisfy the EW $v^2=v_1^2+v_2^2$ constraint and that the corresponding tree-level Yukawa couplings fulfill Eq.~\eqref{eq:leadingY}. Besides, the contributions of the lightest flavors $u,d,c,s$ and $b$ to the effective potential will be neglected keeping only the dominant contribution from the top quark while the VLQ provides a nearly field-independent contribution to the effective potential and thus can be omitted (for more details see below).
              
\section{Vacuum structure and phase transitions at tree level}
\label{sec:structure}

While the Higgs sector of the SM only allows for a second-order EW phase transition in the early universe, a typically rich vacuum structure of 
its multi-scalar extensions offers novel non-trivial possibilities for the EWSB via first-order transitions. In this section, as a first step, we focus 
our attention on the tree-level scalar potential of the SHUT-inspired THDSM model given by Eq.~(\ref{eq:VH}), systematically studying its 
vacuum configurations, (meta)stability conditions, its phase diagram as well as possible transitions between the different vacua.

\subsection{The limit of vanishing scalar singlet}
\label{Sec:THDM}

Let us start by investigating a simpler THDM limit of the considering THDSM model. In particular, taking $\Phi \equiv 0$, the tree-level potential 
\eqref{eq:VH} with the doublets $H_{1,2}$ replaced by the translation-invariant classical background fields $\phi_1$ and $\phi_2$ takes the simple form
\begin{equation}
\label{eq:2HDM}
V_{\rm tree}[\phi_1,\phi_2] = \frac{m_1^2}{2} \phi_1^2
+ \frac{m_2^2}{2} \phi_2^2
+ \frac{\lambda_1}{8} \phi_1^4
+ \frac{\lambda_2}{8} \phi_2^4
+ \frac{\lambda_{12}}{4} \phi_1^2 \phi_2^2
\end{equation}
with $\lambda_{12} \equiv \lambda_{3}+\lambda_{3}'$. The conditions that warrant boundedness from below (BFB) read
\begin{align}
\label{eq:UFB1}
\begin{aligned}
\lambda_1, \lambda_2 > 0 \,, \quad
\lambda_{12} > - \sqrt{\lambda_1 \lambda_2}\,.
\end{aligned}
\end{align}
In what follows, it will be convenient to employ the shorthand notations
\begin{align}
\begin{aligned}
C_{ij} &\equiv m_i^2 \lambda_{ij} - m_j^2 \lambda_i \,,  \\
L_{ij} &\equiv \lambda_i \lambda_j - \lambda_{ij}^2\,,
\end{aligned}
\end{align}
where $i,j \in \{1,2\}$, $\lambda_{ij} = \lambda_{ji}$, $C_{ij} \neq C_{ji}$.

Since the potential \eqref{eq:2HDM} solely depends on the real fields $\phi_{i}$, we derive its extremum conditions (ECs) 
by solving the minimization equations $\mathrm{d}V/\mathrm{d} \phi_{i} = 0$, whose solutions provide the real and positive VEVs that we denote as 
$(v_1,v_2)$ in terms of the model parameters. We are therefore left with four possible extrema: $(0,0)$, $(v_1,0)$, $(0,v_2)$, $(v_1,v_2)$. Given that 
the potential is form-invariant w.r.t. interchange $\phi_1 \leftrightarrow \phi_2$, we can focus on the three possibilities $(0,0)$, $(v_1,0)$ and $(v_1,v_2)$, 
without any loss of generality. In what follows, we provide for each of the phases the value of the potential $V^{\rm ext}_{\rm tree}$, the ECs and 
the mass spectrum ${\rm Sp}(M)$, with $M=M[\phi_1,\phi_2]$ being the Hessian matrix\footnote{Note that in the current section we are only interested in studying 
the Higgs vacuum of the theory. Therefore, the Hessian matrix is calculated only in terms of the classical background fields. Numerical analysis of the full 
spectrum for the complete potential will be done in the following sections below.}.
\begin{itemize}
\item {\bf $(0,0)$ case:}
\begin{eqnarray}
V^{\rm ext}_{\rm tree} = 0\,, \qquad {\rm Sp}(M) = \left\{m_1^2\,, m_2^2 \right\} \,.
\end{eqnarray}
\item {\bf $(v_1,0)$ case:}
\begin{eqnarray}
V^{\rm ext}_{\rm tree} = -\frac{m_{1}^4}{2 \lambda _{1}} \,, \quad {\rm Sp}(M) =
\left\{-2 m_1^2\,, \frac{-C_{12}}{\lambda_1}\right\} \,, \quad {\rm EC}: \;\;  
v_1 = \sqrt{-2 m_1^2/ \lambda_1} \,.
\end{eqnarray}
Note that in this case there is no mixing between different scalars.
\item {\bf $(v_1,v_2)$ case:}
\begin{eqnarray} \nonumber
V^{\rm ext}_{\rm tree} = \frac{m_2^4 \lambda _1 + m_1^4 \lambda _2 - 2 m_1^2 m_2^2 \lambda_{12}}{-2 L_{12}}\,, \quad
{\rm EC}: \Big\{ v_1 = \sqrt{\frac{2 C_{21}}{L_{12}}} \,, \;\; v_2 = \sqrt{\frac{2  C_{12}}{L_{12}}} \Big\} \,, \\
{\rm Sp}(M) = \frac{1}{L_{12}} \left( \lambda_1 C_{21} + \lambda_2 C_{12}
\mp \sqrt{ ( \lambda_1 C_{21} + \lambda_2 C_{12} )^2 - 4C_{12} C_{21} L_{12}} \right) \,.
\label{eq:2HDMcond}
\end{eqnarray}
\end{itemize}
As always, if the eigenvalues of $M$ are all strictly positive we have a minimum but not necessarily the global one. Following the standard terminology, 
we then refer to a local minimum as a metastable one as opposed to the global stable minimum.

\begin{figure}[!ht]
\centering
\includegraphics[width=\linewidth]{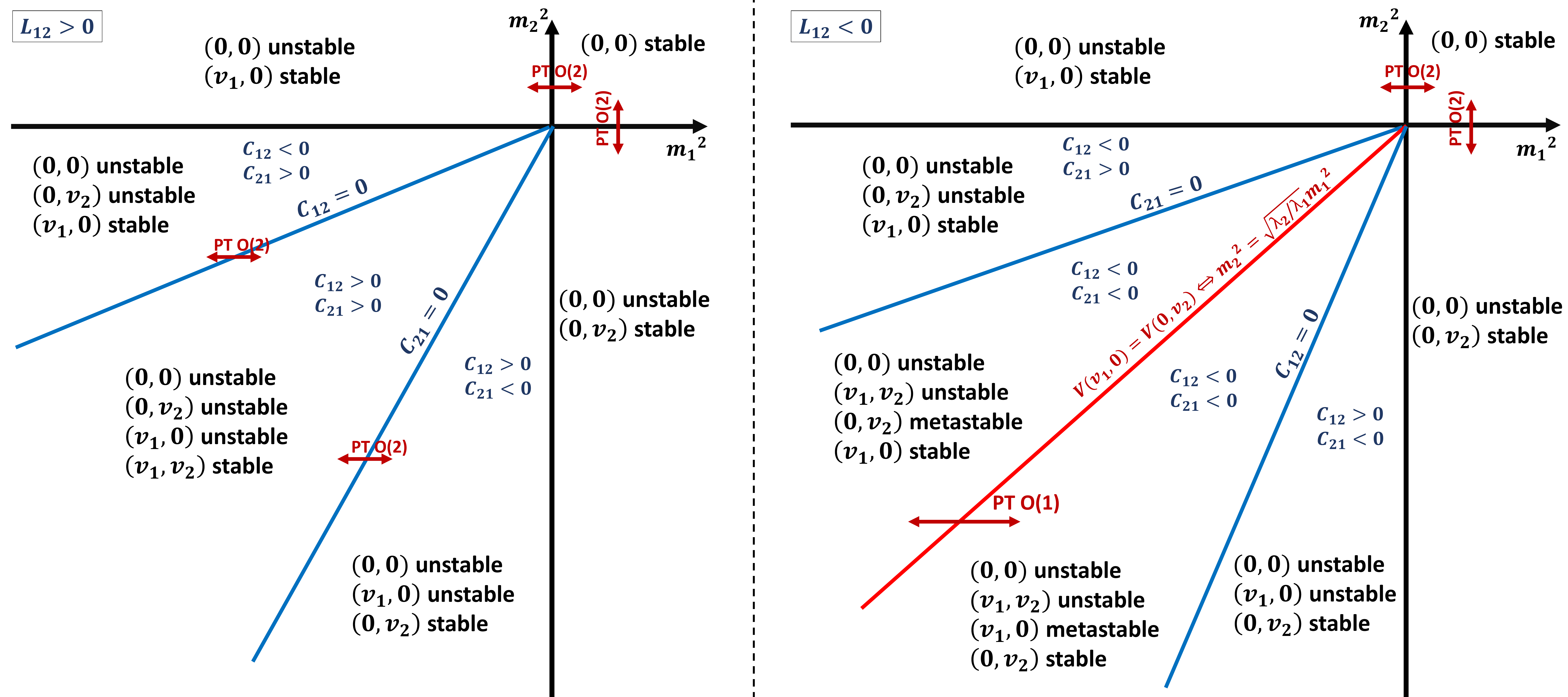}
\caption{\label{fig:phase_diagramme_2HDM} 
\it Phase diagram for the THDM limit of the considering THDSM scenario. Here, \textbf{stable} means that the minimum is global whereas \textbf{metastable} 
that the minimum is local. The red arrows indicate second-order $O(2)$ or first-order $O(1)$ phase transitions. In the LM scenario, $L_{12} > 0$ (left), 
there are only second-order phase transitions at tree level, whereas in the HM scenario, $L_{12} < 0$ (right), there is a first-order phase transition realised 
across the red line.}
\end{figure}

The structure of the scalar potential \eqref{eq:2HDM} allows for a mixing between $\phi_1$ and $\phi_2$ whose strength if given by the $\lambda_{12}$ 
(and thus $L_{12}$) parameter. Consequently, we have two scenarios in the phase diagram, as displayed in Fig.~\ref{fig:phase_diagramme_2HDM}:
\begin{itemize}
\item The low-mixing (LM) scenario with $L_{12} > 0$ in the following three cases:
\begin{enumerate}
\item $C_{12} > 0$ and $C_{21} < 0$,
\item $C_{12} < 0$ and $C_{21} > 0$,
\item $C_{12} > 0$ and $C_{21} > 0$. 
\end{enumerate} 
For the first two cases, the system either ``sits'' above the upper blue line, i.e.~in the $(v_1,0)$-minimum, or below the lower blue line corresponding to 
the $(0,v_2)$-minimum, and such phases cannot coexist simultaneously and are both stable. For the third case, the $(v_1,v_2)$ minimum exists between 
the blue lines and is the only stable solution.
\item The high-mixing (HM) scenario with $L_{12} < 0$ in the following three cases:
\begin{enumerate}
\item $C_{12} < 0$ and $C_{21} > 0$,
\item $C_{12} > 0$ and $C_{21} < 0$,
\item $C_{12} < 0$ and $C_{21} < 0$. 
\end{enumerate} 
While for the first two cases the $(v_1,0)$ and $(0,v_2)$ minima cannot coexist similarly to the LM scenario, they become simultaneously compatible 
for the third case between the blue lines such that the stability of the extrema $(v_1,0)$ and $(0,v_2)$ depends on the sign of the $C_{ij}$ parameters. 
When both minima are present, the global minimum is determined by the lowest value of the vacuum potential. In the third case, there is also an unstable 
extremum with both non-zero VEVs $(v_1,v_2)$ which implies
\begin{equation}
\lambda_{1} C_{21} + \lambda_{2} C_{12}
+ \sqrt{ \( \lambda_1 C_{21} +\lambda_2 C_{12} \)^2 - 4 C_{12} C_{21} L_{12}} > 0 \,,
\label{instability_v1v20_0}
\end{equation}
yielding a negative eigenvalue in Eq.~\eqref{eq:2HDMcond}.
\end{itemize}
Note, for both LM and HM scenarios the $(0,0)$ extremum is stable in the first quadrant with $m_1^2,\,m_2^2 > 0$ as is seen in 
Fig.~\ref{fig:phase_diagramme_2HDM}.

The above properties of the LM and HM scenarios are summarized in the phase diagrams in Fig.~\ref{fig:phase_diagramme_2HDM}, left and right 
panels, respectively\footnote{These diagrams are similar to the ones drawn in Ref.~\citep{Ginzburg:2010wa} for the case of the IDM, where the HM scenario
corresponds to the case $R>1$, and the LM scenario -- to the case $-1<R<1$, with $R$ being the resulting mixing parameter of the theory.}. 

In the HM scenario (right panel in Fig.~\ref{fig:phase_diagramme_2HDM}), a first-order phase transition occurs when the system crosses the phase border 
$m_2^2 = \sqrt{\lambda_2/\lambda_1}\, m_1^2$ which is represented by a red line. Such a transition is basically the inversion of two different stable minima, 
i.e.~$(v_1,0) \leftrightarrow (0,v_2)$. In the LM scenario, there are only second-order phase transitions similar to the crossover occurring in the SM. 
Even if such transitions may become first order at one loop due to the effect of cubic terms in the quantum and thermal corrections, they are less likely to be strong. 
It is then a representative feature of the considering model having a first-order transition readily at tree-level, which, as we will see, may well be a strong one 
without relying on quantum effects. We note also that a transition between the LM and HM scenarios is mathematically expected when $L_{12}$ evolves 
from positive (negative) to negative (positive) values. Furthermore, keeping the squared masses fixed, i.e.~considering a given point in the phase diagrams in 
Fig.~\ref{fig:phase_diagramme_2HDM}, a first-order phase transition may happen in the LM scenario when the blue lines $C_{12} = 0$ and $C_{21} = 0$ 
move due to variations of $L_{12}$ between positive and negative values. However, thermal effects do not play a significant role in the evolution of the quartic 
couplings making this type of transitions insignificant unless a higher degree of fine tuning is concerned.

\subsection{The phase diagram of the complete THDSM potential} 
\label{Sec:complete}

Adding the complex singlet scalar field to the previously discussed toy-model THDM potential \eqref{eq:2HDM}, the complete tree-level THDSM potential 
in terms of translation-invariant classical fields reads
\begin{align}
\label{eq:2HDMS}
\begin{aligned}
V[\phi_1, \phi_2, \phi_s ] =& 
  \frac{m_1^2}{2} \phi_1^2
+ \frac{m_2^2}{2} \phi_2^2 
+ m_s^2 \phi_s^2
+ \frac{\lambda_1}{8} \phi_1^4
+ \frac{\lambda_2}{8} \phi_2^4  \\
+& \frac{\lambda_s}{2} \phi_s^4
+ \frac{\lambda_{12}}{4} \phi_1^2 \phi_2^2 
+ \frac{\lambda_{s1}}{2} \phi_1^2 \phi_s^2
+ \frac{\lambda_{s2}}{2} \phi_2^2 \phi_s^2 \,,
\end{aligned}
\end{align}
where $\phi_s\equiv|\tilde{\phi}_s|=\sqrt{[{\rm Re}\tilde{\phi}_s]^2+[{\rm Im}\tilde{\phi}_s]^2}$.
A straightforward extension of the BFB conditions found earlier for the case of vanishing singlet in Eq.~(\ref{eq:UFB1})
can be summarized as follows
\begin{align}
\lambda_1, \lambda_2, \lambda_s > 0 \,, \qquad
\lambda_{12} > - \sqrt{\lambda_1 \lambda_2}\,, \qquad
\lambda_{s1} > - \sqrt{\lambda_1 \lambda_s}/2\,, \qquad
\lambda_{s2} > - \sqrt{\lambda_2 \lambda_s}/2\,.
\end{align}
Now, there are eight possible extrema with the following structure: (0,0,0), ($v_1$,0,0), (0,$v_2$,0), (0,0,$v_s$), ($v_1$,$v_2$,0), ($v_1$,0,$v_s$), (0,$v_2$,$v_s$), 
($v_1$,$v_2$,$v_s$), where $v_1$, $v_2$, $v_s$ are the classical background fields $\phi_1$, $\phi_2$ and $\phi_s$ at zeroth temperature $T=0$, respectively. Due to the symmetries of the potential, this set can be reduced down to four distinct cases with the following properties: 
\begin{itemize}
\item {\bf $(0,0,0)$ case:}
\begin{eqnarray}
V^{\rm ext}_{\rm tree} = 0\,, \qquad {\rm Sp}(M) = \left\{m_1^2\,,m_2^2\,,2m_s^2 \right\} \,.
\end{eqnarray}
\item {\bf ($v_1$,0,0) case:}
\begin{eqnarray}
V^{\rm ext}_{\rm tree} =  -\frac{m_1^4}{2\lambda_1} \,, \quad {\rm EC}: \;\; v_1 = \frac{-2 m_1^2}{\lambda_1} \,, \quad
{\rm Sp}(M) = \left\{-2 m_1^2\,, \frac{-C_{12}}{\lambda_1}\,,\frac{-2 C_{1s}}{\lambda_1}\right\}
\end{eqnarray}
\item {\bf ($v_1$,$v_2$,0) case:}
\begin{eqnarray} \nonumber
V^{\rm ext}_{\rm tree} &=& - \frac{m_2^4 \lambda_1 + m_1^4 \lambda_2 - 2 m_1^2 m_2^2 \lambda_{12}}{2L_{12}} \,, \quad 
{\rm EC}:  \Big\{ v_1 = \sqrt{\frac{2 C_{21}}{L_{12}}} \,, v_2 = \sqrt{\frac{2 C_{12}}{L_{12}}} \Big\} \,, \\ \nonumber
{\rm Sp}(M) &=& \Big\{ \frac{1}{L_{12}} ( \lambda_1 C_{21} + \lambda_2 C_{12}
\mp \sqrt{ (\lambda_1 C_{21} + \lambda_2 C_{12})^2 - 4C_{12} C_{21} L_{12}})\,, \\
& \quad & \frac{2}{L_{12}} (\lambda_{s1} C_{21} + \lambda_{s2} C_{12} + m_s^2 L_{12}) \Big\} \,.
\label{eq_v1v20}
\end{eqnarray}
\item {\bf ($v_1$,$v_2$,$v_s$) case:}
\begin{eqnarray}
\label{eq:ECv1v2vs}
{\rm EC}: \;\;
\begin{cases}
	\xi v_1^2 = - 2 \lambda_{12} C_{s2} - 2 \lambda_{s1} C_{2s} - 2 m_1^2 L_{s2} \,, \\
	\xi v_2^2 = - 2 \lambda_{12} C_{s1} - 2 \lambda_{s2} C_{1s} - 2 m_2^2 L_{s1} \,, \\
	\xi v_s^2 = - \lambda_{s1} C_{21} - \lambda_{s2} C_{12} - m_s^2 L_{12} \,,
\end{cases}
\end{eqnarray}
where
\begin{equation} 
\label{eq:xi}
\xi = \lambda_1 \lambda_2 \lambda_s + 2 \lambda_{12} \lambda_{s1} \lambda_{s2} - \lambda_s \lambda_{12}^2 - 
\lambda_2 \lambda_{s1}^2 - \lambda_1 \lambda_{s2}^2\,.
\end{equation}
Then, the eigenvalues of the Hessian matrix are the roots of a cubic polynomial, $x^3 + b x^2 + c x + d$, with
\begin{align}
\label{eq:b}
b&=-(v_1^2 \lambda_1 + v_2^2 \lambda_2 + 4 v_s^2 \lambda_s)\,, \\
\label{eq:c}
c&= v_1^2 v_2^2 L_{12} + 4 v_1^2 v_s^2 L_{s1} + 4 v_2^2 v_s^2 L_{s2} \,,\\
\label{eq:d}
d&= -4 v_1^2 v_2^2 v_s^2 \xi \,.
\end{align}
\end{itemize}

For completeness we analyze all possible regimes which will be distinguished according to the signs of the $L_{ij}$ and $\xi$ parameters, 
as summarized in Tab.~\ref{tab:all_regimes}. The following conditions apply
\begin{align}
\begin{aligned}
\forall ij: \; \; L_{ij} < 0 &\Rightarrow \xi < 0  \\
\forall ij: \; \; L_{ij} > 0 &\Rightarrow \xi > 0
\end{aligned}
\end{align}
and
\begin{align}
\begin{aligned}
L_{ij} <0 &\Rightarrow C_{ij} < 0 \text{ or } C_{ji} < 0 \\
L_{ij} >0 &\Rightarrow C_{ij} > 0 \text{ or } C_{ji} > 0
\end{aligned}
\end{align}
and if e.g.~$L_{12} < 0$ then $(v_1,v_2,0)$ exists if and only if $C_{21} < 0$ and $C_{12} < 0$, because of Eq.~\eqref{eq_v1v20}. However, as noted before in Eq.~\eqref{instability_v1v20_0}, 
this implies that
\begin{equation}
\frac{2}{L_{12}} ( \lambda_1 C_{21} + \lambda_2 C_{12} + \sqrt{(\lambda_1 C_{21} + \lambda_2 C_{12})^2 - 4 C_{12}C_{21}L_{12}} < 0 \,,
\label{instability_v1v20}
\end{equation}
and therefore $(v_1,v_2,0)$ is always unstable.

The same reasoning holds for $(v_1,0,v_s)$ and $(0,v_2,v_s)$ configurations. Therefore, if e.g. $(v_1,v_2,0)$ is to be 
a stable minimum, then we necessarily need $L_{12} > 0$, which means that $\lambda_{s1} C_{21} + \lambda_{s2} C_{12} + m_s^2 L_{12} > 0$ is required. However, the 
third extremum condition for the minimum $(v_1,v_2,v_s)$ in Eq:~\eqref{eq:ECv1v2vs} can only be fulfilled if $\xi < 0$ for which $(v_1,v_2,v_s)$ becomes unstable. 
The minima $(v_1,v_2,0)$ and $(v_1,v_2,v_s)$ are therefore always incompatible, with the same reasoning applying for $(v_1,0,v_s)$ and $(0,v_2,v_s)$. On the contrary, 
in general, nothing prevents the compatibility of $(v_1,v_2,v_s)$ with $(v_1,0,0)$, $(0,v_2,0)$ and $(0,0,v_s)$ as well as $(v_1,0,v_s)$ with $(0,v_2,v_s)$.

Finally, the existence of the minimum $(v_1,v_2,0)$ in the case $L_{12} > 0$ requires $C_{12},C_{21} > 0$ which is incompatible with the minima $(v_1,0,0)$ and $(0,v_2,0)$, 
but not with $(0,0,v_s)$. However, if e.g.~$L_{12} > 0$ then either $C_{12} > 0$ or $C_{21} > 0$ which results in a non-compatibility of the minima $(v_1,0,0)$ and $(0,v_2,0)$.
\begin{table}
\centering
\begin{tabular}{|cccc|c|} \hline
$L_{12}$ & $L_{s1}$ & $L_{s2}$ & $\xi$ & Regime \\\hline
$-$ & $-$ & $-$ & $+$ & Impossible \\
$-$ & $-$ & $-$ & $-$ & HM regime (HMR) \\
$+$ & $-$ & $-$ & $-$ & HM regime-1 (HMR-1)\\
$+$ & $-$ & $-$ & $+$ & HM regime-2 (HMR-2)\\\hline
$+$ & $+$ & $+$ & $-$ & Impossible \\
$+$ & $+$ & $+$ & $+$ & LM regime (LMR) \\
$-$ & $+$ & $+$ & $+$ & LM regime-1 (LMR-1) \\
$-$ & $+$ & $+$ & $-$ & LM regime-2 (LMR-2)\\\hline
\end{tabular}
\caption{\label{tab:all_regimes} \it Definition of all possible regimes.}
\end{table}

\subsubsection{High-mixing transitions}
\label{Sec:HMR}

\begin{itemize}

\item {\bf The HMR case}

Similarly to what was found for the THDM limit of our model in Sect.~\ref{Sec:THDM}, we define the HM regime (HMR) by 
\begin{eqnarray}
\xi < 0\,, \qquad L_{12},\, L_{s1},\, L_{s2} < 0 \,,
\end{eqnarray}
such that all the mixing parameters are negative. In this regime and in the case of three non-zeroth VEVs ($v_1$,$v_2$,$v_s$), the negative $\xi$ parameter 
defined in Eq.~\eqref{eq:xi} implies $d > 0$ and $c < 0$ given by Eqs.~\eqref{eq:d} and \eqref{eq:c}, respectively. This means that, if there are three real roots, 
at least one of them is negative, thus the considered extremum $(v_1,v_2,v_s)$ is always unstable.

In the case of vanishing singlet VEV, the extremum $(v_1,v_2,0)$ exists if and only if $C_{12}\,,C_{21} <0$. As noted before in Eq.~\eqref{instability_v1v20}, 
this implies that $(v_1,v_2,0)$ is always unstable as well. So, the HMR clearly forbids the existence of stable minima with two non-zeroth Higgs VEVs 
in the model. 

The properties of the remaining vacua (such as their existence and stability) are summarized in Tab.~\ref{tab:minima_high_mixing_regime}, where the cases 
of two coexisting ones are shown in rows five to seven and three coexisting ones are in the last row. For such a class of scenarios, the global minima are determined 
by the lowest values of the tree-level potential $V^{\rm min}_{\rm tree}$ for each VEVs setting
\begin{equation}
V^{\rm min}_{\rm tree}[v_1,0,0] = - \frac{m_1^2}{2 \lambda_1}\,, \quad  
V^{\rm min}_{\rm tree}[0,v_2,0] = - \frac{m_2^2}{2 \lambda_2}\,, \quad
V^{\rm min}_{\rm tree}[0,0,v_s] = - \frac{m_s^2}{2 \lambda_s} \,.
\end{equation}

In the considering HMR case, the previously shown 2D phase diagrams of Fig.~\ref{fig:phase_diagramme_2HDM} correspond to the projections of the 3D phase diagram 
of the complete THDSM potential onto the planes defined by the origin and two axis. The transition line becomes a transition plane and any cosmological scenario 
can be followed through its projections. An illustrative way to represent such scenario is to draw a diagram $(H_1,H_2,\Phi)$ with arrows indicating the successive 
transitions from, e.g.~the minimum $(0,0,v_s)$ (denoted as $\Phi$) to the minimum $(v_1,0,0)$ (denoted as $H_1$). An example of such diagram is given in
Fig.~\ref{fig:transition_pattern} (a.1), for the case of the transition: $(0,0,v_s) \to (v_1,0,0) \to (0,v_2,0)$ (or $\Phi \to H_1 \to H_2$, in short). We represent 
the tree-level first-order phase transitions by red arrows whereas blue arrows indicate the second-order transitions. We notice that the complete list of first-order
transitions in the HMR case reads: $H_1 \leftrightarrow H_2$, $H_1 \leftrightarrow \Phi$, $H_2 \leftrightarrow \Phi$ (see also Fig.~\ref{fig:pyramids} in a representation 
explained below).
\begin{table}
\centering
\begin{tabular}{|c|c|c|c|c|c|} \hline
$m_1^2$ & $m_2^2$ & $m_s^2$ & Conditions on $C_{ij}$ & Minimum \\\hline
$+$&$+$&$+$& - & (0,0,0) \\\hline
$-$&$+$&$+$& - & ($v_1$,0,0)\\\hline
$+$&$-$&$+$& - & (0,$v_2$,0)\\\hline
$+$&$+$&$-$& - & (0,0,$v_s$)\\\hline
$+$&$-$&$-$& $C_{2s}<0$&(0,$v_2$,0) \\
$+$&$-$&$-$& $C_{s2}<0$&(0,0,$v_s$) \\\hline
$-$&$+$&$-$& $C_{1s}<0$&($v_1$,0,0) \\
$-$&$+$&$-$& $C_{s1}<0$&(0,0,$v_s$) \\\hline
$-$&$-$&$+$& $C_{12}<0$&($v_1$,0,0) \\
$-$&$-$&$+$& $C_{21}<0$&(0,$v_2$,0) \\\hline
$-$&$-$&$-$& $C_{12}<0$ and $C_{1s}<0$&($v_1$,0,0) \\
$-$&$-$&$-$& $C_{21}<0$ and $C_{2s}<0$&(0,$v_2$,0) \\
$-$&$-$&$-$& $C_{s1}<0$ and $C_{s2}<0$&(0,0,$v_s$) \\\hline
\end{tabular}
\caption{\label{tab:minima_high_mixing_regime} \it The list of stable and metastable minima in the HMR case. The $+$ and $-$ symbols denote the sign 
of the corresponding mass squared parameters in the tree-level potential. Remark: $L_{ij}<0 \Rightarrow C_{ij}<0$ or $C_{ji}<0$, ensuring that there always 
exists at least one stable minimum.}
\end{table}

Clearly, with more than two scalar fields the structure of the tree-level vacuum rather involved, even in the simplest case with the accidental $\mathbb{Z}_2$ symmetry. 
In particular, we can expect a coexistence of several metastable states during the phase transition, as shown in the last four rows of Tab.~\ref{tab:minima_high_mixing_regime}. 
Furthermore, we can expect that there is a range of temperatures where the probability to roll down from a higher value of the potential into two different metastable states 
with smaller values of the potential is similar, so that two different first-order phase transitions may be realized, leading to the nucleation of different types of vacuum bubbles 
at the same scale. We can even expect the nucleation of bubbles of metastable or stable vacua inside the bubbles of a metastable vacuum, in a slow two-step transition. 
Such exotic processes are not expected to affect baryogenesis due to the sphaleron suppression. We will further explore the interesting and unusual properties of such 
``nested'' bubbles scenarios in a forthcoming work.
\begin{figure}[!ht]
\centering
\includegraphics[width=0.8\linewidth]{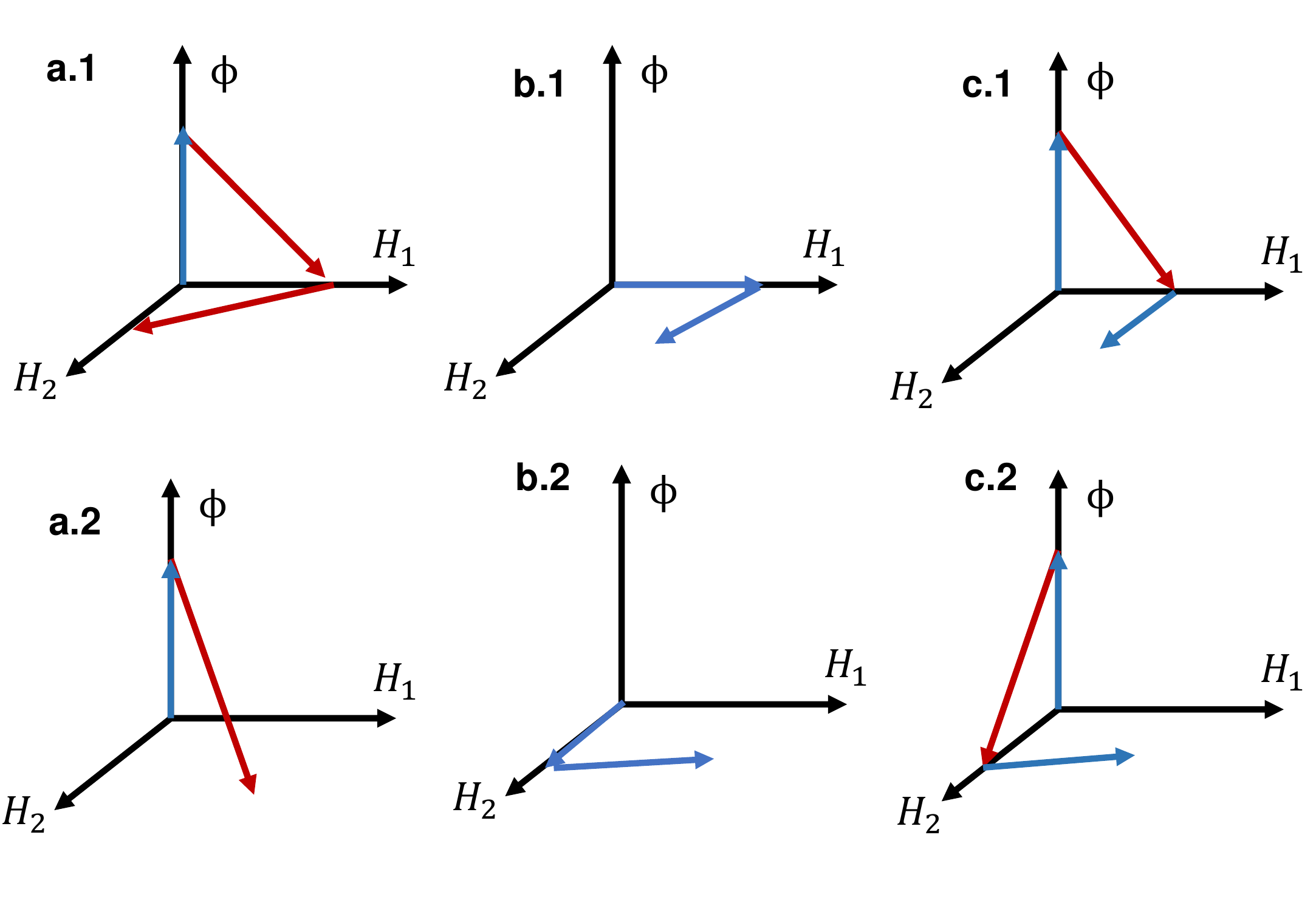}
\caption{\label{fig:transition_pattern} \it An example of the transition patterns in the HMR-1 (a.1) and HMR-2 (others). Red arrows denote the first-order 
phase transitions whereas blue ones represent second-order (or weak first-order) ones. Here, $H_1$, $H_2$, $\Phi$ are shorthand notations for the VEVs 
configurations $(v_1,0,0)$, $(0,v_2,0)$, $(0,0,v_s)$, respectively. A point in the plane $(H_1,H_2)$ corresponds to the configuration $(v_1,v_2,0)$.}
\end{figure}

As discussed earlier in Sec.~\ref{sec:MD} a realistic model involving non-zero $u$ and $d$ quark masses requires non-vanishing values for both Higgs VEVs $v_1$ and $v_2$ at zero temperature. In the HMR case, one of the $v_{1,2}$ VEVs is zero at $T=0$ yielding massless $u$ and $d$ quarks before any approximations on Yukawa couplings are taken into consideration. So, strictly speaking the stable vacuum solutions in the HMR are not compatible with a phenomenologically viable SM spectrum. In the current analysis, however, the $u$ and $d$ quark masses are neglected. So, the only valid motivation to go beyond the considered HMR case is to consider more general and representative vacuum configurations and transitions involving both $v_{1,2}$ Higgs VEVs such as $(0,0,v_s) \to (v_1,v_2,0)$.\\

\item {\bf The HMR-1 case}

Consider now what we denote as HM regime-1 (HMR-1) defined by
\begin{align} 
\label{eq:SBHMR}
\begin{aligned}
\xi < 0\,, \qquad
L_{12} > 0 \,, \qquad
L_{s1}\,, \, L_{s2} < 0\,,
\end{aligned}
\end{align}
with one positive mixing parameter. The projection of the 3D phase diagram onto the $(H_1,H_2)$ plane yields the 2D phase diagram of the LM regime 
of the THDM potential in Fig.~\ref{fig:phase_diagramme_2HDM} (left), whereas the projections onto the planes $(H_1,\Phi)$ and $(H_2,\Phi)$ give 2D phase diagrams 
similar to those discussed earlier in the HM regime of the THDM potential in Fig.~\ref{fig:phase_diagramme_2HDM} (right). The transition lines in these two diagrams 
are, in fact, intersections of the same transition plane, as illustrated in Fig.~\ref{fig:transition_plane}. As a consequence, it is possible to cross the transition plane without 
crossing the transition lines, as e.g.~for the step $(0,0,v_s) \to (v_1,v_2,0)$ illustrated in Fig.~\ref{fig:transition_pattern} (a.2). This leaves us the following complete list of first-order transitions in the HMR-1 case: 
$(v_1,0,0) \leftrightarrow (0,0,v_s)$, $(0,v_2,0) \leftrightarrow (0,0,v_s)$, $(0,0,v_s) \leftrightarrow (v_1,v_2,0)$ (see also Fig.~\ref{fig:pyramids} for a representation 
explained below).
\begin{figure}[!ht]
\centering
\includegraphics[width=0.5\linewidth]{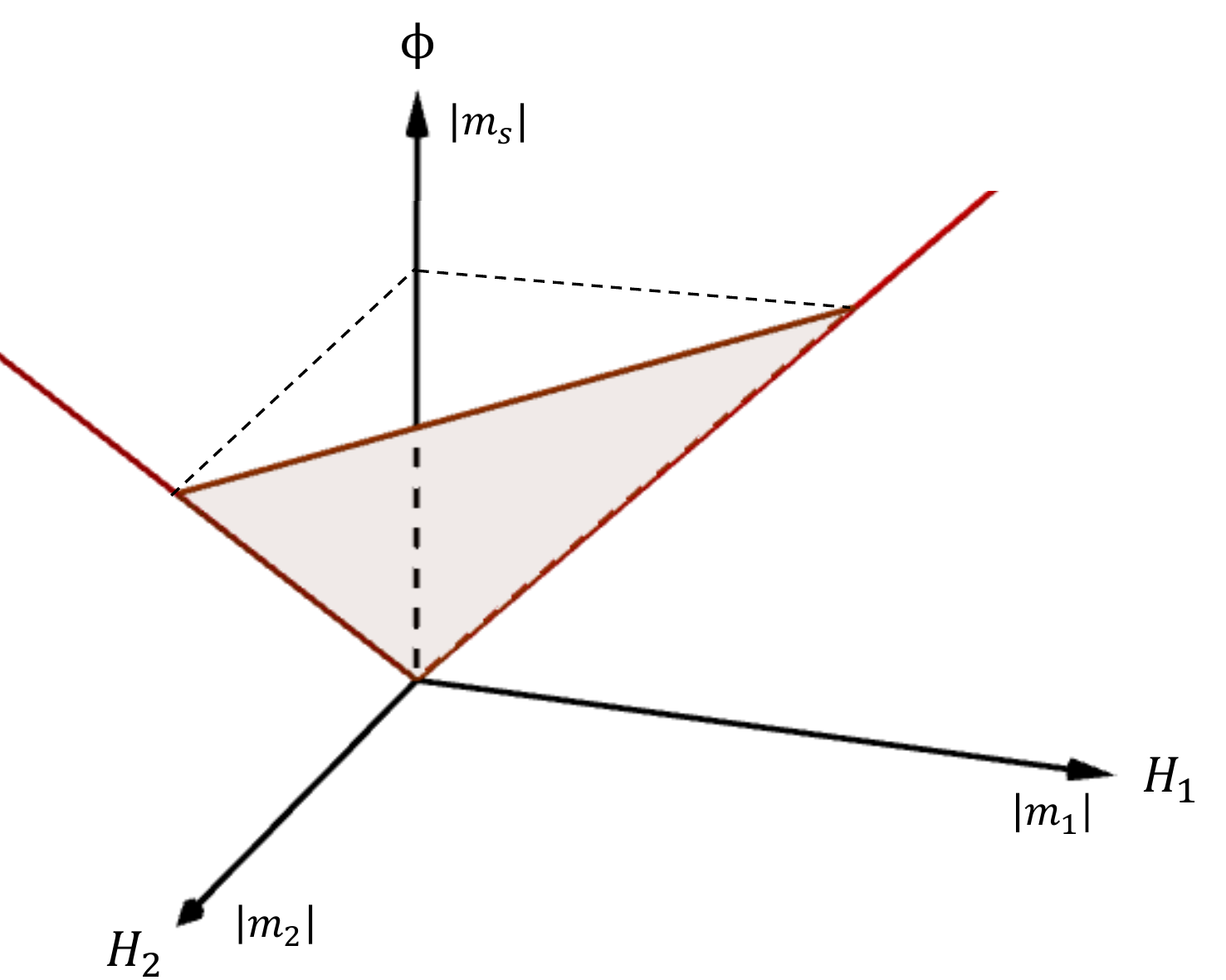}
\caption{\label{fig:transition_plane} \it Schematic representation of the 3D phase diagram in the case of the HMR-1. The transition lines in red belong to the planes 
$(H_1,\Phi)$ and $(H_2,\Phi)$ and correspond to the transition lines of the 2D phase diagram shown in Fig.~\ref{fig:phase_diagramme_2HDM} (right) for the case of the HR 
regime in the THDM potential. There is no transition line in the plane $(H_1,H_2)$ since $L_{12}>0$: we retrieve in this plane the 2D phase diagram of the LM regime 
shown in Fig.~\ref{fig:phase_diagramme_2HDM} (left).}
\end{figure}

Our study of the phase transitions was so far dedicated to a tree-level analysis of the scalar potential. Note that one-loop corrections may promote the tree-level second-order 
phase transitions to the first-order ones, in particular, due to the effect of cubic terms proportional to $m^3(\phi)/T^3$ emerging due to finite temperature corrections. 
However, such transitions are typically not strong. On the other hand, the first-order transitions already present at tree-level are expected to be strong. From the allowed first order phase transition found above, we can then identify four 
types of realistic transition patterns and corresponding tree-level expectations as follows:
\begin{itemize}
\item $(0,0,0) \to (v_1,0,0) \to (v_1,v_2,0)$: can be of the second (or of the weak first) order (diagram b.1 or b.2 in Fig.~\ref{fig:transition_pattern}). 
\item $(0,0,0) \to (0,0,v_s) \to (v_1,0,0) \to (v_1,v_2,0)$: second step is of the first order (diagram c.1 or c.2 in Fig.~\ref{fig:transition_pattern}).
\item $(0,0,0) \to (0,0,v_s) \to (v_1,v_2,0)$: second step is of the first order (diagram a.2 in Fig.~\ref{fig:transition_pattern}). 
In our numerical results in Sec.~\ref{sec:Num} below we found this transition to be of the first order.
\item $(0,0,0) \to (v_1,0,0) \to (0,0,v_s) \to (v_1,v_2,0)$: an example of a marginal process which rarely emerges in simulations but is similar to the others. 
The second and third steps are of the first order and, similarly to the previous one, our numerical analysis below also shows that such transitions are the first-order ones.
\end{itemize}

\item {\bf The HMR-2 case}

This case is similar to the HMR-1 except for a positive $\xi$, i.e.
\begin{align} 
\begin{aligned}
\xi > 0\,, \qquad
L_{12} > 0 \,, \qquad
L_{s1}\,, \, L_{s2} < 0\,.
\end{aligned}
\end{align}
and is denoted as the HM regime-2 or, HMR-2, in short. This regime offers a possibility for a minimum in the three-VEV $(v_1,v_2,v_s)$ configuration.
One should therefore add to the list of transitions in the HMR-1 given above the following possibly first-order transitions
\begin{eqnarray} 
\nonumber
(v_1,v_2,v_s) \longleftrightarrow (v_1,0,0) \,,  \quad (v_1,v_2,v_s) \longleftrightarrow (0,v_2,0) \,, \quad (v_1,v_2,v_s) \longleftrightarrow (0,0,v_s) \,,
\end{eqnarray}
which, as before, are building blocks for the physical pattern of the transitions. 
\end{itemize}

\subsubsection{Low-mixing transitions}
\label{Sec:LMR}

\begin{itemize}

\item {\bf The LMR case}

For the LM regime (LMR) the corresponding signs on the theory parameters are given in Tab.~\ref{tab:all_regimes}. 
In this case, there are no a priori forbidden minima. The possible first-order phase transitions are as follows:
\begin{eqnarray}
(v_1,0,0) \longleftrightarrow  (0,v_2,v_s) \,, \qquad
(0,v_2,0) \longleftrightarrow  (v_1,0,v_s) \,, \nonumber \\
(0,0,v_s) \longleftrightarrow  (v_1,v_2,0)  \,, \qquad
(v_1,v_2,0) \longleftrightarrow  (0,v_2,v_s) \,, \nonumber \\
(v_1,v_2,0) \longleftrightarrow  (v_1,0,v_s) \,, \qquad
(v_1,0,0) \longleftrightarrow  (v_1,v_2,v_s) \,, \nonumber \\
(0,v_2,0) \longleftrightarrow  (v_1,v_2,v_s) \,,  \qquad
(0,0,v_3) \longleftrightarrow  (v_1,v_2,v_s) \,. 
\end{eqnarray}

\item {\bf The LMR-1 case}

Similarly to what was done earlier to define the set of HM regimes, we now consider that, in a scenario where a LM dominates, we get two directions highly mixed 
such that e.g.~$L_{12} < 0$. As such we denote this regime as the LM regime-1, or LMR-1 shortly. In this case the minima $(v_1,0,0)$ and $(0,v_2,0)$ are compatible but 
$(v_1,0,0)$ and $(0,0,v_s)$ -- are not. This is due to vacuum stability requiring, respectively, that $C_{1s} < 0$ and $C_{s1} < 0$, which is excluded when $L_{s1} > 0$. 
Therefore, the allowed tree-level first-order phase transitions in this case are
\begin{align}
(v_1,0,0) &\longleftrightarrow (v_1,v_2,v_s) \,, \quad (0,v_2,0) \longleftrightarrow (v_1,v_2,v_s) \,, \quad
(0,0,v_s) \longleftrightarrow (v_1,v_2,v_s) \,, \notag \\
& \qquad (v_1,0,0) \longleftrightarrow (0,v_2,v_s) \,, \quad (0,v_2,0) \longleftrightarrow (v_1,0,v_s) \,, \label{LMR-PTs} \\
& \qquad (v_1,0,v_s) \longleftrightarrow (0,v_2,v_s) \,, \quad (v_1,0,0) \longleftrightarrow (0,v_2,0) \,. \notag
\end{align}

\item {\bf The LMR-2 case}

This regime, which we denote as the low mixing regime-2 (LMR-2), is similar to the previous one except for $\xi <0$. Therefore, the difference for the LMR-1 is that the minimum 
$(v_1,v_2,v_s)$ is always unstable which means that all transitions involving it are discarded.

\end{itemize}

\subsection{Pyramidal representation of the tree-level vacuum structure}
\label{Sec:pyramid}

\begin{figure}
\centering
\includegraphics[width=1\linewidth]{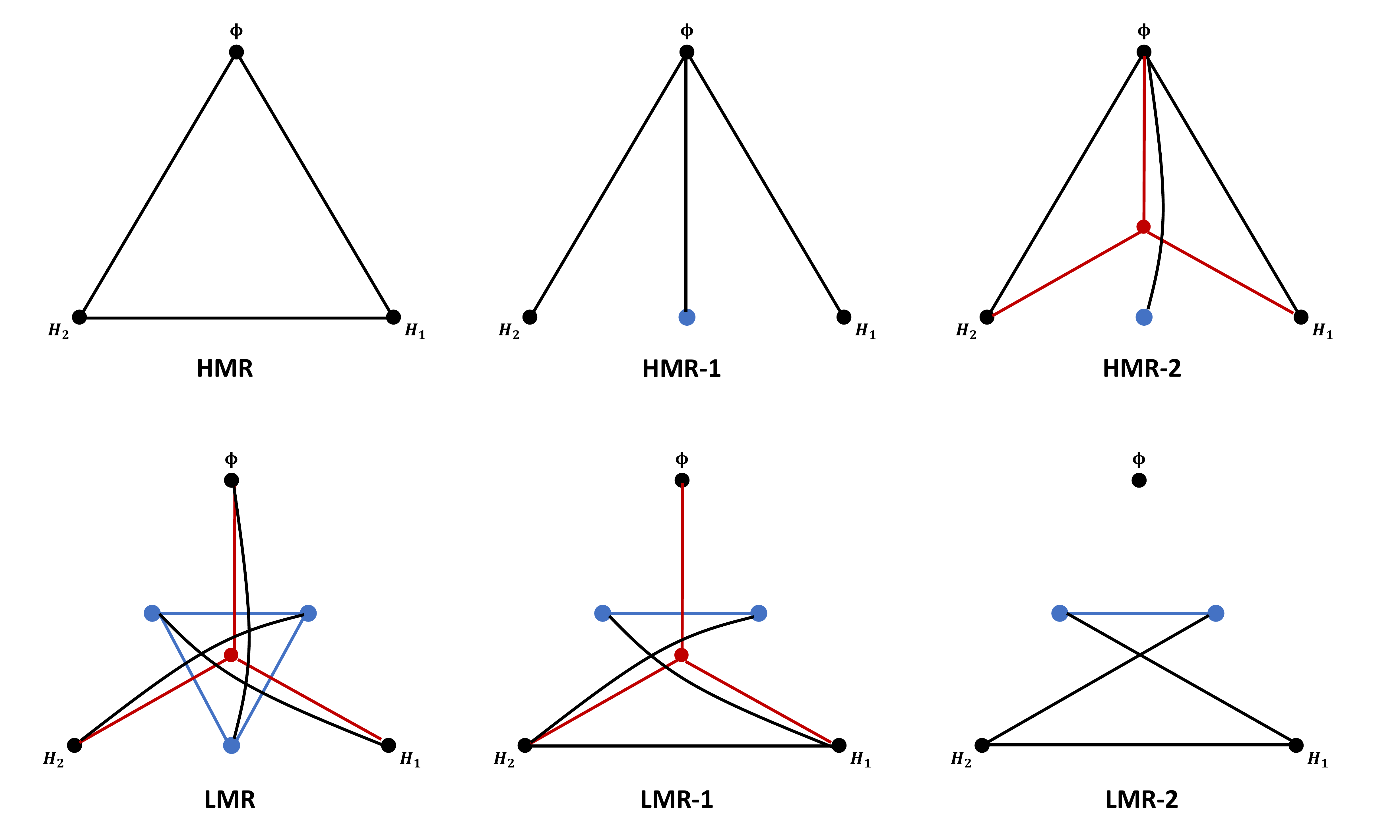}
\caption{\label{fig:pyramids} \it Pyramidal representations of the tree-level vacuum structure in all possible regimes. 
Dots represent the minima while lines denote the tree-level first-order phase transitions. The colors are just for 
improved readability and have no physical meaning here.}
\end{figure}

It is instructive for the reader to represent all previously described tree-level first-order phase transitions in a visual manner. In this section, 
we introduce a geometrical representation constructed upon the following steps:
\begin{itemize}
\item Build a 3D basis and label three vectors $H_1$, $H_2$ and $\Phi$.
\item Draw one point on each axis: they represent the minima $(v_1,0,0)$, $(0,v_2,0)$, $(0,0,v_s)$ respectively.
\item If e.g.~$L_{12} > 0$ then place a point on the plane $(H_1,H_2)$. It represents the minimum $(v_1,v_2,0)$. The same for the other two planes.
\item If $\xi > 0$ then place a point in the middle of the graph: it represents the minimum $(v_1,v_2,v_s)$.
\end{itemize}
In the following we will derive simple rules to link these points, i.e.~to visualize the possible transitions between the various minima.
Instead of drawing 3D Cartesian coordinates it is more convenient to draw a pyramid viewed from the top as in Fig.~\ref{fig:pyramids}. 
The vertices of the triangular basis correspond to the minima $(v_1,0,0)$, $(0,v_2,0)$, $(0,0,v_s)$, points on edges correspond to the minima 
$(v_1,v_2,0)$, $(v_1,0,v_s)$, $(0,v_2,v_s)$, and the point at the center (the ``top'' of the pyramid) represents the minimum $(v_1,v_2,v_s)$. 
For our purposes in this section the minimum $(0,0,0)$ is not indicated since there are no first-order phase transitions from such a minimum 
at tree-level.

Using such an illustrative pyramidal representation, we show how to construct all possible tree-level first-order phase transitions 
by linking the mentioned points within the pyramid.
\begin{itemize}
\item We have seen that if e.g.~$(v_1,v_2,0)$ configuration exists, then it is not compatible with neither $(v_1,0,0)$ nor $(0,v_2,0)$. It was also observed that $(v_1,0,0)$ 
and $(0,v_2,0)$ are also not compatible. This is translated into the following rule: \textit{if two vertices are separated by a point in the edge, then they 
are not linked. However, vertices are always linked with the opposite edge, if it exists, and it is always possible to link two edge points.}
\item We have seen that if e.g. $(v_1,v_2,0)$ is a stable minimum, then $(v_1,v_2,v_s)$ cannot be stable. This is translated into the following rule: 
\textit{points on edges are never linked to the point at the center.}
\item We have seen that in general nothing prevents the compatibility of $(v_1,v_2,v_s)$ with $(v_1,0,0)$, $(0,v_2,0)$, $(0,0,v_s)$ configurations. 
This is translated into the following rule: \textit{vertices are always linked to the point at the center (if the latter exists).}
\end{itemize}
The pyramidal representation of the six regimes are shown in Fig.~\ref{fig:pyramids}. By symmetry, it exhaustively covers the tree-level vacuum structure of the scalar 
THDSM potential \eqref{eq:2HDMS}. By definition, the building blocks of a pyramidal diagram are one-step transitions that can be part of multi-step transition scenarios. 
For instance, one can realize $(v_1,0,0) \to (v_1,v_2,v_s)$ transition directly, or compose a multi-step transition like $(v_1,0,0) \to (0,v_2,0) \to 
(v_1,v_2,v_s)$ (LMR-1 case), or any other path following the corresponding pyramidal diagram in Fig.~\ref{fig:pyramids}.

\begin{figure}
\centering
\includegraphics[width=1\linewidth]{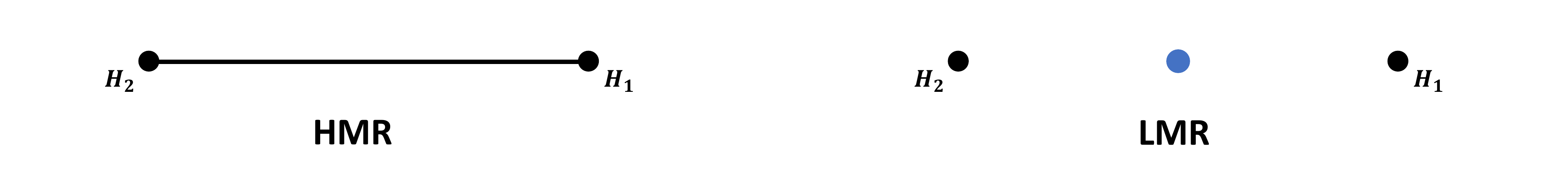}
\caption{\label{fig:pyramids2HDM} \it Pyramidal representation of the tree-level vacuum structure of the THDM limit discussed in Sec.~\ref{Sec:THDM}. 
In the HM regime (left), there is a first-order phase transition between the minima $(v_1,0)$ ($H_1$) and $(0,v_2)$ ($H_2$), while in the LM regime (right), 
the stable minimum $(v_1,v_2)$ (blue dot) prevents such a transition to occur.}
\end{figure}

In a sense, the pyramidal representation unifies previous studies of simple scalar extensions of the SM. Indeed, well-known $\mathbb{Z}_2$-symmetric extensions of the SM 
with three fields or less fit into this classification. In e.g.~Ref.~\citep{Vaskonen:2016yiu}, a study of the HM regime in two dimensions is done, where the EWPT is realized 
in two-steps, whereas Ref.~\citep{Beniwal:2017eik,Dorsch:2013wja} proposes a study of the LM regime, where the transition takes place in a single step and 
is only of the first order due to one-loop effects. The distinction between both regimes is pointed out in Refs.~\citep{Li:2014wia,Ginzburg:2010wa} in terms of tree-level 
and one-loop transitions. Finally, recent studies in two distinct versions of THDSM extensions of the SM \citep{Alanne:2016wtx,Chao:2017vrq} consider the transition 
$(0,0,v_s) \to (v_1,v_2,0)$. In general, our classification for the phase transitions fits well into distinct variants of THDSM-type of models despite the existence of diverse 
variants with respect to their transformation laws or the presence of soft-symmetry breaking terms. Moreover, the THDM limit discussed in Sec.~\ref{Sec:THDM} can 
also be described in this classification. In particular, as represented in Fig.~\ref{fig:pyramids2HDM} we have two vertices corresponding to the extrema $(v_1,0)$ and 
$(0,v_2)$ which can either be linked if there is no point between them (left), i.e.~$(v_1,v_2)$ is not a minimum, or not linked if there is a $(v_1,v_2)$ minimum (right). 
As observed before, while in the latter case the situation with no link should be interpreted as the absence of a first-order phase transition corresponding to the LM regime 
in Fig.~\ref{fig:phase_diagramme_2HDM} for $L_{12} > 0$, the former one represents the HM regime, $L_{12}<0$, and a first-order transition is possible between 
$(v_1,0)$ and $(0,v_2)$ vacuum configurations.

In our model, any cosmologically relevant scenario with a sequence of transitions should end up in either the $(v_1,v_2,0)$ or $(v_1,v_2,v_s)$ vacua at $T=0$. 
Furthermore, for a scenario to be of interest for baryogenesis, the relevant phase transition should start from a high-symmetric vacuum where the Higgs doublets 
do not have VEVs. Otherwise, the sphaleron processes that induce the baryon creation become way too suppressed. This can be realized in all regimes except 
the HMR.

In what follows, we are concentrated mostly on the HMR-1 case which is the simplest physically consistent scenario although it is not motivated by any specific physics
argument so far. All other regimes discussed above (except for the HMR case which is simpler but unphysical) are also consistent but are much more complicated as 
can be seen from their pyramidal diagrams in Fig.~\ref{fig:pyramids}. Indeed, they display a lot of possibilities including the most complicated vacuum structure with 
three non-zeroth VEVs $(v_1,v_2,v_s)$ (except for the LMR-2 case where it is unstable) that may be worth studying in the future. In particular, it would be very 
interesting to investigate the transition represented by a blue line, between e.g. $(0,v_2,v_s)$ and $(v_1,0,v_s)$ minima.
          
\section{The electroweak phase transition}
\label{sec:EWPT}

\subsection{One-loop corrections to the effective potential at finite temperatures}
\label{Sec:one-loop-T}

In Sec.~\ref{sec:structure} we have performed a systematic zero temperature analysis of the tree-level scalar potential in order to characterize the different regimes for 
which the first-order phase transitions are possible. However, for a comprehensive understanding of the phase transition phenomena we need to analyze how different 
vacua have evolved in a cooling Universe. Therefore, it is necessary to introduce the temperature dependence to the effective potential which, to leading order in 
perturbation theory has the following form \citep{Quiros:1999jp,Curtin:2016urg}
\begin{equation}
V^{(1)}(T) = V_{\rm tree} + V_{\rm CW} + \Delta V^{(1)}(T)\,.
\end{equation}
Here, the tree-level contribution, $V_{\rm tree}$, is given in Eq.~\eqref{eq:2HDMS}. The second part, $V_{\rm CW}$, is the well-known zero-temperature 
Coleman-Weinberg potential
\begin{equation}
V_{\rm CW} = \sum_i (-1)^F n_i \frac{m_i^4}{64 \pi^2} \left( \log\left[ \frac{m_i^2(\phi_\alpha)}{\Lambda^2}\right] - c_i \right) \,,
\end{equation}
where $F=0(1)$ for bosons (fermions), $m_i^2(\phi_\alpha)$ is the $\phi_\alpha$-field dependent mass of the particle $i$, $n_i$ is the number of degrees of freedom 
for the particle $i$, $\Lambda$ is a renormalization scale which will be taken to be the EW scale $v \equiv \sqrt{v_1^2 + v_2^2} = 246~{\rm GeV}$ relevant for 
the EW phase transition, and, in the $\overline{\rm MS}$-renormalization scheme, the constant $c_i$ is equal to $3/2$ for scalars and fermions, and to $5/6$ for 
vector degrees of freedom. Finally, the necessary thermal corrections $\Delta V^{(1)}(T)$ read \citep{Quiros:1999jp}:
\begin{equation}
\Delta V^{(1)}(T) = \frac{T^4}{2 \pi^2} \left\{ \sum_{b} n_b J_B\left[\frac{m_i^2(\phi_\alpha)}{T^2}\right] - \sum_{f} n_f J_F\left[\frac{m_i^2(\phi_\alpha)}{T^2}\right] \right\}\,,
\label{finite_T_correction}
\end{equation} 
where $J_B$ and $J_F$ are the thermal integrals for bosons and fermions, respectively, given by
\begin{align} \label{eq:JBJF}
J_{B/F}(y^2) = \int_0^\infty \d x \, x^2 \log\left( 1 \mp \exp [ - \sqrt{x^2 + y^2}] \right)\,.
\end{align}
Relevant details about the computation of these corrections, in particular, the resummation procedure \citep{Curtin:2016urg}, are given below and 
in Appendix~\ref{sec:app1}.

\subsection{Thermal corrections}
\label{sec:thermal_masses}

The field-dependent masses for the scalar sector at zero temperature are the eigenvalues of the Hessian matrix, $M_{\alpha,\beta}^2 = 
\left. \frac{\partial^2 V}{\partial \phi_\alpha \partial \phi_\beta} \right|_{\phi_\alpha}$, with the potential $V$ taken to be the $\mathbb{Z}_2$-symmetric 
version of the one given in Eq.~\eqref{eq:VH}, for simplicity. The field-dependent gauge boson masses read
\begin{align}
\begin{aligned}
m_Z^2(\phi_\alpha) &= \frac{\phi_1^2 + \phi_2^2}{4}(g_Y^2+g_L^2) \,,  \\
m_W^2(\phi_\alpha) &= \frac{\phi_1^2 + \phi_2^2}{4} g_L^2 \,,
\end{aligned}
\end{align}
where $g_L$ and $g_Y$ are the $\SU{2}{L}$ and $\U{Y}$ gauge couplings of the SM, respectively. For the $t$ and $D$ quark masses,
we adapt the corresponding relations from Eq.~\eqref{eq:Mq-simp} where only the top-quark mass is field-dependent,
\begin{eqnarray}
\label{top}
m_t (\phi_2) \simeq \frac{\phi_2}{\sqrt{2}} |Y_{3\,2}^u| = y_t' \frac{\phi_2}{\sqrt{2}} \, = \( y_t \frac{v}{v_2} \) \frac{\phi_2}{\sqrt{2}},
\end{eqnarray}
with $y_t$ the SM Yukawa coupling for the top quark, and $v \equiv \sqrt{v_1^2 + v_2^2} = 246~{\rm GeV}$. The VLQ mass is fixed by the mass parameter $m_D$ to the leading order as in Eq.~\eqref{eq:Mq-simp}. The latter implies that the VLQ contribution only adds a constant to the potential and can be discarded. We have checked in the numerical computation that the field-dependent contributions beyond the leading order to the mass of the VLQ are indeed completely negligible for $m_D > 1$ TeV.

In the high temperature limit, the thermal integrals can be expanded in powers of $y \equiv m/T$ such as
\begin{align} 
\label{eq:JBJF_expansion}
\begin{aligned}
J_{B}(y^2) &\overset{y \ll 1}{\simeq} \frac{- \pi^4}{45} + \frac{\pi^2}{12} y^2 - \frac{\pi}{6} y^3 + \mathcal{O}(y^4) \,, \\
J_{F}(y^2) &\overset{y \ll 1}{\simeq} \frac{7 \pi^4}{360} - \frac{\pi^2}{24} y^2 + \mathcal{O}(y^4) \,.
\end{aligned}
\end{align}

At order $y^2$, the thermal corrections reduce to
\begin{align}
\label{eq:y2}
\Delta V^{(1)}(T) = \frac{T^2}{24} \left\{ \Tr\left( M_{\alpha,\beta}^2 \right) + \sum_{i=W,Z} n_i m_i^2(\phi_\alpha) + 
\sum_{i=t} \frac{n_i}{2} m_i^2(\phi_\alpha) \right\} \,,
\end{align}
where we have dropped field-independent terms. 

For the scalar sector contribution (first term in Eq.~\eqref{eq:y2}), provided that the trace is basis invariant we can consider the diagonal elements of the 
mass matrix in the gauge basis. Therefore, the following 10 non-physical field-dependent masses for the scalar degrees of freedom contributing 
to the trace read
\begin{align}
\label{diagonal_terms}
\begin{aligned}
m_{h_1}^2(\phi_1,\phi_2,\phi_s) &= m_1^2 + \frac{\lambda_{12}}{2} \phi_2^2 + \lambda_{s1} \phi_s^2 + \frac{3}{2} \phi_1^2 \lambda_1  \,, \\
m_{\eta_1,\chi_1,\chi_1'}^2(\phi_1,\phi_2,\phi_s) &=  m_1^2 + \frac{\lambda_{12}}{2} \phi_2^2 + \lambda_{s1} \phi_s^2 + \frac{1}{2} \phi_1^2 \lambda_1  \,, \\
m_{h_2}^2(\phi_1,\phi_2,\phi_s) &= m_2^2 + \frac{\lambda_{12} \phi_1^2}{2} + \lambda_{s2} \phi_s^2 + \frac{3}{2} \phi_2^2 \lambda_2  \,, \\
m_{\eta_2,\chi_2,\chi_2'}^2(\phi_1,\phi_2,\phi_s) &= m_2^2 + \frac{\lambda_{12}}{2} \phi_1^2 + \lambda_{s2} \phi_s^2 + \frac{1}{2} \phi_2^2 \lambda_2  \,, \\
m_{S_R}^2(\phi_1,\phi_2,\phi_s) &= 2 m_s^2 + \lambda_{s1} \phi_1^2 + \lambda_{s2} \phi_2^2 + 6 \phi_s^2 \lambda_s  \,, \\
m_{S_I}^2(\phi_1,\phi_2,\phi_s) &= 2 m_s^2 + \lambda_{s1} \phi_1^2 + \lambda_{s2} \phi_2^2 + 2 \phi_s^2 \lambda_s \,,
\end{aligned}
\end{align}
in terms of the norms of the classical background fields $\phi_{1,2,s}$. The corrections \eqref{eq:y2} do not change the shape 
of the potential but make the bare masses to run with the temperature. Those thermal masses will play an important role when including the higher order terms, 
through the so-called resummation procedure. In Eq.~\eqref{eq:y2}, the numbers of degrees of freedom for the vector bosons and for the top (anti)quark are
\begin{equation}
n_W = 6, \qquad n_Z = 3, \qquad n_t = 12\,.
\end{equation}

The effective potential at $T \neq 0$, $V_{\rm eff}(\phi_1,\phi_2,\phi_s,T) = V_{\rm tree} + \Delta V^{(1)}(T)$ is then given by
\begin{align}
\begin{aligned}
V_{\rm eff}(\phi_1,\phi_2,\phi_s,T) &=
\mu_1^2(T) \frac{\phi_1^2}{2}
+ \mu_2^2(T) \frac{\phi_2^2}{2}
+ \mu_s^2(T) \phi_s^2
+ \frac{\lambda_1}{2} \frac{\phi_1^4}{4}
+ \frac{\lambda_2}{2} \frac{\phi_2^4}{4}
+ \frac{\lambda_s}{2} \phi_s^4 \,, \\
&
+ \lambda_{12} \frac{\phi_1^2}{2} \frac{\phi_2^2}{2}
+ \lambda_{s1} \frac{\phi_1^2}{2} \phi_s^2
+ \lambda_{s2} \frac{\phi_2^2}{2} \phi_s^2 \,,
\end{aligned}
\end{align}
where the thermal masses read
\begin{align}
\mu_1^2(T) = m_1^2 + c_1 T^2 \,,\qquad
\mu_2^2(T) = m_2^2 + c_2 T^2 \,, \qquad
\mu_s^2(T) = m_s^2 + c_s T^2 \,,
\end{align}
in terms of $c_1$, $c_2$ and $c_3$ coefficients
\begin{align} \label{coeff_thermalmasses}
\begin{aligned}
c_1 &= \frac{1}{48} \( 9 g_L^2 + 3 g_Y^2 + 16 \lambda_1 + 8 \lambda_{12} + 8 \lambda_{s1} \) \,, \\
c_2 &= \frac{1}{48} \( 9 g_L^2 + 3 g_Y^2 + 12 y_t'^2 + 16 \lambda_2 + 8 \lambda_{12} + 8 \lambda_{s2} \) \,, \\
c_s &= \frac{1}{6} \( 2\lambda_s + \lambda_{s1} + \lambda_{s2} \) \,.
\end{aligned}
\end{align}
Note, the Yukawa coupling $y_t'$ does not appear in $c_1$ since in the considering THDSM model the top quark is coupled only to $H_2$. 

\subsection{Electroweak transition in the HMR-1 case: an analytic discussion}
\label{Sec:analytical_results_HMR1}

In the $(m/T)$-expansion at order $(m/T)^2$, the only correction to the tree-level potential is due to the mass parameters' dependence on $T^2$. 
This implies that, as temperature decreases, the evolving squared-mass parameters are connected through linear relations. In the first 
approximation, a cosmological evolution can therefore be represented as a straight line in the phase diagram $(m_1^2,\,m_2^2,\,m_s^2)$.

Consider, for example, the phase transitions in the HMR-1 case considered above and which will also be investigated numerically in what follows. 
Defining the critical temperatures for the first step in the transition patterns $(0,0,0) \to (0,0,v_s) \to \cdots$ (see e.g.~diagrams a.2, c.1, c.2 
in Fig.~\ref{fig:transition_pattern}), which is the second-order ($O(2)$), as $\tilde{T}_1, \tilde{T}_2, \tilde{T}_s$ such that
\begin{align}
\mu_1^2(\tilde{T}_1) = 0 \,,\qquad
\mu_2^2(\tilde{T}_2) = 0 \,, \qquad
\mu_s^2(\tilde{T}_s) = 0 \,,
\end{align}
this transition is realized if the following conditions hold:
\begin{align}
\tilde{T}_s > \tilde{T}_1 \qquad {\text{and}} \qquad \tilde{T}_s > \tilde{T}_2\,.
\label{condition_phibreaksfirst}
\end{align}
Indeed, since the first step is of the second order, the transition only occurs when the other minimum appears (the two minima do not co-exist).
In terms of mass parameters and the coefficients introduced in Eq.~\eqref{coeff_thermalmasses}, such conditions are equivalent to
\begin{align}
\frac{- m_s^2}{c_s} > \frac{- m_1^2}{c_1} \qquad {\text{and}} \qquad \frac{- m_s^2}{c_s} > \frac{- m_2^2}{c_2} \,, \qquad m_{1,2}^2,\,m_s^2 <0 \,.
\end{align}
These are of particular interest since they correspond to the patterns whose second step in the breaking chain contains a first-order phase transition at tree-level, which is likely to be strong. In this case the transition does not occur when the other minimum appears but when it becomes 
the global one (when both minima are at the same value of the potential). In particular, for the patterns
\begin{itemize}
\item [(A)]: $(0,0,0) \overset{O(2)}{\to} (0,0,v_s) \overset{O(1)}{\to} (v_1,0,0) \overset{O(2)}{\to} (v_1,v_2,0)$, see Fig.~\ref{fig:transition_pattern} (diagram c.1),
\item[(B)]: $(0,0,0) \overset{O(2)}{\to} (0,0,v_s) \overset{O(1)}{\to} (0,v_2,0) \overset{O(2)}{\to} (v_1,v_2,0)$, see Fig.~\ref{fig:transition_pattern} (diagram c.2),
\item [(C)]: $(0,0,0) \overset{O(2)}{\to} (0,0,v_s) \overset{O(1)}{\to} (v_1,v_2,0)$, see Fig.~\ref{fig:transition_pattern} (diagram a.2),
\end{itemize}
the critical temperature for the $O(1)$ phase transition, $T_c$, can then be analytically found by equating the vacuum potentials for each pattern, namely,
\begin{align}
\begin{aligned}
V(0,0,v_s;T_c) = V(v_1,0,0;T_c) \qquad \Rightarrow \qquad
T_c^2 = \frac{\sqrt{\lambda_1} m_s^2 - \sqrt{\lambda_s} m_1^2}{\sqrt{\lambda_s} c_1 - \sqrt{\lambda_1} c_s}
\label{eq_analytical_A}
\end{aligned}
\end{align}
for pattern (A),
\begin{align}
\begin{aligned}
V(0,0,v_s;T_c) = V(0,v_2,0;T_c) \qquad \Rightarrow \qquad
T_c^2 = \frac{\sqrt{\lambda_2} m_s^2 - \sqrt{\lambda_s} m_2^2}{\sqrt{\lambda_s} c_2 - \sqrt{\lambda_2} c_s}
\label{eq_analytical_B}
\end{aligned}
\end{align}
for pattern (B), and
\begin{align}
\begin{aligned}
V(0,0,v_s;T_c) = V(v_1,v_2,0;T_c) \qquad \Rightarrow \qquad A T_c^4 + 2B T_c^2 + C = 0
\label{eq_analytical_C}
\end{aligned}
\end{align}
for pattern (C), where
\begin{align}
A &=  c_s^2 (\tilde{\lambda}_1 \tilde{\lambda}_2 - 1 ) / \tilde{\lambda}_s - \tilde{\lambda}_1 c_2^2 - 
\tilde{\lambda}_2 c_1^2 + 2 c_1 c_2 \,, \notag \\
B &= m_s^2 c_s (\tilde{\lambda}_1 \tilde{\lambda}_2 - 1) / \tilde{\lambda}_s - \tilde{\lambda}_1 c_2 m_2^2 - 
\tilde{\lambda}_2 c_1 m_1^2 + m_1^2 c_2 + m_2^2 c_1 \,, \notag \\
C &= m_s^4 (\tilde{\lambda}_1 \tilde{\lambda}_2 - 1)/\tilde{\lambda}_s - \tilde{\lambda}_1 m_2^4 - 
\tilde{\lambda}_2 m_1^4 + 2 m_1^2 m_2^2 \,, \notag
\end{align}
with $\tilde{\lambda}_i \equiv \lambda_i/\lambda_{12}$ for $i=1,2,s$. In our numerical analysis we only need to compute the first step in order to know when the transition begins (critical temperature of the second order phase transition) and which minima are involved, and then the second step in order to determine when the first-order transition occurs, i.e.~the critical temperature. The third step in the chains (A), (B) and (C) is of the second order such that the same consideration is applied as for the first step. The corresponding critical temperatures are not relevant 
for the following discussion and thus are not computed.

Note that the condition \eqref{condition_phibreaksfirst} is sufficient but not necessary. In fact, exotic breaking patterns as $(0,0,0) \to (0,v_2,0) \to (0,0,v_s) 
\to (v_1,v_2,0)$ may also happen as well as certain patterns with symmetry restoration. However, such type of processes are rather rare in our numerical scans and we 
do not discuss them further on.

So far we have considered the one-loop thermal corrections only up to order $(m/T)^2$, while the higher-order thermal corrections may, in principle, be important. 
In general, in addition to the running of the physical couplings and masses with temperature, the shape of the potential may also be altered, offering a possibility 
to promote a transition that is of the second order at tree level (e.g.~in the SM) to a first-order one at one loop. This is a result of a negative higher power correction 
proportional to $(m/T)^3$ in the high temperature expansion \eqref{eq:JBJF_expansion} of Eq.~\eqref{eq:JBJF}, which generates negative cubic terms in the field space 
creating a potential barrier between the existent tree-level minima. However, as it was noted in Refs.~\citep{Li:2014wia,Alanne:2016wtx} when considering the transitions 
already present at tree-level, the higher order corrections are not likely to significantly change the shape of the potential and thus the transition patterns, while 
the values of the critical temperatures and the VEVs may be shifted. 

In section \ref{sec:Num} we will use the results derived here to perform a comparison between the analytical expectations and a numerical scan that accounts for higher order corrections. A good agreement will be found, showing that, when considering tree-level first order phase transitionss, the analytical procedure developed here is robust and can be useful in the search of interesting regions of the parameter space.

\subsection{Bubble nucleation}
\label{Sec:bubble}

The phase transition is a dynamical process corresponding to a non-trivial non-perturbative solution of the field equations of motion, the instanton 
\citep{Linde1983,Dine:1992wr}. Such process can occur either through a thermal jump, in the high temperature limit, or by quantum tunneling at 
low temperature \citep{Linde1983}. Both processes can be described by the same formalism as classical motion in Euclidean space \citep{Coleman:1977py}. 
In particular, such a solution is described by the action,
\begin{equation}
\hat{S}_3 = 4 \pi \int_0^\infty \mathrm{d}r \, r^2 \left\{ \frac{1}{2} \left( \frac{\mathrm{d}\hat{\phi}}{\mathrm{d}r} \right)^2 + V(\hat{\phi}) \right\} \,,
\end{equation}
which can be assumed to be $\mathrm{O(3)}$-symmetric in the high temperature limit \citep{Linde1983}. Here, $\hat{\phi}$ represents a solution of 
the Euclidean equation of motion found by computing the path that minimizes the energy \citep{Coleman:1977py,Wainwright:2011kj}.

In the high temperature limit, the probability for $\hat{\phi}$ to be realized reads \citep{Linde1983}
\begin{equation}
\Gamma \sim T^4 \( \frac{\hat{S}_3}{2 \pi T} \)^{3/2} \exp \left( - \hat{S}_3 / T \right) \,.
\end{equation}
The temperature at which the transition does actually occur is not the critical temperature but, instead, the nucleation temperature, which can be much 
smaller than the former one. While the critical temperature is defined by the temperature corresponding to the degenerate minima, the nucleation 
temperature is the temperature where the probability to realize one transition per cosmological horizon volume is one, that is,
\begin{align}
\Gamma/H^4 \sim 1 \,,
\end{align}
with $H$ being the Hubble constant. Following Refs.~\citep{Dine:1992wr,Quiros:1999jp}, this translates into 
\begin{equation}
\hat{S}_3 / T_n \sim -4 \log\(T_n/m_P \) \sim 140 \,,
\label{def_nucleation_temperature}
\end{equation}
where $m_P$ is the reduced Planck mass. This last relation defines the nucleation temperature. If the solution of Eq.~\eqref{def_nucleation_temperature} with 
respect to $T_n$ is much smaller than $T_c$, which is common in the case of strong first-order phase transitions, it is denoted as supercooling. On the other hand, 
if Eq.~\eqref{def_nucleation_temperature} has no solution then the transition never occurs \citep{Kurup:2017dzf} and the system gets trapped in a supercooled 
metastable state until zero temperature limit is reached in the course of cosmological expansion. Once at $T=0$, the transition can still take place through quantum 
tunneling, however, such processes are beyond the scope of this article and are not further discussed here. 

Finally, due to the supercooling property, we often witness a change in the transition pattern: while the critical temperature was defined by the possibility to transit 
from e.g.~the phase $(0,0,v_s)$ to the phase $(v_1,0,0)$, the nucleation process actually happens between the phase $(0,0,v_s)$ and the phase $(v_1,v_2,0)$. This can 
be understood by noting that at $T_n \ll T_c$, the phase $(v_1,0,0)$ is likely to have already changed into $(v_1,v_2,0)$.

\begin{figure}[!ht]
\centering 
\includegraphics[width=0.7\linewidth]{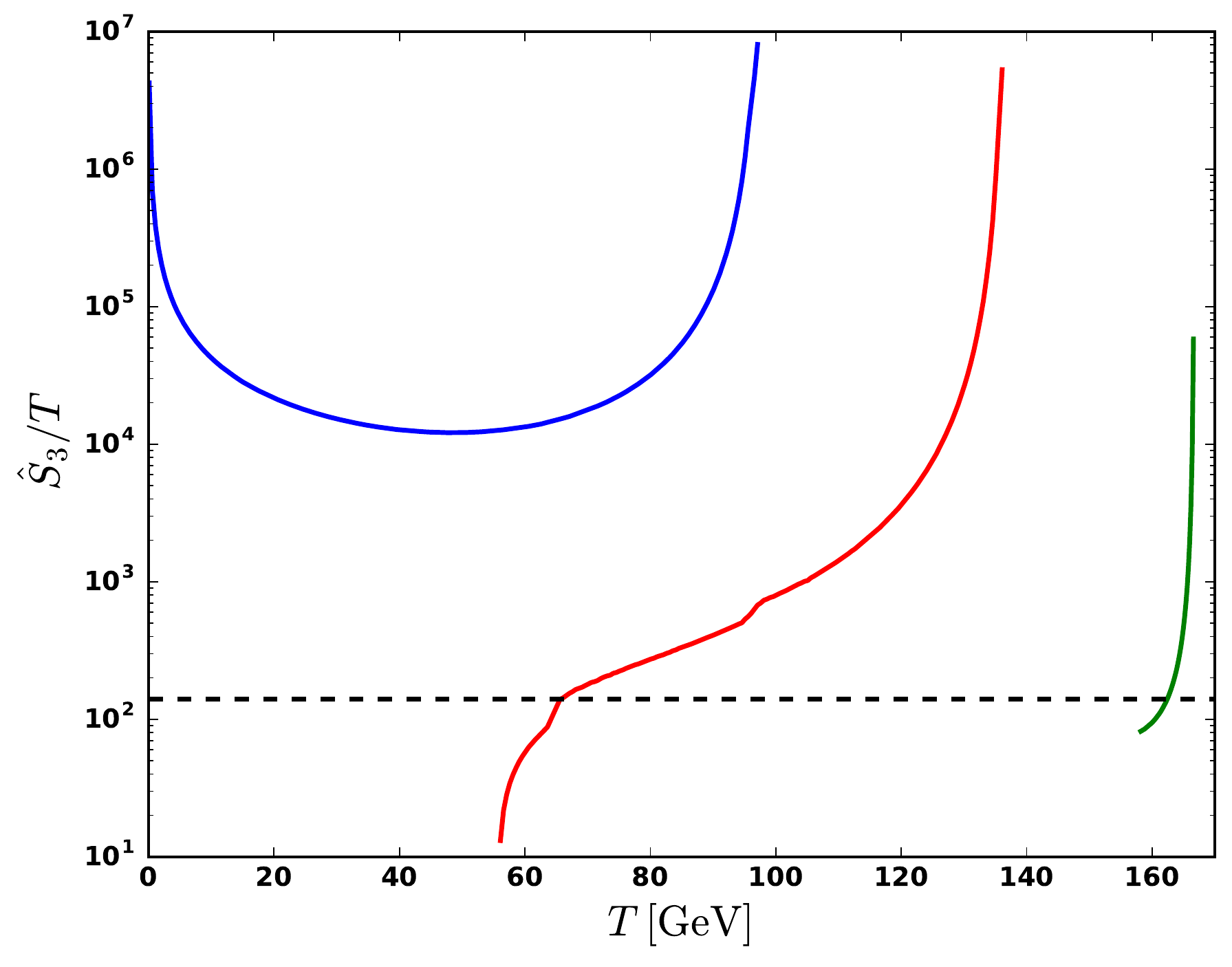}
\caption{\label{fig:supercooling} \it Blue curve: trapped system. Red curve: supercooled transition. Green curve: transition with $T_n \simeq T_c$. 
These lines correspond to the individual parameter space points of the numerical scans specified in Tab.~\ref{tab:parameters_scans}. Blue curve: $M_{h_2}= 275~{\rm GeV}$, 
$\lambda_2 = 1.8$. Red curve: $M_{h_2}= 275~{\rm GeV}$, $\lambda_2 = 2.8$. Green curve: $M_{h_2}= 200~{\rm GeV}$ and $M_s= 350~{\rm GeV}$. The dashed 
line represents the $\hat{S}_3/T = 140$ condition: when it is crossed, the process of bubbles' nucleation occurs.}
\end{figure}

Examples of first order phase transition are shown in Fig.~\ref{fig:supercooling} for three representative combinations of the physical scalar masses $M_{h_2}$, $M_s$ 
and the coupling $\lambda_2$ (for a more detailed discussion of the scalar mass spectra and parameters, see Appendix~\ref{sec:app2}). The nucleation process never 
occurs in the case of the blue curve which is an example of a trapped system. For the case of the red curve, supercooled nucleation happens when $T_n \ll T_c$ (intersection with the dashed line) and, 
for the case of the green curve, nucleation of vacuum bubbles takes place at $T_n \simeq T_c$ without experiencing the supercooling effect.

\section{Numerical analysis} 
\label{Sec:numan}

\subsection{Procedure} 
\label{sec:procedure}

In this Section, we perform a numerical study of the phase transitions in our model using the public package CosmoTransitions \citep{Wainwright:2011kj} to analyse 
the tunneling solutions. Details of the implementation can be found in Appendix \ref{sec:app1}. Here we provide an overview of the steps followed in our calculations:
\begin{enumerate}
\item In each of the scans we use the physical masses and VEVs as input parameters. In particular, we set the SM Higgs boson mass $M_h = 125.09~{\rm GeV}$ and 
the EW breaking VEV $v = 246~{\rm GeV}$. Then, we perform three independent scans by randomly generating two independent parameters while keeping the remaining 
ones fixed as shown in Tab.~\ref{tab:parameters_scans}.
\item We invert the constraint system given in Eq.~\eqref{eq:constraints} of Appendix \ref{sec:app2} to obtain $m_1^2$, $m_2^2$, $m_s^2$ and $\lambda_{12}$. 
We check whether the solution sits in the HMR-1 case and that the minimum at $T=0$ is $(v_1,v_2,0)$. If not, we randomly generate another parameter point and 
repeat the process until both conditions are simultaneously satisfied.
\item We determine the shift in theory parameters necessary to balance the effect of the one-loop corrections. For this effect, using the CW potential, we extract its 
counterterms as detailed in Appendix \ref{sec:app1}.
\item We then compute the full effective potential at $T \neq 0$. Using CosmoTransitions we track the possible minima in terms of the temperature. For each point 
in field space, we determine the numerical eigenvalues of the matrices $A$ and $G$ given in Eqs.~\eqref{eq:A} and \eqref{eq:G}, respectively, in order to extract 
$M_s$ and $M_{h_2}$. Then we calculate the $J_{B/F}$ functions in its exact form and add up all contributions.
\item Finally, we extract the critical temperature, the nucleation temperature and classify the transition, saving this information. Then, we are able to identify the transition 
pattern and to compute the strength of the transition, as well as the nucleation temperature by solving $\hat{S_3}/T_n = 140$, where $\hat{S_3}$ is obtained using 
CosmoTransitions without assuming the thin-wall approximation. Note that, as pointed out in Ref.~\citep{Alanne:2016wtx}, the thin-wall approximation is not realistic 
for the supercooled transitions.
\end{enumerate}
\begin{table}
\centering
\begin{tabular}{|l|ccccccc|} \hline
& $M_{h_2}/{\rm GeV}$ & $M_s/{\rm GeV}$ &  $\lambda_1$ & $\lambda_2$ & $\lambda_s$ & $\lambda_{s_1}$ & $\lambda_{s_2}$ \\\hline
Scan $(M_{h_2},\lambda_2)$ & $[150,400]$ & 150 & 0.6 & $[1,4]$ & 0.3 & 0.8 & 1 \\\hline
Scan $(M_s,M_{h_2})$ & $[150,400]$ & $[150,500]$ & 0.5 & 2.5 & 2.5 & 3 & 3\\\hline
Scan $(M_s,\lambda_{s_1},\lambda_{s_2})$ & 300 & $[140,500]$ & 1 & 2 & 1 & $[1,4]$ & $\lambda_{s_1}$ \\\hline
\end{tabular} 
\caption{\label{tab:parameters_scans} 
\it The physical mass and coupling parameters used in the numerical scans.}
\end{table}

\subsection{Results for the HMR-1 case} 
\label{sec:Num}

A major quest of our study is to determine whether the strong first-order phase transitions can happen 
in the considering model. This gains a particular relevance in the context of baryogenesis and production of gravitational waves as they can only occur if 
the phase transition is strong enough. The usual criterion for defining a \textit{strong first-order phase transition} widely used in the literature is by means 
of the following ratio
\begin{equation}
\frac{v_c}{T_c} \gtrsim 1 \,,
\label{strongPT_critical}
\end{equation}
where $v_c = v(T_c)$ is the value of the EW scale $v \equiv \sqrt{v_1^2+v_2^2}$ in the broken phase at the critical temperature $T_c$.
This criterion follows from the requirement of sphaleron suppression in the broken phase. Note that there is no very precise consensus on this value, which 
we found to vary between 0.6 and 1.5. In our case, due to the possibility of supercooling as discussed above, that is, $T_n \ll T_c$, we replace the critical 
temperature by the nucleation temperature \citep{Kurup:2017dzf}
\begin{equation}
\frac{v_n}{T_n} \gtrsim 1 \,,
\label{strongPT_nucleation}
\end{equation}
where $v_n=v(T_n)$. In fact, as bubbles are nucleated at $T_n$, only then we can speak of baryon number violation at the boundary as well as gravitational
waves production. Therefore, we consider that Eq.~\eqref{strongPT_nucleation} is indeed the correct criterion to apply in our study. Since the VEVs do not 
significantly vary with temperature, the difference between the more realistic criterion \eqref{strongPT_nucleation} and the usual criterion 
\eqref{strongPT_critical} is mostly due to a difference between the critical and the nucleation temperatures. 
In particular, noting that the nucleation temperature is smaller, the transition is expected to be stronger. We will show that in the case of a strong first-order 
transition involving the phase $(0,0,v_s)$, the nucleation temperature is typically two times smaller than the critical temperature. We will therefore consider 
such transition to be strong, and thus relevant for baryogenesis, if
\begin{equation}\label{eq:vcTc}
\frac{v_c}{T_c} > 0.5\,.
\end{equation}
Given the lack of a precise consensus in the literature about this value we consider \eqref{eq:vcTc} adequate for our study.
\begin{figure}[!ht]
\centering 
\includegraphics[width=1\linewidth]{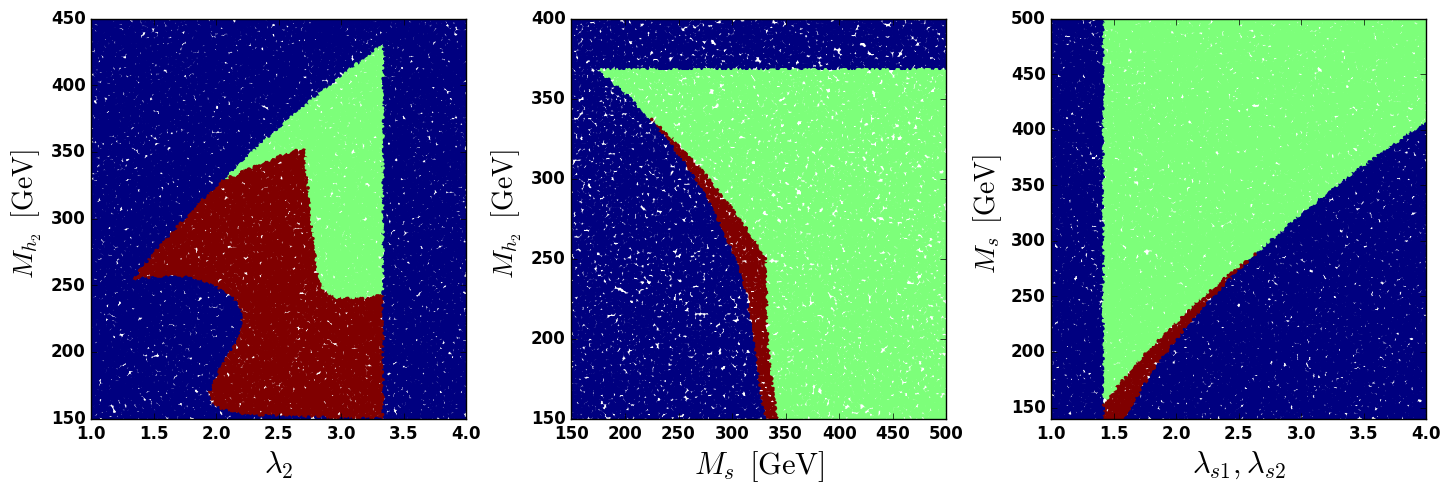}
\caption{\label{fig:scan_treelvl}
\it Tree-level expectations obtained by implementation of the analytical results at order $(m/T)^2$ found above. Left: scan $(M_{h2},\lambda_2)$. 
Center: scan $(M_s,M_{h2})$. Right: scan $(M_s,\lambda_{s1},\lambda_{s2})$. The blue region represents the solutions where $T=0$ vacuum 
is not the EW one, or which are not in the HMR-1 case. The green region represents the solutions where the EWPT is of the second order at tree level. 
The red region represents the solutions with first-order phase transitions at tree-level, hence, they are expected to be strong.}
\end{figure}

In what follows we analyze three representative scans over the parameter space (see Tab.~\ref{tab:parameters_scans}) in order to study interesting features 
of the considering SHUT inspired model. In Fig.~\ref{fig:scan_treelvl} we show the tree-level expectations for each scan, i.e.~using the analytical results 
to order $(m/T)^2$ in the high temperature expansion, which allow us to discriminate
\begin{itemize}
\item[i)] the cases where the vacuum $(v_1,v_2,0)$ is not developed at $T=0$ or when the HMR-1 case is not satisfied (blue region in Fig.~\ref{fig:scan_treelvl});
\item[ii)] the cases where we expect a transition pattern starting from $(0,0,0) \to (0,0,v_s)$ stage, therefore, the EWPT is of the first order at tree-level 
(red region in Fig.~\ref{fig:scan_treelvl});
\item[iii)] and the cases where we expect a transition pattern starting by $(0,0,0) \to (v_1,0,0)$ or $(0,0,0) \to (0,v_2,0)$ stages, meaning that, apart from marginal 
cases, the EWPT is of the second order at tree level (green region in Fig.~\ref{fig:scan_treelvl}).
\end{itemize}
In the following, we refer to the patterns $(0,0,0) \to (0,0,v_s) \to ...$ as the \textit{strong patterns} since they are already of the first order at tree level, and to the other 
patterns (not involving the minimum $(0,0,v_s)$) as the \textit{weak patterns} as they are only of the second order at tree level.

The advantage of tree-level calculations with the corresponding results shown in Fig.~\ref{fig:scan_treelvl} is that they are fast to perform on an ordinary computer.
We use these results to quickly determine the potentially interesting regions of the parameter space. In general, the green regions (weak patterns) are characterized 
by very weak transitions, which owe their first-order character to the CW and thermal corrections only, in particular, due to the effect of loop-induced cubic terms. 
The red regions (strong patterns) correspond to usually very strong transitions, and often \textit{too strong} in a sense that, as already pointed out in 
Refs.~\citep{Alanne:2016wtx,Kurup:2017dzf}, they suffer from supercooling which, in some cases, may prevent the bubble nucleation process.

Figs.~\ref{fig:scan_Mh2l2}, \ref{fig:scan_MpMh2} and \ref{fig:scan_Mplp12} show the results of the full computation including the higher order thermal corrections, 
the one-loop CW quantum contributions \eqref{HO_corrections} and the counterterms \eqref{CT_corrections}. Each figure shows the type of the transition 
(i.e.~the transition pattern), the strength of the transition $v_c/T_c$, the critical temperature and the nucleation temperature.

\subsubsection*{Parameter scan I: $(M_{h_2},\,\lambda_2)$}

\begin{figure}[!ht]
\centering 
\includegraphics[width=1\linewidth]{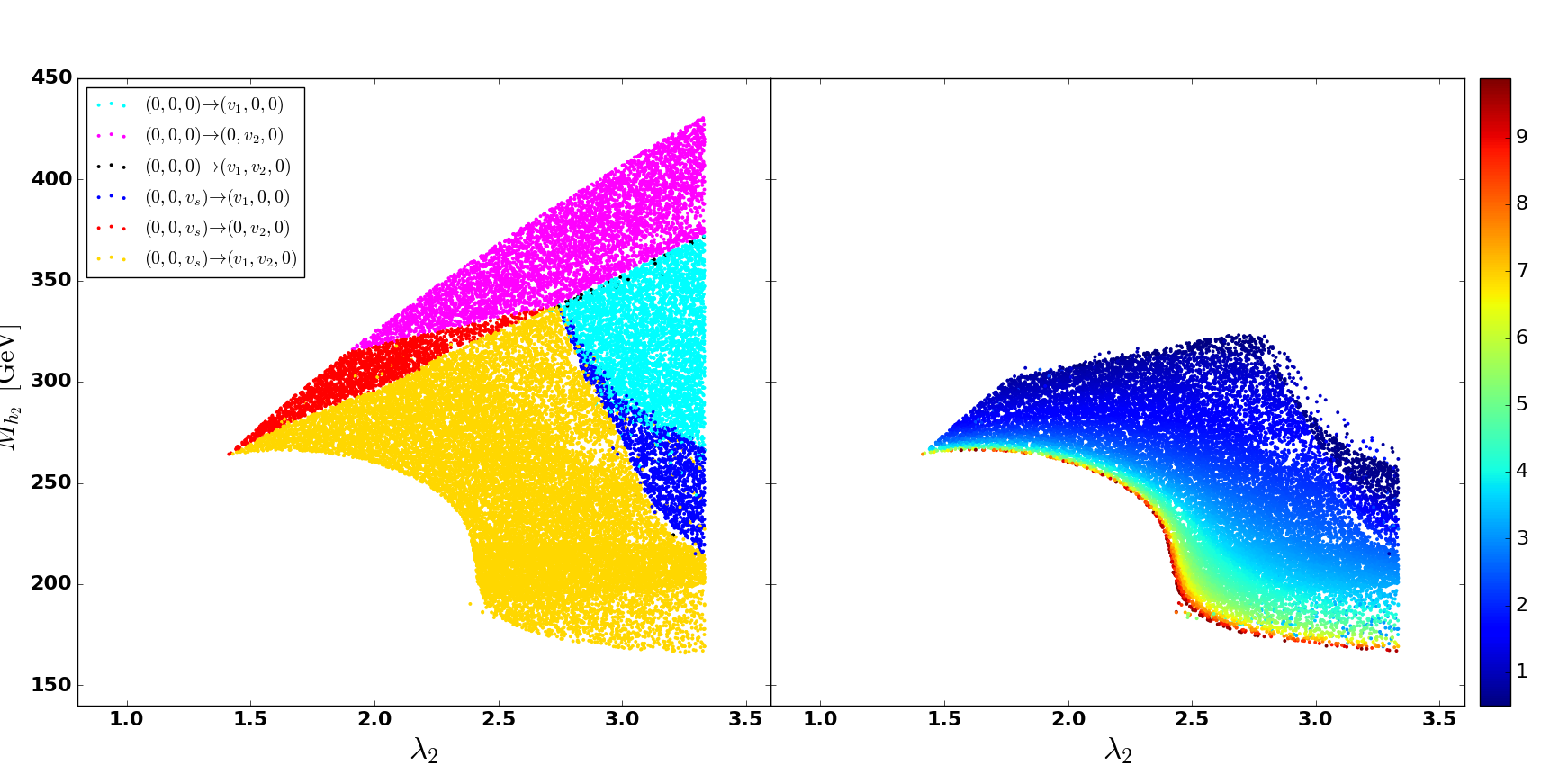}
\includegraphics[width=1\linewidth]{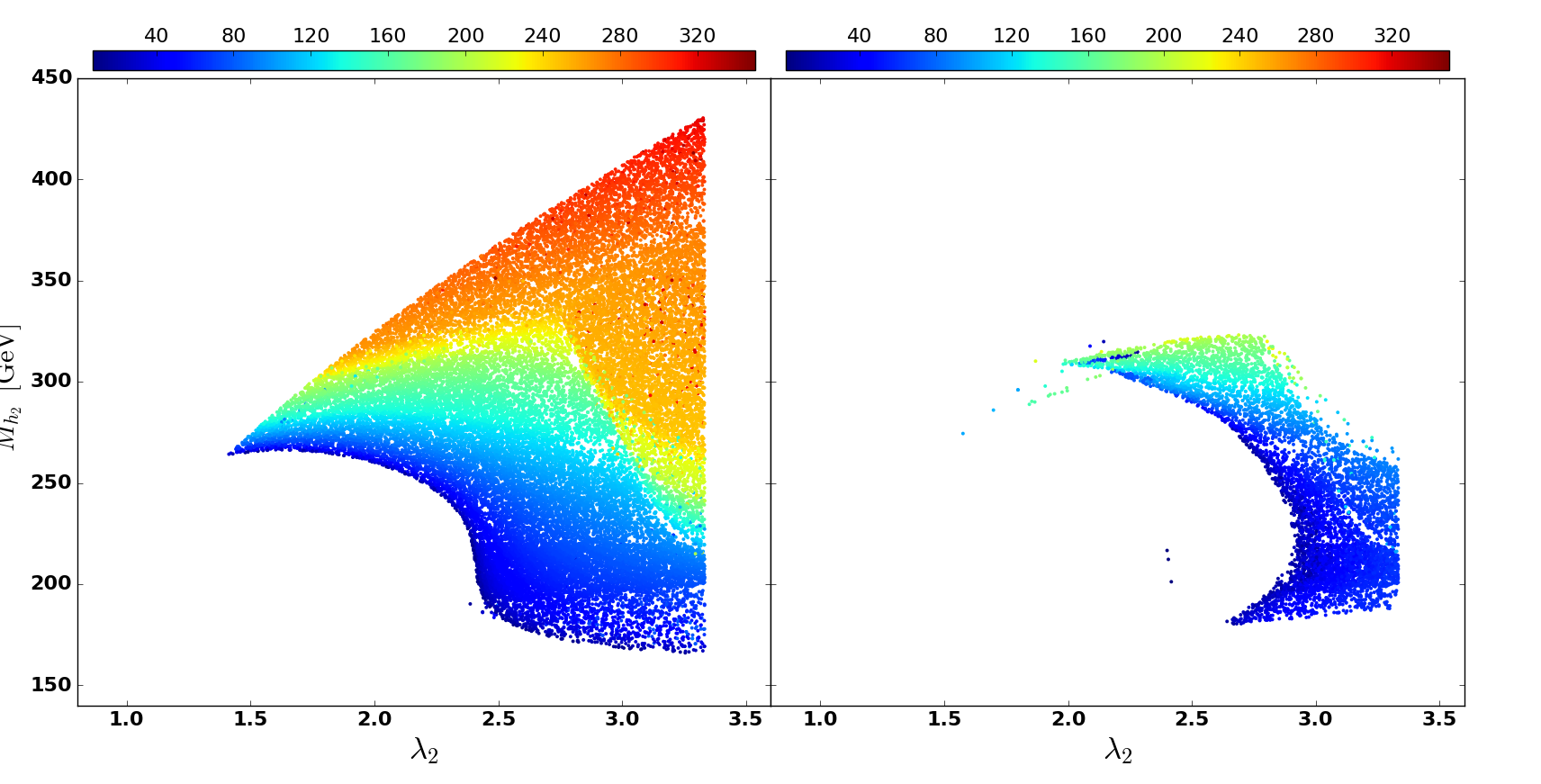}
\caption{\label{fig:scan_Mh2l2}\it 
\textbf{Top-left panel:} identification of the different types of transitions. 
\textbf{Top-right panel:} identification of strong first-order phase transitions. 
The colour scale on these panels indicates the ratio $v_c/T_c$. 
\textbf{Bottom-left panel:} the same as top-left panel but 
with dependence on the critical temperature represented by the colour scale. 
\textbf{Bottom-right:} identification of the strong first-order phase transitions 
where nucleation of bubbles occurs, with dependence on the nucleation temperature 
represented by the colour scale.}
\end{figure}

This first scan is limited to a slice in the parameter space that corresponds to the tree-level predictions in the left panel of Fig.~\ref{fig:scan_treelvl}. According to 
Tab.~\ref{tab:parameters_scans}, it is characterized by mixing couplings of order $\lambda_{ij} \sim 1$. Note that the $\lambda_{12}$ parameter, also of order one, 
is extracted from the conditions derived in Appendix~\ref{sec:app2}. We find a good agreement between the tree-level expectations and the characteristics of transitions 
computed accounting for the full thermal and loop corrections. In particular, while the green region in Fig.~\ref{fig:scan_treelvl} corresponds to the light blue and pink 
regions of the top-left panel of Fig.~\ref{fig:scan_Mh2l2}, the red region in Fig.~\ref{fig:scan_treelvl} corresponds to the red, blue, and yellow regions of the same top-left 
panel of Fig.~\ref{fig:scan_Mh2l2}. However, a discrepancy occurs for small values of the $H_2$ mass, $M_{h_2}$: while at $T=0$ the minimum $(v_1,v_2,0)$ 
is expected at tree level, the thermal corrections may switch it to another irrelevant minimum, as e.g.~$(0,0,v_s)$ or even $(0,v_2,v_s)$. This can be seen in the sparser 
region at the bottom of the top-left panel. Such regions with relatively light BSM scalars ($m < 200~\mathrm{GeV}$) may, in addition, be well constrained by collider 
and flavor physics measurements which should then be taken into consideration in a more dedicated future study.

Referring now to the top-right panel of Fig.~\ref{fig:scan_Mh2l2}, we introduce a colour scale to represent the ratio $v_c/T_c$, providing a measure of how strong a first-order 
phase transition can be. Here, we reject all points where the first-order transition is weak, that is $v_c/T_c < 0.5$, thus not interesting for baryogenesis. Without surprise, 
the identified transitions $(0,0,0) \to (v_1,0,0) \to (v_1,v_2,0)$, (light blue region on the top-left panel) and $(0,0,0) \to (0,v_2,0) \to (v_1,v_2,0)$ (pink region on top-left 
panel) are indeed too weak and do not fulfill the criterion for the sphaleron suppression. Such regions are now almost absent as we can see on the top-right panel of 
Fig.~\ref{fig:scan_Mh2l2}. On the other hand, the transition $(0,0,0) \to (0,0,v_s) \to (v_1,v_2,0)$ (yellow region on the top-left panel) can be very strong with 
$v_c/T_c$ predominately around 2-3 but also reaching larger values up to 10.

On the bottom-left panel of Fig.~\ref{fig:scan_Mh2l2} we investigate the dependence on the critical temperature represented by the colour scale. There is a clear 
correlation between the type of phase transition, hence its strength, and the critical temperature. The stronger the transition, the lower the critical temperature. 
In particular, for the pattern $(0,0,0) \to (0,0,v_s) \to (v_1,v_2,0)$, the critical temperature is usually below $100~{\rm GeV}$, while for the weak patterns it 
is mostly between $250$ and $350~{\rm GeV}$.

Finally, for the bottom-right panel of Fig.~\ref{fig:scan_Mh2l2}, we impose a second cut where, among all the points satisfying the criterion for a strong first-order 
transition, the very strong ones where supercooling prevents nucleation, i.e.~when $\hat{S}_3/T = 140$ has no solution, are rejected. For the allowed transitions 
the nucleation temperature is much smaller than the critical temperature, which corresponds to the supercooled states represented by the red line 
in Fig.~\ref{fig:supercooling}. We are then left with an interesting smaller region of the parameter space where the strength of the transition is $v_c/T_c \sim 1$ 
and nucleation of bubbles is realized making such transitions possible candidates for baryogenesis.

We can now confront the numerical results obtained in Fig.~\ref{fig:scan_Mh2l2} with what would be expected in the high temperature regime. In particular, we use 
the results obtained in Sec.~\ref{Sec:analytical_results_HMR1} at order $(m/T)^2$, and perform an analytical scan using the following procedure:
\begin{itemize}
\item[i)] Compute the critical temperatures of the three strong patterns using the analytical expressions in Eqs.~\eqref{eq_analytical_A}, 
\eqref{eq_analytical_B} and \eqref{eq_analytical_C}. There are at most four well-defined temperatures (two possibilities for pattern (C)).
\item[ii)] For each pattern, check that the initial and final phases exist and are stable, using the analytical expressions derived in Sec.~\ref{sec:structure}.
\item[iii)] Compare the critical temperatures of the existing patterns and select the larger one.
\end{itemize}
\begin{figure}
\centering 
\includegraphics[width=\linewidth]{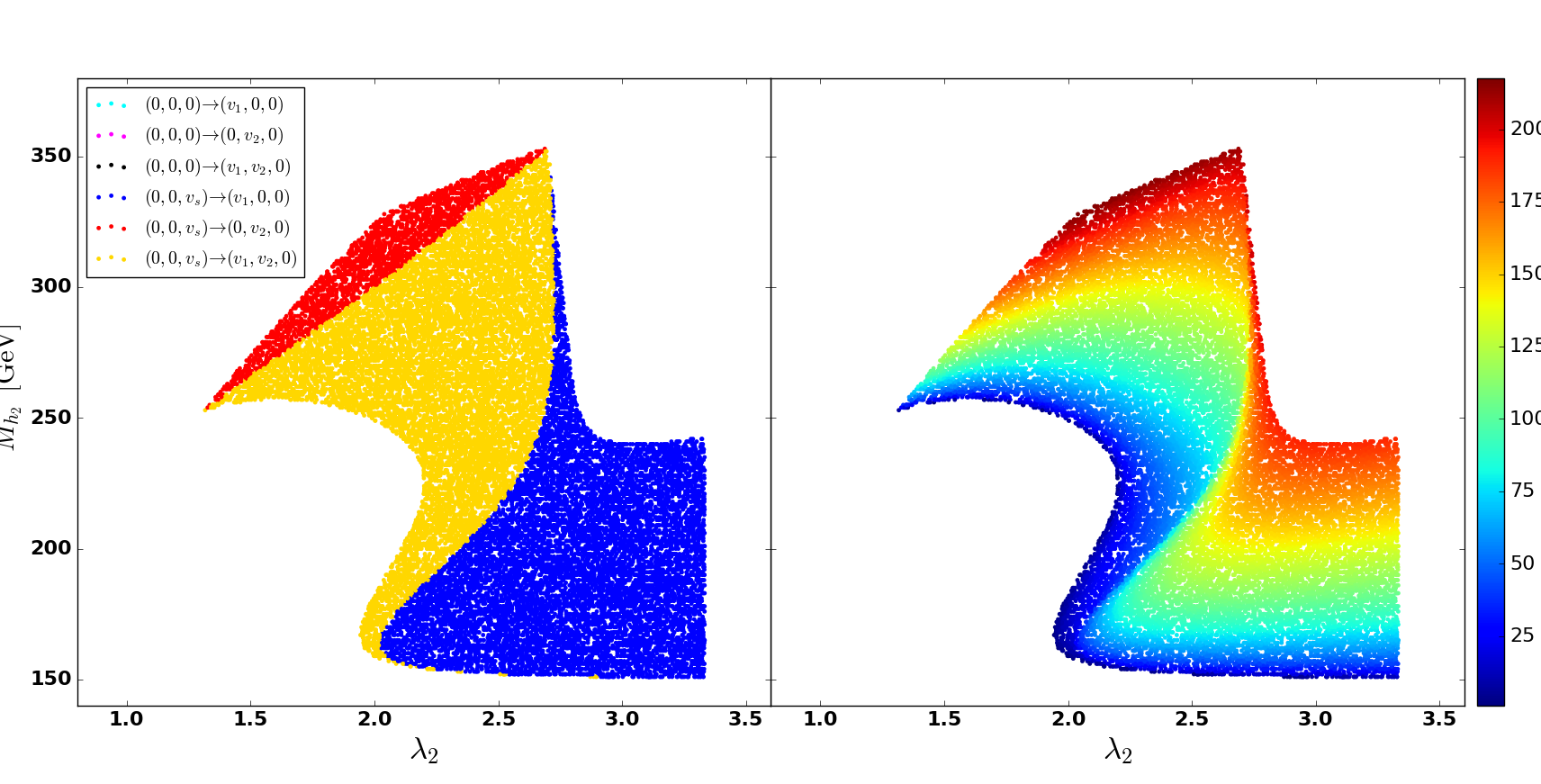}
  \caption{\label{fig:scan_analytical} 
  \it Analytical scan. 
  \textbf{Left panel:} identification of different types of the phase transitions. 
  \textbf{Right panel:} the same but with dependence on critical temperature 
  reflected in the color scale.}
\end{figure}
The results displayed in Fig.~\ref{fig:scan_analytical}, which only concern the strong patterns, show a good agreement with the numerical scan for $M_{h_2} > 250$ GeV. However, some discrepancies with respect to the type of transition are noticed
for smaller masses, which also includes a significant region between $2.0 \lesssim \lambda_2 \lesssim 2.5$ that is absent in the numerical scan. This clearly shows the limitations of 
the high temperature expansion, where we have $m(\phi)^2 \sim -M^2 + c T^2$, with $m(\phi)$ being the field-dependent mass that appears in the 
$(m(\phi)/T)$-expansion, and $M$ the physical mass (e.g.~$M_{h_2}$). In fact, for the case of large $M_{h_2}$, the field-dependent masses are small at the transition stage,
and the critical temperature is large, hence the expansion in $m(\phi)/T \sim \mathcal{O}(0.1)$ is valid and yields a good agreement with the numerical results. 
On the other hand, for small $M_{h_2}$ we have $m(\phi)/T > 1$ due to large field-dependent masses. Therefore, when a transition of type (A) (blue dots) is analytically 
expected, it does not actually occur as a consequence of an overestimation of the critical temperature by the high-temperature expansion. In reality, the system has to ``wait'' 
for a smaller temperature when the phase $(v_1,0,0)$ becomes $(v_1,v_2,0)$, so that the transition $(0,0,v_s) \to (v_1,0,0)$ (blue dots) becomes $(0,0,v_s) \to (v_1,v_2,0)$ (yellow dots). This is the reason 
why we see several yellow dots in the numerical results not accounted for by the high-temperature expansion. Furthermore, it may also happen that the transition never occurs, which is why a lot of points in the range $2.0 \lesssim \lambda_2 \lesssim 2.5$ are lost. This example suggests that it is indeed important to consider the higher-order thermal corrections 
even for strong transitions already present at order $(m/T)^2$.

In summary, for the case of mixing couplings of order one, the first-order phase transition already present at tree level tend to be too strong preventing nucleation 
to occur as a result of supercooled metastable states. On the other hand, the first-order phase transition generated at one loop are too weak to be of interest 
for baryogenesis.  The problem of too weak transitions is well-known, as e.g.~in the SM, while the problem of too strong transitions was recently pointed out by 
Ref.~\citep{Alanne:2016wtx} focusing on the transition pattern $(0,0,0) \to (0,0,v_s) \to (v_1,v_2,0)$ and by Ref.~\citep{Kurup:2017dzf} in the $\mathbb{Z}_2$-symmetric 
real-singlet extended SM, where the transition is also present at tree level. 

\subsubsection*{Parameter scan II: $(M_s,\,M_{h_2})$}

\begin{figure}[!ht]
\centering 
\includegraphics[width=1\linewidth]{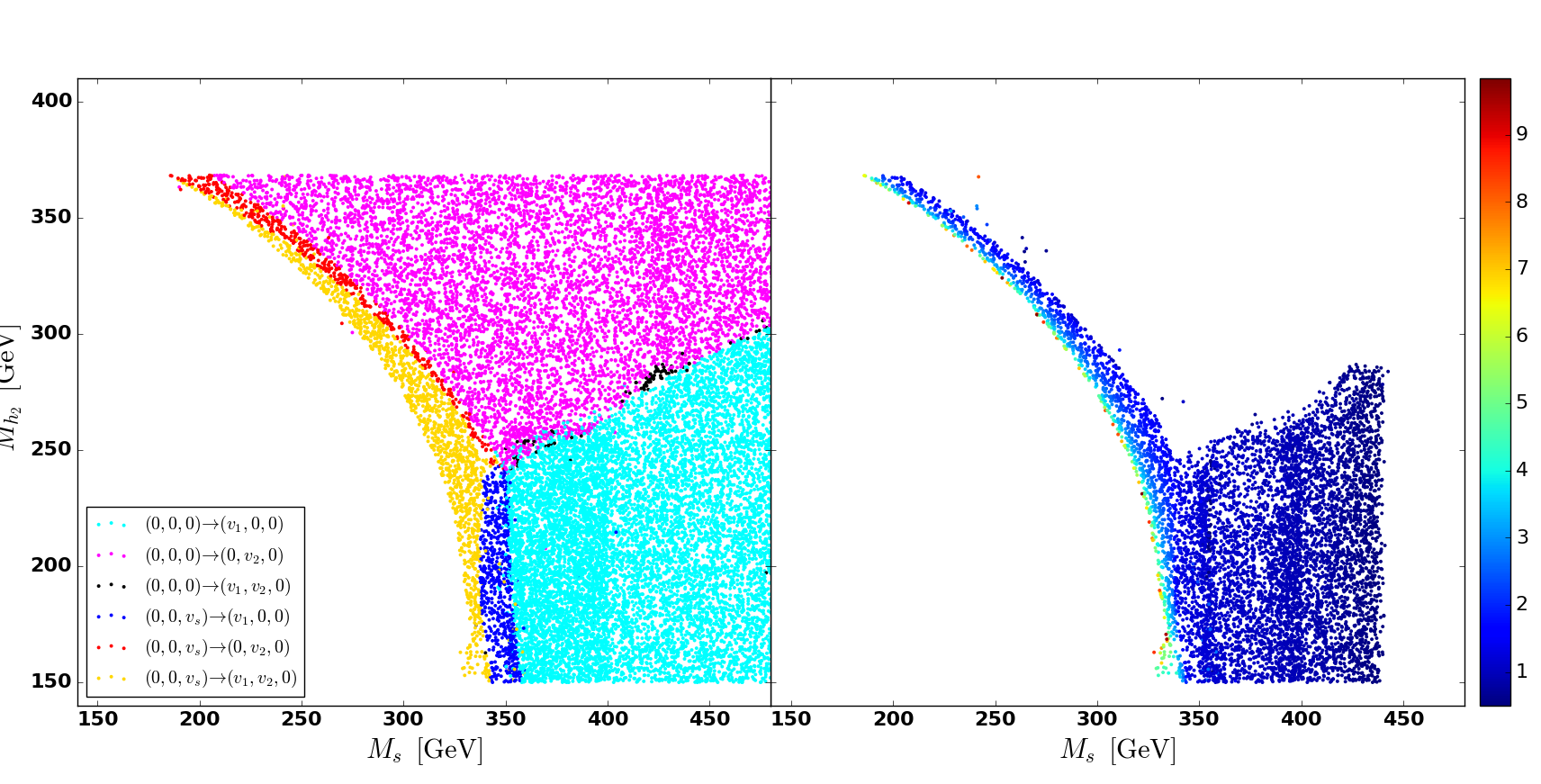}
\includegraphics[width=1\linewidth]{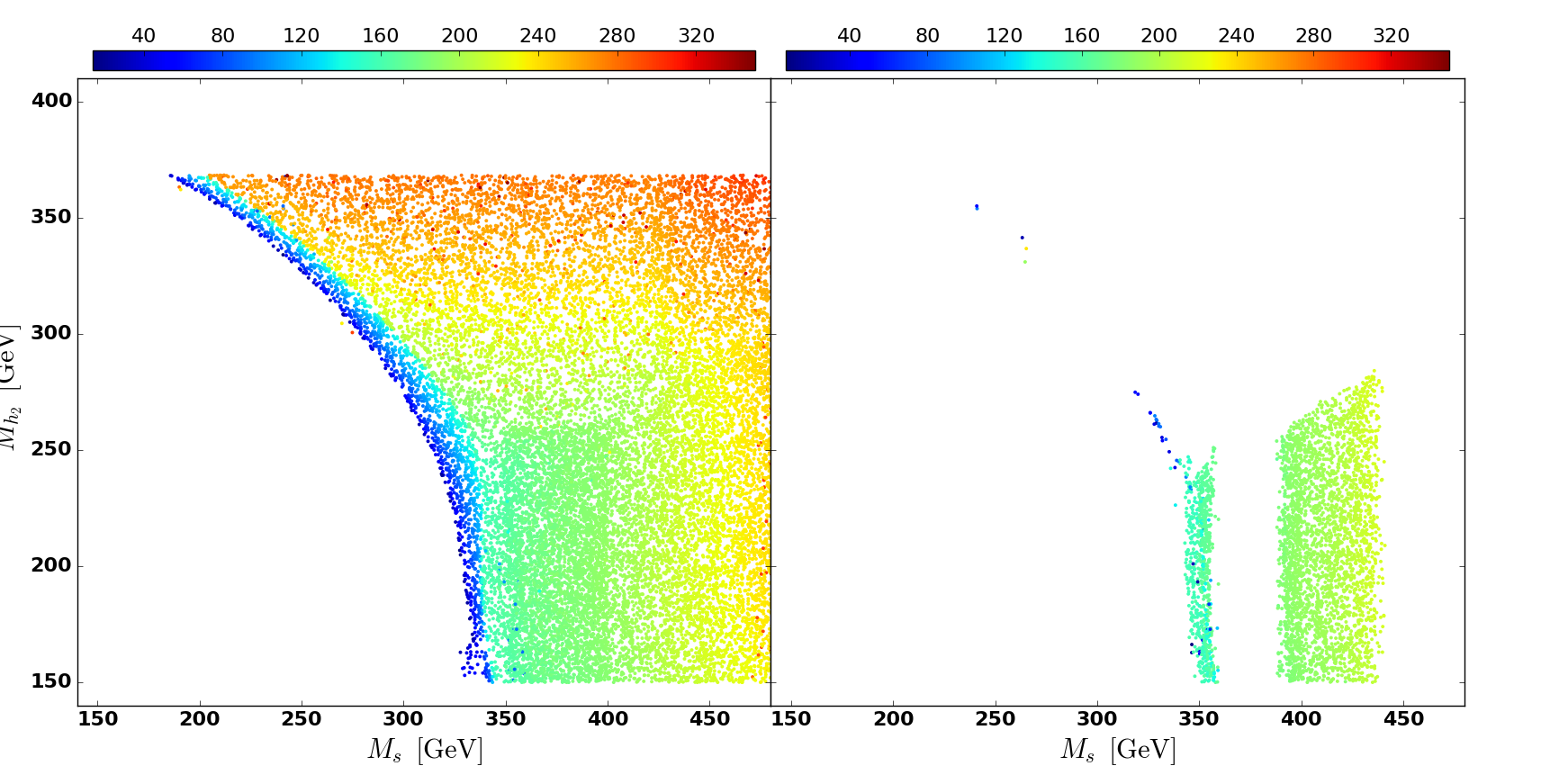}
\caption{\label{fig:scan_MpMh2}\it 
\textbf{Top-left panel:} identification of different types of the phase transitions. 
\textbf{Top-right panel:} identification of strong first-order phase transitions. The colour scale indicates the ratio $v_c/T_c$. 
\textbf{Bottom-left panel:} the same but with dependence on the critical temperature represented by the colour scale. 
\textbf{Bottom-right:} identification of strong first-order phase transitions where nucleation of bubbles occurs.}
\end{figure}

For the second scan we study how the thermal corrections affect the tree-level predictions shown in the middle panel of Fig.~\ref{fig:scan_treelvl}. Comparing with 
the top-left and the bottom-left panels of Fig.~\ref{fig:scan_MpMh2}, where the colours have the same meaning as in Fig.~\ref{fig:scan_Mh2l2}, we observe that 
the full computation agrees with the tree-level one. As it was already observed, the patterns $(0,0,0) \to (0,v_2,0) \to (v_1,v_2,0)$ (pink region), as well as many of 
the points corresponding to the pattern $(0,0,0) \to (v_1,0,0) \to (v_1,v_2,0)$ (light blue region), are too weak and absent in both top-right and bottom-right panels in 
Fig.~\ref{fig:scan_MpMh2}. Recall that, in addition to the weakly first-order phase transitions, the supercooling scenarios without nucleation of bubbles are also removed 
in the bottom-right panel, which only contains points where nucleation of bubbles takes place. However, there is a region where the pattern $(0,0,0) \to (v_1,0,0) 
\to (v_1,v_2,0)$ is strong enough to fit the criterion $v_c/T_c \sim 1$, where most of the points have the nucleation temperature close to the critical one.


The key differences between the current scan $(M_s,\,M_{h_2})$ and the previous one $(M_{h_2},\,\lambda_2)$ are the values of the couplings and the mass of 
the singlet $\Phi$. Here, the singlet is rather strongly coupled to the Higgs bosons originating from the Higgs doublets due to a much larger mixing than before, 
which explains why the pattern $(0,0,0) \to (v_1,0,0) \to (v_1,v_2,0)$ being of the second order at tree level, may become a strongly first-order transition once the 
loop and thermal contributions are included. As a result of such large couplings, the higher order corrections are clearly non-negligible. With this analysis, we have shown 
that the strong first-order phase transitions generated at one loop are expected in such multi-scalar models as the considering THDSM and need to be carefully accounted for. 
In other words, in order to fully characterize the phase transitions, it is not sufficient to solely rely on the strong patterns already present at tree level and summarized 
in Fig.~\ref{fig:pyramids}.

\subsubsection*{Parameter scan III: $(M_s,\,\lambda_{s1},\,\lambda_{s2})$}

\begin{figure}[!ht]
\centering 
\includegraphics[width=1\linewidth]{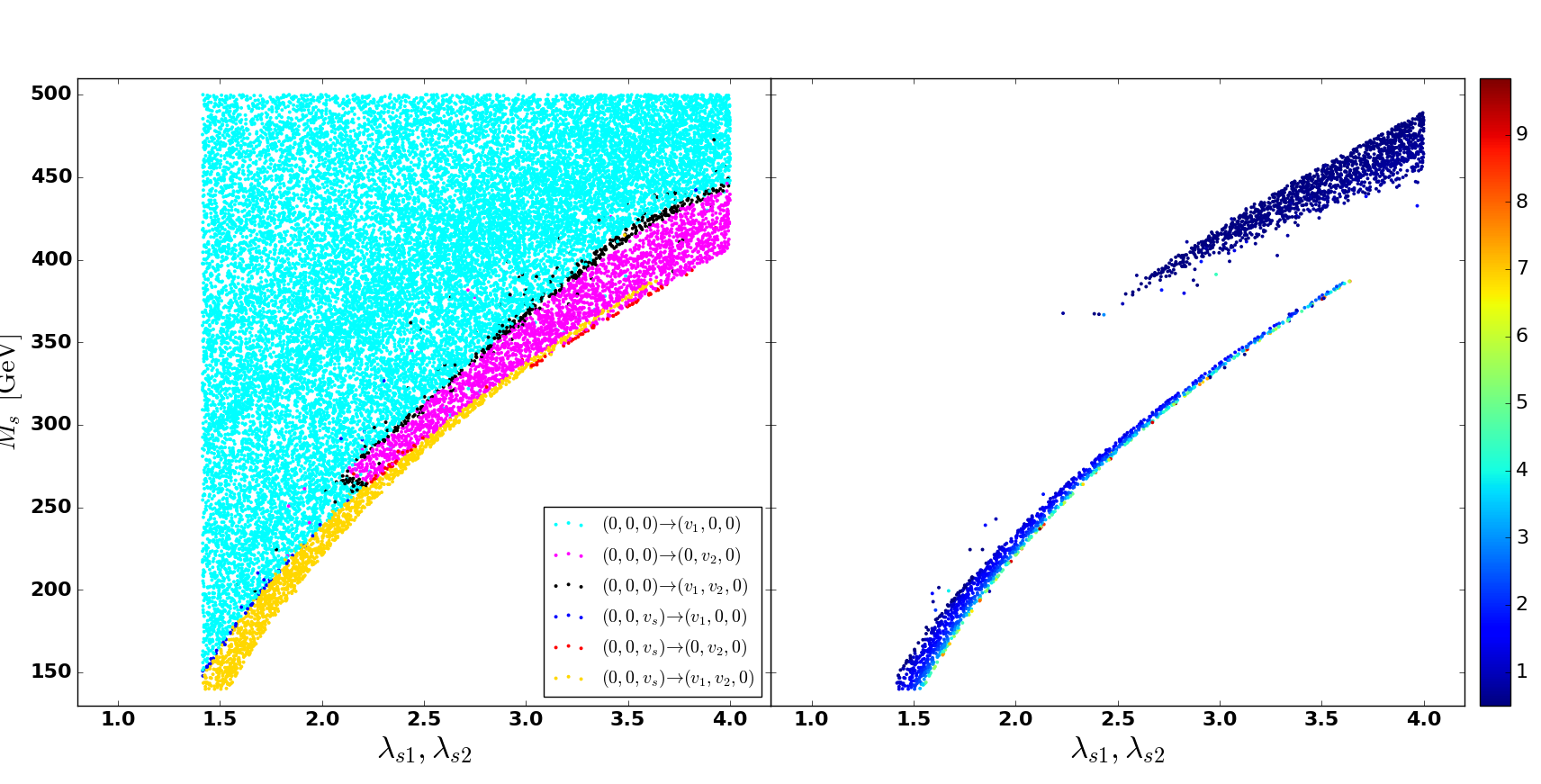}
\includegraphics[width=1\linewidth]{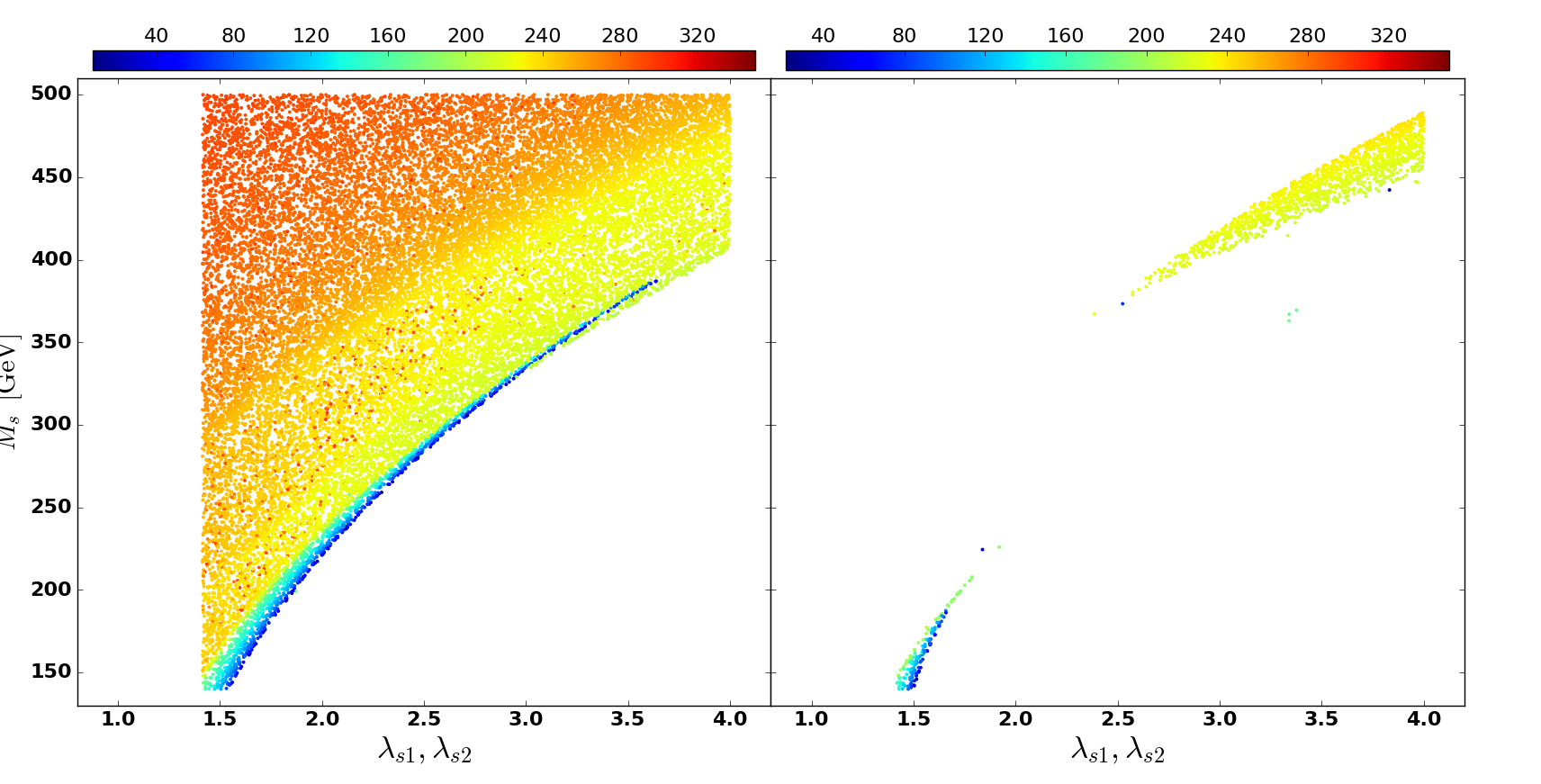}
\caption{\label{fig:scan_Mplp12}\it \textbf{Top-left panel:} identification of different types of the phase transitions. 
\textbf{Top-right panel:} identification of strong first-order phase transitions. The colour scale indicates the ratio $v_c/T_c$. 
\textbf{Bottom-left panel:} the same but with dependence on the critical temperature represented by the colour scale. 
\textbf{Bottom-right:} identification of strong first-order phase transitions where nucleation of bubbles occurs.}
\end{figure}
In the previous scan we have verified that larger values of the mixing parameters $\lambda_{s1}$ and $\lambda_{s2}$ have a significant impact on promoting 
the second-order phase transitions at tree level to strongly first-order ones at quantum level. Besides, we have observed that when those parameters are considered 
to be small, the interesting strong first-order phase transitions need to be present readily at tree level. For completeness, here we perform a third scan where 
we study the dependence on $M_s$, $\lambda_{s1}$ and $\lambda_{s2}$ parameters.

As it can be seen in Fig.~\ref{fig:scan_Mplp12}, for small values of the singlet mass $M_s$ and its mixing couplings, those patterns where the phase transitions 
are of the weakly first order or of the second order at tree level (pink and light blue regions) remain weak after thermal and loop corrections are included, thus removed 
from both right panels of Fig.~\ref{fig:scan_Mplp12}. On the other hand, larger values of the mixing parameters, where the strength of the couplings between the singlet 
and the doublet states becomes larger, results in strongly first-order phase transitions after thermal and loop corrections are applied, hence favourable for baryogenesis. 
So, it is typically easier to find a strong first-order phase transition when the mixing couplings are large providing an enhanced effect from additional scalars.
This can be noticed in the bottom-right panel of Fig.~\ref{fig:scan_Mplp12}, where the large mixing solutions are concentrated in the light-green triangular shaped region. 
As was mentioned above, for this type of points, the nucleation temperature tends to approach the critical one as one can see by comparing both panels on the bottom. 
In Fig.~\ref{fig:scan_Mplp12} (bottom-right panel) it is also possible to observe the supercooled scenarios with the strong first-order transitions for both small singlet scalar 
mass and small mixing couplings. This region matches that of the first scan and, as expected, is in agreement with the results discussed above.

Performing now a comparison with the second scan, if we go back to Fig.~\ref{fig:scan_MpMh2}, which represents the results of a scenario where the mixing couplings 
are large, i.e.~$\lambda_{s1} = \lambda_{s2} = 4$, we see that the strong tree-level patterns (yellow, blue and red bands of the top-left panel) disappear when 
$M_s \gtrsim 350~{\rm GeV}$. A similar trend is observed in the top-left panel of Fig.~\ref{fig:scan_Mplp12} where the same bands survive almost up to 
$M_s \simeq 400~{\rm GeV}$ but, in agreement with Fig.~\ref{fig:scan_MpMh2}, fade away when $\lambda_{s1}$ and $\lambda_{s2}$ approach $4$. 
However, in both Figs.~\ref{fig:scan_MpMh2} and \ref{fig:scan_Mplp12} we see that large mixing couplings enhance the loop contributions whose effects 
transform weak tree-level patterns into strongly first-order phase transitions. All in all, we observe the same type of behavior as already discussed in 
the previous two scans.

\begin{figure}
\centering 
\includegraphics[width=1\linewidth]{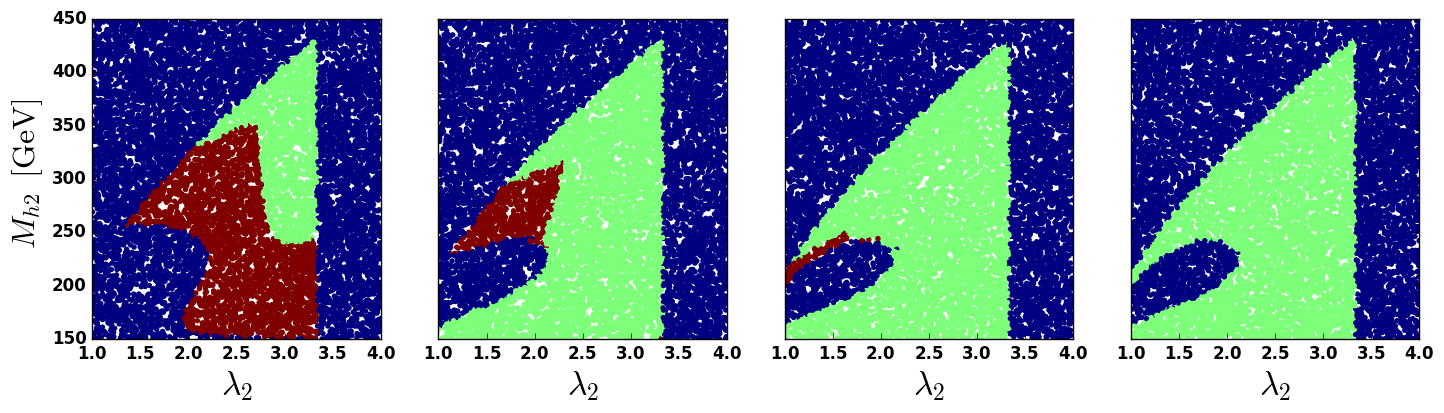}
\caption{\label{fig:scan_tree_lvl_Ms}
\it Tree-level expectation for varying $M_s$. Other parameters are fixed as in Tab.~\ref{tab:parameters_scans} used for the previous 
scan $(M_{h_2},\lambda_2)$. From left to right: $M_s = 150~{\rm GeV}$, $160~{\rm GeV}$, $180~{\rm GeV}$, $200~{\rm GeV}$. 
The color scheme is the same as in Fig.~\ref{fig:scan_treelvl}.}
\end{figure}

It was the presence of strong first-order phase transitions at tree level that leads us to fix $M_s = 150~{\rm GeV}$ for the first scan, such that we could have 
plenty of interesting points to investigate. To clearly see this effect, if we recover the tree-level expectations for the first scan, but now choose four different 
values for the singlet mass, $M_s= 150~\mathrm{GeV}$, $160~\mathrm{GeV}$, $180~\mathrm{GeV}$ and $200~\mathrm{GeV}$, we observe again that 
with increasing $M_s$, the strong patterns at tree level tend to disappear as shown in Fig.~\ref{fig:scan_tree_lvl_Ms}. This behavior confirms that for larger 
values of $M_s$ we do need the thermal and loop corrections to trigger the strongly first-order phase transitions.

In summary, even knowing that strong patterns at tree-level are highly favoured by lighter singlets, we note that from the results of the second and third scans above 
that the dependence on $\lambda_{s1}$ and $\lambda_{s2}$ plays an important role in promoting the tree-level weak patterns to strongly first-order phase transitions 
at one loop. Therefore, we can conclude that, while a light singlet on its own favours the possibility for strongly first-order phase transitions, a heavy singlet is also 
not a problem due to the effect of loop and thermal corrections. In particular, it is due to thermal corrections that go beyond the leading $(m/T)^2$ order that 
such effects can be enhanced. This conclusion goes beyond the discussion in Ref.~\citep{Alanne:2016wtx} where it is claimed that a heavy singlet is not compatible 
with baryogenesis. In fact, Ref.~\citep{Alanne:2016wtx} only considers the leading contributions to the scalar potential, with which it is not possible to transform 
the tree-level second-order phase transitions into the one-loop strongly first-order ones, as the shape of the potential is unaffected 
(see also Sect.~\ref{Sec:analytical_results_HMR1}). 

Finally, the existing experimental data may put strong constraints on the masses and interactions of light scalars. In a perspective, it would be interesting to check 
if these constraints are significant or not for baryogenesis in the considering THDSM. On the other hand, it is worth noticing that for the case of small mixing parameters 
making the light states hardly observable, baryogenesis can still be triggered by supercooled transitions.

\subsubsection*{Nucleation temperature}

\begin{figure}[!ht]
\centering
\includegraphics[width=0.48\linewidth]{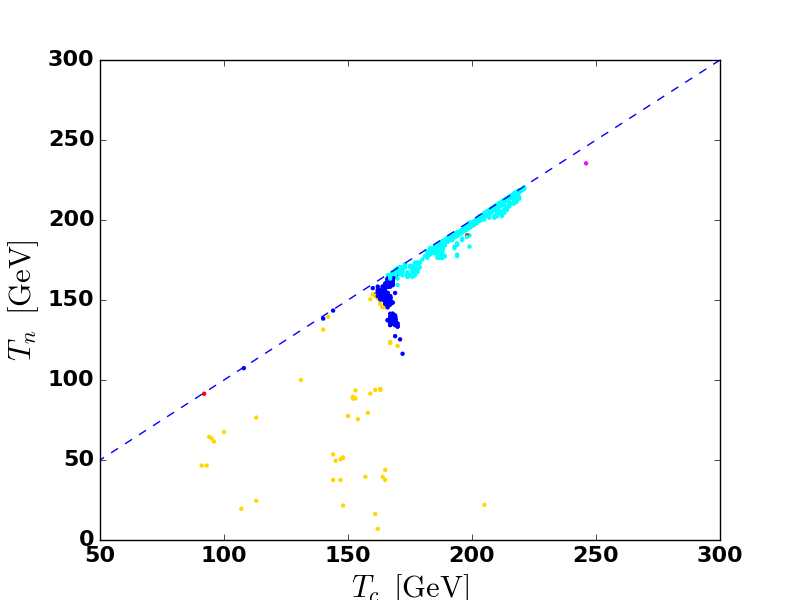}
\includegraphics[width=0.48\linewidth]{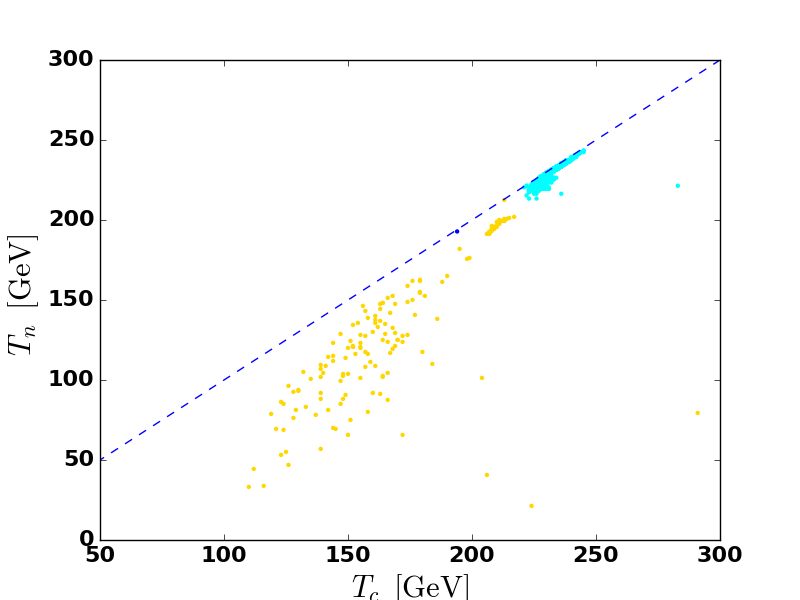}
\includegraphics[width=0.48\linewidth]{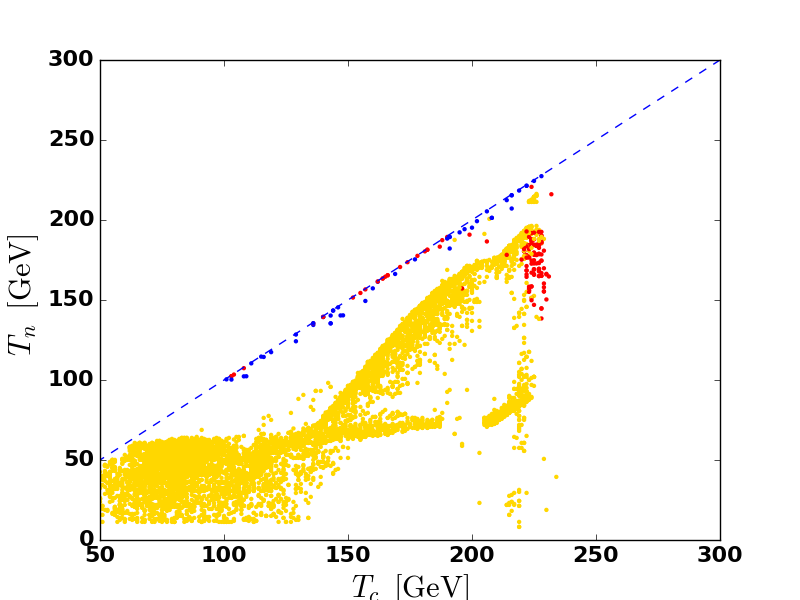}
\includegraphics[width=0.48\linewidth]{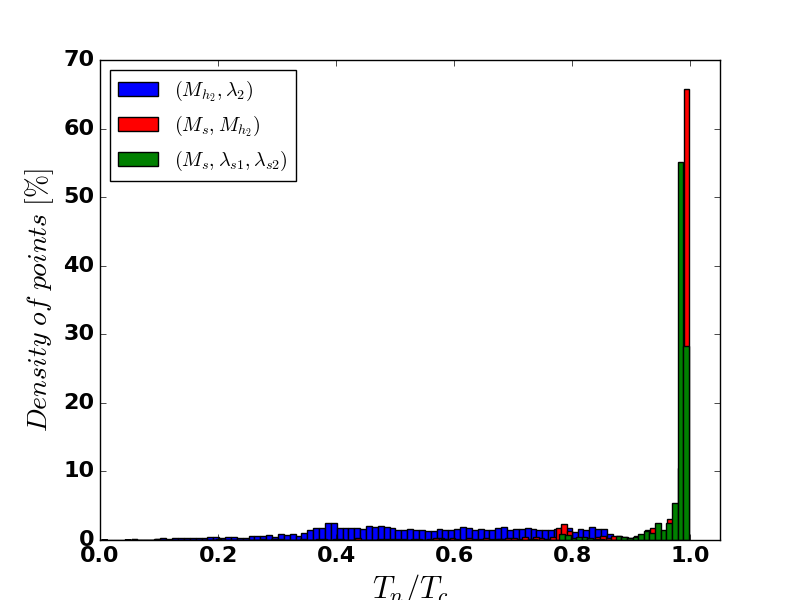}
\caption{\label{fig:TcTn_supercooling} \it 
The nucleation temperature versus the critical temperature. 
\textbf{Top-left:} results for the $(M_s,M_{h_2})$ scan. 
\textbf{Top-right:} results for the $(M_s,\lambda_{s1},\lambda_{s2})$ scan. 
\textbf{Bottom-left:} supercooling for the case of the $(M_{h_2},\lambda_2)$ scan (strong patterns). 
The dashed line represents the equality $T_n = T_c$. The color code is the same as in Fig.~\ref{fig:scan_Mh2l2} . 
\textbf{Bottom-right:} distribution of the points as function of $T_n/T_c$, for each scan.}
\end{figure}
In Fig.~\ref{fig:TcTn_supercooling} we show the nucleation temperature as a function of the critical temperature for each of the three scans above. 
The top panels show the cases of the $(M_s,M_{h_2})$ (left) and $(M_s,\lambda_{s1},\lambda_{s2})$ (right) scans, where we have seen that the phase transitions 
tend to occur through the weak patterns where supercooling is not expected. This is why most of the points fall on (or close to) the line $T_n = T_c$. However, the strong patterns 
indicated by the yellow points also experience supercooling. Note that none of these cases involves the trapped supercooled states where $v_c/T_c \gtrsim 3$. 
On the other hand, the bottom-left panel, which represents the scan $(M_{h_2},\lambda_2)$, clearly shows the supercooling property of the strong patterns, 
in particular, for the yellow and red points that sit below the $T_n = T_c$ line. In this case, the nucleation temperature can be about two times smaller than 
the critical temperature $T_c$. However, we also have the scenarios along the $T_n = T_c$ line which correspond to the points in the upper edge of the allowed 
region in the bottom-right panel of Fig.~\ref{fig:scan_Mh2l2}, that is, when $M_{h_2} \gtrsim 300~\mathrm{GeV}$. This is summarized in the histogram on 
the bottom-right panel, where it is clear that $T_n /T_c \ll 1$ for the first scan and $T_n /T_c \simeq 1$ for the other two scans. It is also possible to see that 
the first scan clearly does not favour a particular ratio.

\section{Conclusions}
\label{sec:conclusions}

The vacuum structure of multi-scalar extensions of the SM exhibits a dramatic increase in complexity with the growing 
number of Higgs doublets and singlets. This happens even with the Higgs sectors possessing additional continuous and 
discrete symmetries. We have performed one of the first studies of phase transitions in a simple model containing more 
than two scalar fields, and found rather generic features of such transitions that are expected to emerge also in a variety of 
other multi-Higgs models.

As an example, in this work we have identified and classified various types of multi-step phase transitions that occur in 
the simplest low-energy limit of the supersymmetric Trinification-based GUT model containing two Higgs doublets, one 
complex singlet scalar as well as one generation of vector-like quarks. A particular simplicity of the tree-level potential in this 
model due to the presence of an additional (global) family $\U{T}$ both in the Higgs and Yukawa sectors, as well as 
several $\mathbb{Z}_2$ symmetries in the Higgs sector, enables a complete analytic investigation of its tree-level mass spectra and 
phase transitions in various regimes. We have exhaustively investigated the key features of 
the tree-level phase diagram in both cases with and without the complex singlet scalar and have elaborated a novel pyramidal 
representation of the tree-level vacuum structure as a useful geometric tool to identify the first-order phase transitions in such 
multi-scalar models.

The results of our comprehensive tree-level analysis were used as a useful guide in a complete one-loop study of the vacuum 
structure considering the effective potential at finite temperatures. Our numerical analysis accounts for the complete one-loop 
(Coleman-Weinberg) and thermal corrections and does not rely on the thin-wall approximation for the bubble properties.
We have shown that the transitions that are of the first order at tree level are often too strong and suffer from supercooling. 
Nevertheless, significant regions of the parameter space are still relevant for baryogenesis through the bubble nucleation process. 
Besides, we have demonstrated that the tree-level second-order transitions can become of the strongly first order if the singlet scalar 
is strongly coupled to the Higgs doublets.

Our results show that multi-Higgs models display a lot of possibilities to fulfill the third Sakharov condition while the presence of the VLQ sector helps in satisfying the second Sakharov condition, both relevant for efficient baryogenesis. In a forthcoming paper, we will quantitatively study the details of baryogenesis in the considered THDSM and will also discuss its potential for gravitational waves production. The complexity of the scalar sector that we have investigated may also lead to unusual multi-steps transitions where several steps may be of the strongly first order, with interesting astrophysical consequences that will also be studied in a future work.

\acknowledgments
The authors would like to thank C.~Herdeiro, M.~Sampaio, J.~Rosa and M. Ouerfelli for useful discussions in the various stages of this work. A.P.M. is funded 
by the FCT grant SFRH/BPD/97126/2013. R.P. thanks Prof.~C.~Herdeiro for support of the project and hospitality during his visits at Aveiro university. 
R.P. is partially supported by the Swedish Research Council, contract number 621-2013-428 and by CONICYT grant PIA ACT1406. The work in this paper is 
also supported by the CIDMA project UID/MAT/04106/2013. This project has received a financial support Erasmus+ for international mobility from 
Universit\'e Paris-Sud.

\appendix
\section{Coleman-Weinberg potential and thermal corrections at one loop} 
\label{sec:app1}

The full one-loop effective potential reads
\begin{equation}
V^{(1)}(T) = V_{\rm tree} + V_{\rm CW} + \Delta V^{(1)}(T) \,,
\end{equation}
where the Coleman-Weinberg (CW) contribution is
\begin{equation}
V_{\rm CW} = \sum_i (-1)^F g_i \frac{m_i^4}{64 \pi^2} \left( \log\left[ \frac{m_i^2}{\Lambda^2}\right] - c_i \right) \,,
\end{equation}
and $\Delta V^{(1)}(T)$ is the $T$-dependent part of the one-loop correction given by Eq.~\eqref{finite_T_correction}.

We cannot use the previous trick of only computing the trace to obtain the field-dependent masses involved in the one-loop correction
due to non-linear terms. The $10 \times 10$ Hessian matrix reads 
\begin{equation}
\mathcal{M}^2 = \left(
\begin{array}{cc}
 A & 0 \\
 0 & G \\
\end{array}
\right) \,,
\end{equation}
where
\begin{align}\label{eq:A}
A &= \left(
\begin{array}{ccc}
 m_{h_1}^2(\phi_1,\phi_2,\phi_s) & \lambda_{12} \phi_1 \phi_2 & 2 \lambda_{s1} \phi_1 \phi_s\\
 \lambda_{12} \phi_1 \phi_2 & m_{h_2}^2(\phi_1,\phi_2,\phi_s) & 2 \lambda_{s2} \phi_2 \phi_s\\
 2 \lambda_{s1} \phi_1 \phi_s & 2 \lambda_{s2} \phi_2 \phi_s & m_{S_R}^2(\phi_1,\phi_2,\phi_s) \\
\end{array}
\right) \,,
\end{align}
\begin{multline}\label{eq:G}
G = \mathrm{Diag} \(
m_{\chi_1}^2(\phi_1,\phi_2,\phi_s), 
m_{\chi_1}^2(\phi_1,\phi_2,\phi_s),
m_{\chi_1}^2(\phi_1,\phi_2,\phi_s), \right.\\ \left.
m_{\chi_2}^2(\phi_1,\phi_2,\phi_s), 
m_{\chi_2}^2(\phi_1,\phi_2,\phi_s),
m_{\chi_2}^2(\phi_1,\phi_2,\phi_s),
m_{S_I}^2(\phi_1,\phi_2,\phi_s) \) \,.
\end{multline}
The diagonal terms $m_{h_1}^2(\phi_1,\phi_2,\phi_s)$, $m_{h_2}^2(\phi_1,\phi_2,\phi_s)$, $m_{S_R}^2(\phi_1,\phi_2,\phi_s)$, $m_{\chi_1}^2(\phi_1,\phi_2,\phi_s)$, $m_{\chi_2}^2(\phi_1,\phi_2,\phi_s)$ and $m_{S_I}^2(\phi_1,\phi_2,\phi_s)$
are given by Eq.~\eqref{diagonal_terms}.

The last step is to improve the approximation by solving the problem of the perturbative expansion breakdown due to symmetry restoration. 
This is done by resumming the bosonic degrees of freedom \citep{Curtin:2016urg}, i.e.~by replacing the field-dependent masses by
\begin{equation}
m_i^2(\phi_\alpha) \longrightarrow m_i^2(\phi_\alpha) + \Pi_i(\phi_\alpha,T) \,,
\end{equation}
where $\Pi_i(\phi_\alpha,T)$ are the thermal corrections. We employ the truncated dressing of the effective potential, i.e. we compute the thermal masses 
to the leading order, as was already done in Sect.~\ref{sec:thermal_masses}. For the physical fields, $m_i^2(\phi_\alpha) + \Pi_i(\phi_\alpha,T)$ are 
the eigenvalues of
\begin{equation}
A' = A + T^2 \mathrm{Diag} ( c_1 , c_2 , c_s ) \,,
\end{equation}
where $c_1,c_2,c_s$ are given by Eq.~\eqref{coeff_thermalmasses}. For the Goldstone bosons, $m_i^2(\phi_\alpha) + \Pi_i(\phi_\alpha,T)$ are 
the eigenvalues of
\begin{equation}
G' = G + T^2 \mathrm{Diag}(c_1,c_1,c_1,c_2,c_2,c_2,c_s) \,.
\end{equation}
For the gauge bosons, the corrected masses are the eigenvalues of (see Ref.~\citep{Carrington:1991hz})
\begin{equation}
M_{\rm SM} + C_{\rm SM} =
\frac{\phi_1^2+\phi_2^2}{4}
\begin{pmatrix}
g_L^2 & 0 & 0 & 0 \\
0 & g_L^2 & 0 & 0 \\
0 & 0 & g_L^2 & -g_L g_Y \\
0 & 0 & -g_L g_Y & g_Y^2
\end{pmatrix}
+
\frac{11}{6} T^2
\begin{pmatrix}
g_L^2 & 0 & 0 & 0 \\
0 & g_L^2 & 0 & 0 \\
0 & 0 & g_L^2 & 0 \\
0 & 0 & 0 & g_Y^2
\end{pmatrix} \,.
\end{equation}
The photon has zero mass but gets a thermal correction. Only the longitudinal polarizations have to be corrected.

The full one-loop corrections to the effective potential read
\begin{multline}
V_{\rm CW} = \sum_{\mathclap{\substack{\lambda_i(T) \in \\ {\rm Sp}(A') \cup {\rm Sp}(G')}}} \frac{\lambda_i(T)^2}{64 \pi^2} 
\( \log \[ \frac{\lambda_i(T)}{\Lambda^2} \] - 3/2 \)
+ \sum_{\mathclap{\substack{\lambda_i(T) \in \\ {\rm Sp}(M_{\rm SM}+C_{\rm SM})}}} \frac{\lambda_i(T)^2}{64 \pi^2} 
\( \log \[ \frac{\lambda_i(T)}{\Lambda^2} \] - 3/2 \) \\
+ \sum_{\mathclap{\substack{\lambda_i(T) \in \\ {\rm Sp}(M_{\rm SM})}}} 2 \frac{\lambda_i(T)^2}{64 \pi^2} 
\( \log \[ \frac{\lambda_i(T)}{\Lambda^2} \] - 1/2 \)
- 12 \sum_{i = t} \frac{m_i^4(\phi_\alpha)}{64 \pi^2} 
\( \log \[ \frac{m_i^2(\phi_\alpha)}{\Lambda^2} \] - 3/2 \)
\end{multline}
\begin{multline}
\Delta V^{(1)}(T) = \frac{T^4}{2 \pi^2} \left\{ 
\sum_{\substack{\lambda_i(T) \in \\ {\rm Sp}(A') \cup {\rm Sp}(G')}} J_B \[\frac{\lambda_i(T)}{T^2} \]
+ \sum_{\mathclap{\substack{\lambda_i(T) \in \\ {\rm Sp}(M_{\rm SM}+C_{\rm SM})}}} J_B \[\frac{\lambda_i(T)}{T^2} \] 
+ \sum_{\mathclap{\substack{\lambda_i(T) \in \\ {\rm Sp}(M_{\rm SM})}}} 2 J_B \[ \frac{\lambda_i(T)}{T^2} \]
\right.\\ \left.
- 12 \sum_{i = t} J_F \[ \frac{m_i^2(\phi_\alpha)}{T^2} \] \right\} \,.
\label{HO_corrections}
\end{multline}
Here, the top quark contribution is given in terms of its mass determined in Eq.~\eqref{eq:Mq-simp} (for the field-dependent top 
mass, see Eq.~\eqref{top}).

Due to the CW contribution, the VEVs and physical masses are shifted even at $T=0$, which is not good since we want to keep under control 
the experimentally measured Higgs parameters. We should therefore add the counterterms in the potential to match the results derived 
at tree level, in particular, to retrieve the Higgs mass and VEV. The simplest way to do this is to add the following contribution to the one-loop 
effective potential
\begin{multline}
V_{\rm ct} =
\frac{\delta m_1^2}{2} \phi_1^2
+ \frac{\delta m_2^2}{2} \phi_2^2
+ \delta m_s^2 \phi_s^2
+ \frac{\delta \lambda_1}{8} \phi_1^4
+ \frac{\delta \lambda_2}{8} \phi_2^4
+ \frac{\delta \lambda_s}{2} \phi_s^4 \\
+ \frac{\delta \lambda_{12}}{4} \phi_1^2 \phi_2^2 
+ \frac{\delta \lambda_{s1}}{2} \phi_1^2  \phi_s^2
+ \frac{\delta \lambda_{s2}}{2} \phi_2^2 \phi_s^2 \,.
\end{multline}
In the considered scenarios, $\phi_s=0$ at $T=0$. The contribution $V_{\rm ct}$ reduces to
\begin{equation}
V_{\rm ct} =
\frac{\delta m_1^2}{2} \phi_1^2
+ \frac{\delta m_2^2}{2} \phi_2^2
+ \frac{\delta \lambda_1}{8} \phi_1^4
+ \frac{\delta \lambda_2}{8} \phi_2^4
+ \frac{\delta \lambda_{12}}{4} \phi_1^2 \phi_2^2 \,.
\end{equation}
Then, two renormalization conditions are required to shift the VEVs:
\begin{equation}
\left. \frac{\partial (V_{\rm CW} + V_{\rm ct})}{\partial \phi_1} \right|_{\rm VEVs} 
= \left. \frac{\partial (V_{\rm CW} + V_{\rm ct})}{\partial \phi_2} \right|_{\rm VEVs} =0 \,.
\end{equation}
This implies that $V_{\rm tree} +V_{\rm CW}+V_{\rm ct}$ has the same minimum as $V_{\rm tree}$ at $T=0$.
Three additional conditions are required to shift the masses:
\begin{equation}
\left. \frac{\partial^2 (V_{\rm CW} + V_{\rm ct})}{\partial \phi^\alpha \phi^\beta} \right|_{\rm VEVs} = 0 \,.
\end{equation}
Since the matrix is symmetric, this provides three relations bringing the total number of equations to five, hence, $V_{\rm ct}$ 
is completely defined.

There is, however, a problem with this procedure since the Goldstone bosons' contributions introduce IR logarithmic divergences through 
terms proportional to $\log m^2$ at $m=0$ (in vacuum). This indicates that the effective potential method is not a well-defined 
procedure when the Goldstone bosons are present. In fact, those IR divergences cancel when considering  $p \neq 0$ diagrams 
\citep{Camargo-Molina:2016moz} and one can show that dealing with the divergences in the effective potential approach is possible by 
introducing an IR cut-off.

Here, for simplicity and also to speed up the numerical calculations, we will follow the procedure in Ref.~\citep{Jiang:2015cwa} and do not consider 
the conditions on the Hessian matrix. The latter is justified since the shifts of the couplings are rather small. The counterterms are obtained numerically 
by computing the derivatives of the CW potential such that
\begin{align}
\delta m_1^2 &= - \frac{1}{v_1} \left. \frac{\partial V_{\rm CW}}{\partial \phi_1} \right|_{v_1,v_2} \notag \\
\delta m_2^2 &= - \frac{1}{v_2} \left. \frac{\partial V_{\rm CW}}{\partial \phi_2} \right|_{v_1,v_2} \,.
\label{CT_corrections}
\end{align}

\section{Choice of a model from physical parameters} 
\label{sec:app2}

We recall that in the ($v_1$,$v_2$,0) vacuum the physical states read
\begin{gather}
|H_1\rangle^{\rm broken} = \cos\theta |H_1\rangle^{\rm unbroken} + \sin\theta |H_2\rangle^{\rm unbroken} \,, \notag \\
|H_2\rangle^{\rm broken} = -\sin\theta |H_1\rangle^{\rm unbroken} + \cos\theta |H_2\rangle^{\rm unbroken} \,, \notag \\
|\Phi\rangle^{\rm broken} = |\Phi\rangle^{\rm unbroken} \,,
\end{gather}
where
\begin{equation}
\tan 2 \theta = \frac{2 \lambda_{12} \sqrt{C_{12}} \sqrt{C_{21}}}{\lambda_1 C_{21} - \lambda_2 C_{12}} \,.
\end{equation}
The VEVs of the unbroken states are
\begin{equation}
v_1 = \sqrt{\frac{2 C_{21}}{L_{12}}}\,, \qquad 
v_2 = \sqrt{\frac{2 C_{12}}{L_{12}}} \,.
\end{equation}
The VEVs of the broken states are, therefore,
\begin{gather}
v_h = \cos\theta \sqrt{\frac{2 C_{21}}{L_{12}}} + \sin\theta \sqrt{\frac{2 C_{12}}{L_{12}}} \notag \\
v_{h_2} = -\sin\theta \sqrt{\frac{2 C_{21}}{L_{12}}} + \cos\theta \sqrt{\frac{2 C_{12}}{L_{12}}} \,.
\end{gather}
Defining
\begin{gather}
c_{12} \equiv \frac{C_{12}}{L_{12}} \,, \qquad c_{21} \equiv \frac{C_{21}}{L_{12}} \,,
\end{gather}
the physical masses are as follows
\begin{gather}
m^2(|H_1\rangle^{\rm broken}) = \lambda_1 c_{21} + \lambda_2 c_{12}
- \sqrt{ ( \lambda_1 c_{21} + \lambda_2 c_{12} )^2 - 4 c_{12} c_{21} L_{12}} \,,
\notag \\
m^2(|H_2\rangle^{\rm broken}) = \lambda_1 c_{21} + \lambda_2 c_{12}
+ \sqrt{ ( \lambda_1 c_{21} + \lambda_2 c_{12} )^2 - 4 c_{12} c_{21} L_{12}} \,,
\notag \\
m^2(|\Phi\rangle^{\rm broken}) =
2 \lambda_{s1} c_{21} + 2 \lambda_{s2} c_{12} + 2 m_s^2 \,.
\end{gather}

We identify the lightest state $|H_1\rangle^{\rm broken}$ with the SM Higgs boson of mass $M_h$. The heaviest state $|H_2\rangle^{\rm broken}$ is 
the additional Higgs boson of mass $M_{h_2}$ and the unmixed state is the complex singlet scalar of mass $M_s$. We end up with the following 
mass relations:
\begin{align} 
\label{eq:constraints}
\begin{aligned}
M_h^2 &= \lambda_1 c_{21} + 2 \lambda_2 c_{12} \\
&\phantom{====} - (\lambda_1 \lambda_2 - \lambda_{12}^2) \sqrt{ ( \lambda_1  c_{21} + 
\lambda_2 c_{12} )^2 - 4 c_{12} c_{21} (\lambda_1 \lambda_2 - \lambda_{12}^2)} \,, \\
M_h^2 + M_{h_2}^2 &= 2 \lambda_1 c_{21} + 2 \lambda_2 c_{12} \,,\\
v^2 &= v_h^2 + v_{h_2}^2 = 2 c_{21} + 2 c_{12} \,, \\
M_s^2 - 2 m_s^2 &= 2 \lambda_{s1} c_{21} + 2 \lambda_{s2} c_{12} \,,
\end{aligned}
\end{align}
where $M_h = 125.09$ GeV and $v = 246$ GeV.
We take as input the following parameters:
\begin{equation*}
M_{h_2} \,, \quad M_s \,, \quad 
\lambda_1 \,, \quad \lambda_2 \,, \quad \lambda_s \,, \quad
\lambda_{s1} \,, \quad \lambda_{s2} \,,
\end{equation*}
and we want to express the bare parameters of the potential as functions of these input parameters.
The resolution of the second and third relations of Eq.~\eqref{eq:constraints} gives the quantities $c_{12}$ and $c_{21}$:
\begin{gather}
c_{12} = \frac{1}{2} \frac{ M_h^2+M_{h_2}^2-\lambda_1 v^2 } {\lambda_2-\lambda_1} \notag \\
c_{21} = \frac{1}{2} \frac{ M_h^2+M_{h_2}^2-\lambda_2 v^2 } {\lambda_1-\lambda_2} \,.
\end{gather}
Then the numerical resolution of the first relation gives $\lambda_{12}$ and the last relation gives $m_s$. 
Finally, the definitions of $c_{12}$ and $c_{21}$ allow to retrieve the mass squared parameters $m_1$ and $m_2$.
All parameters are then unambiguously fixed and the mixing angle can be easily computed, as well as the VEVs 
of the physical fields $v_h$ and $v_{h_2}$.

\newpage

\bibliographystyle{JHEP}
\bibliography{bib}

\providecommand{\href}[2]{#2}\begingroup\raggedright\begin{thebibliography}{10}

\bibitem{Sakharov:1967dj}
A.~D. Sakharov, {\it {Violation of CP Invariance, c Asymmetry, and Baryon
  Asymmetry of the Universe}},  {\em Pisma Zh. Eksp. Teor. Fiz.} {\bf 5} (1967)
  32--35. [Usp. Fiz. Nauk161,61(1991)].

\bibitem{Dine:2003ax}
M.~Dine and A.~Kusenko, {\it {The Origin of the matter - antimatter
  asymmetry}},  {\em Rev. Mod. Phys.} {\bf 76} (2003) 1,
  [\href{http://xxx.lanl.gov/abs/hep-ph/0303065}{{\tt hep-ph/0303065}}].

\bibitem{Klinkhamer1984}
F.~R. Klinkhamer and N.~S. Manton, {\it A saddle-point solution in the
  weinberg-salam theory},  {\em Phys. Rev. D} {\bf 30} (Nov, 1984) 2212--2220.

\bibitem{Arnold1987}
P.~Arnold and L.~McLerran, {\it Sphalerons, small fluctuations, and
  baryon-number violation in electroweak theory},  {\em Phys. Rev. D} {\bf 36}
  (Jul, 1987) 581--595.

\bibitem{Morrissey:2012db}
D.~E. Morrissey and M.~J. Ramsey-Musolf, {\it {Electroweak baryogenesis}},
  {\em New J. Phys.} {\bf 14} (2012) 125003,
  [\href{http://xxx.lanl.gov/abs/1206.2942}{{\tt 1206.2942}}].

\bibitem{White-book}
G.~A. White, {\em A Pedagogical Introduction to Electroweak Baryogenesis}.
\newblock 2053-2571. Morgan and Claypool Publishers, 2016.

\bibitem{Xiao:2015tja}
M.-L. Xiao and J.-H. Yu, {\it {Electroweak baryogenesis in a scalar-assisted
  vectorlike fermion model}},  {\em Phys. Rev.} {\bf D94} (2016), no.~1 015011,
  [\href{http://xxx.lanl.gov/abs/1509.02931}{{\tt 1509.02931}}].

\bibitem{Gavela:1993ts}
M.~B. Gavela, P.~Hernandez, J.~Orloff, and O.~Pene, {\it {Standard model CP
  violation and baryon asymmetry}},  {\em Mod. Phys. Lett.} {\bf A9} (1994)
  795--810, [\href{http://xxx.lanl.gov/abs/hep-ph/9312215}{{\tt
  hep-ph/9312215}}].

\bibitem{Konstandin:2003dx}
T.~Konstandin, T.~Prokopec, and M.~G. Schmidt, {\it {Axial currents from CKM
  matrix CP violation and electroweak baryogenesis}},  {\em Nucl. Phys.} {\bf
  B679} (2004) 246--260, [\href{http://xxx.lanl.gov/abs/hep-ph/0309291}{{\tt
  hep-ph/0309291}}].

\bibitem{Aad:2012tfa}
{\bf ATLAS} Collaboration, G.~Aad {\em et~al.}, {\it {Observation of a new
  particle in the search for the Standard Model Higgs boson with the ATLAS
  detector at the LHC}},  {\em Phys. Lett.} {\bf B716} (2012) 1--29,
  [\href{http://xxx.lanl.gov/abs/1207.7214}{{\tt 1207.7214}}].

\bibitem{Chatrchyan:2012xdj}
{\bf CMS} Collaboration, S.~Chatrchyan {\em et~al.}, {\it {Observation of a new
  boson at a mass of 125 GeV with the CMS experiment at the LHC}},  {\em Phys.
  Lett.} {\bf B716} (2012) 30--61,
  [\href{http://xxx.lanl.gov/abs/1207.7235}{{\tt 1207.7235}}].

\bibitem{Rummukainen:1998as}
K.~Rummukainen, M.~Tsypin, K.~Kajantie, M.~Laine, and M.~E. Shaposhnikov, {\it
  {The Universality class of the electroweak theory}},  {\em Nucl. Phys.} {\bf
  B532} (1998) 283--314, [\href{http://xxx.lanl.gov/abs/hep-lat/9805013}{{\tt
  hep-lat/9805013}}].

\bibitem{Cline:2012hg}
J.~M. Cline and K.~Kainulainen, {\it {Electroweak baryogenesis and dark matter
  from a singlet Higgs}},  {\em JCAP} {\bf 1301} (2013) 012,
  [\href{http://xxx.lanl.gov/abs/1210.4196}{{\tt 1210.4196}}].

\bibitem{Li:2014wia}
T.~Li and Y.-F. Zhou, {\it {Strongly first order phase transition in the
  singlet fermionic dark matter model after LUX}},  {\em JHEP} {\bf 07} (2014)
  006, [\href{http://xxx.lanl.gov/abs/1402.3087}{{\tt 1402.3087}}].

\bibitem{Vaskonen:2016yiu}
V.~Vaskonen, {\it {Electroweak baryogenesis and gravitational waves from a real
  scalar singlet}},  {\em Phys. Rev.} {\bf D95} (2017), no.~12 123515,
  [\href{http://xxx.lanl.gov/abs/1611.02073}{{\tt 1611.02073}}].

\bibitem{Beniwal:2017eik}
A.~Beniwal, M.~Lewicki, J.~D. Wells, M.~White, and A.~G. Williams, {\it
  {Gravitational wave, collider and dark matter signals from a scalar singlet
  electroweak baryogenesis}},  \href{http://xxx.lanl.gov/abs/1702.06124}{{\tt
  1702.06124}}.

\bibitem{Kurup:2017dzf}
G.~Kurup and M.~Perelstein, {\it {Dynamics of Electroweak Phase Transition In
  Singlet-Scalar Extension of the Standard Model}},  {\em Phys. Rev.} {\bf D96}
  (2017) 015036, [\href{http://xxx.lanl.gov/abs/1704.03381}{{\tt 1704.03381}}].

\bibitem{Barger:2008jx}
V.~Barger, P.~Langacker, M.~McCaskey, M.~Ramsey-Musolf, and G.~Shaughnessy,
  {\it {Complex Singlet Extension of the Standard Model}},  {\em Phys. Rev.}
  {\bf D79} (2009) 015018, [\href{http://xxx.lanl.gov/abs/0811.0393}{{\tt
  0811.0393}}].

\bibitem{Jiang:2015cwa}
M.~Jiang, L.~Bian, W.~Huang, and J.~Shu, {\it {Impact of a complex singlet:
  Electroweak baryogenesis and dark matter}},  {\em Phys. Rev.} {\bf D93}
  (2016), no.~6 065032, [\href{http://xxx.lanl.gov/abs/1502.07574}{{\tt
  1502.07574}}].

\bibitem{Chiang:2017nmu}
C.-W. Chiang, M.~J. Ramsey-Musolf, and E.~Senaha, {\it {Standard Model with a
  Complex Scalar Singlet: Cosmological Implications and Theoretical
  Considerations}},  \href{http://xxx.lanl.gov/abs/1707.09960}{{\tt
  1707.09960}}.

\bibitem{Chen:2017qcz}
C.-Y. Chen, J.~Kozaczuk, and I.~M. Lewis, {\it {Non-resonant Collider
  Signatures of a Singlet-Driven Electroweak Phase Transition}},  {\em JHEP}
  {\bf 08} (2017) 096, [\href{http://xxx.lanl.gov/abs/1704.05844}{{\tt
  1704.05844}}].

\bibitem{Branco:2011iw}
G.~C. Branco, P.~M. Ferreira, L.~Lavoura, M.~N. Rebelo, M.~Sher, and J.~P.
  Silva, {\it {Theory and phenomenology of two-Higgs-doublet models}},  {\em
  Phys. Rept.} {\bf 516} (2012) 1--102,
  [\href{http://xxx.lanl.gov/abs/1106.0034}{{\tt 1106.0034}}].

\bibitem{Barger:2007im}
V.~Barger, P.~Langacker, M.~McCaskey, M.~J. Ramsey-Musolf, and G.~Shaughnessy,
  {\it {LHC Phenomenology of an Extended Standard Model with a Real Scalar
  Singlet}},  {\em Phys. Rev.} {\bf D77} (2008) 035005,
  [\href{http://xxx.lanl.gov/abs/0706.4311}{{\tt 0706.4311}}].

\bibitem{Costa:2014qga}
R.~Costa, A.~P. Morais, M.~O.~P. Sampaio, and R.~Santos, {\it {Two-loop
  stability of a complex singlet extended Standard Model}},  {\em Phys. Rev.}
  {\bf D92} (2015) 025024, [\href{http://xxx.lanl.gov/abs/1411.4048}{{\tt
  1411.4048}}].

\bibitem{Costa:2017gky}
R.~Costa, M.~O.~P. Sampaio, and R.~Santos, {\it {NLO electroweak corrections in
  general scalar singlet models}},  {\em JHEP} {\bf 07} (2017) 081,
  [\href{http://xxx.lanl.gov/abs/1704.02327}{{\tt 1704.02327}}].

\bibitem{Turok:1990zg}
N.~Turok and J.~Zadrozny, {\it {Electroweak baryogenesis in the two doublet
  model}},  {\em Nucl. Phys.} {\bf B358} (1991) 471--493.

\bibitem{Turok:1991uc}
N.~Turok and J.~Zadrozny, {\it {Phase transitions in the two doublet model}},
  {\em Nucl. Phys.} {\bf B369} (1992) 729--742.

\bibitem{Funakubo:1993jg}
K.~Funakubo, A.~Kakuto, and K.~Takenaga, {\it {The Effective potential of
  electroweak theory with two massless Higgs doublets at finite temperature}},
  {\em Prog. Theor. Phys.} {\bf 91} (1994) 341--352,
  [\href{http://xxx.lanl.gov/abs/hep-ph/9310267}{{\tt hep-ph/9310267}}].

\bibitem{Davies:1994id}
A.~T. Davies, C.~D. froggatt, G.~Jenkins, and R.~G. Moorhouse, {\it
  {Baryogenesis constraints on two Higgs doublet models}},  {\em Phys. Lett.}
  {\bf B336} (1994) 464--470.

\bibitem{Cline:1995dg}
J.~M. Cline, K.~Kainulainen, and A.~P. Vischer, {\it {Dynamics of two Higgs
  doublet CP violation and baryogenesis at the electroweak phase transition}},
  {\em Phys. Rev.} {\bf D54} (1996) 2451--2472,
  [\href{http://xxx.lanl.gov/abs/hep-ph/9506284}{{\tt hep-ph/9506284}}].

\bibitem{Laine:2000rm}
M.~Laine and K.~Rummukainen, {\it {Two Higgs doublet dynamics at the
  electroweak phase transition: A Nonperturbative study}},  {\em Nucl. Phys.}
  {\bf B597} (2001) 23--69,
  [\href{http://xxx.lanl.gov/abs/hep-lat/0009025}{{\tt hep-lat/0009025}}].

\bibitem{Fromme:2006cm}
L.~Fromme, S.~J. Huber, and M.~Seniuch, {\it {Baryogenesis in the two-Higgs
  doublet model}},  {\em JHEP} {\bf 11} (2006) 038,
  [\href{http://xxx.lanl.gov/abs/hep-ph/0605242}{{\tt hep-ph/0605242}}].

\bibitem{Lee:1973iz}
T.~D. Lee, {\it {A Theory of Spontaneous T Violation}},  {\em Phys. Rev.} {\bf
  D8} (1973) 1226--1239. [,516(1973)].

\bibitem{Lee:1974jb}
T.~D. Lee, {\it {CP Nonconservation and Spontaneous Symmetry Breaking}},  {\em
  Phys. Rept.} {\bf 9} (1974) 143--177. [,124(1974)].

\bibitem{Weinberg:1976hu}
S.~Weinberg, {\it {Gauge Theory of CP Violation}},  {\em Phys. Rev. Lett.} {\bf
  37} (1976) 657.

\bibitem{Bodeker:2004ws}
D.~Bodeker, L.~Fromme, S.~J. Huber, and M.~Seniuch, {\it {The Baryon asymmetry
  in the standard model with a low cut-off}},  {\em JHEP} {\bf 02} (2005) 026,
  [\href{http://xxx.lanl.gov/abs/hep-ph/0412366}{{\tt hep-ph/0412366}}].

\bibitem{Fromme:2006wx}
L.~Fromme and S.~J. Huber, {\it {Top transport in electroweak baryogenesis}},
  {\em JHEP} {\bf 03} (2007) 049,
  [\href{http://xxx.lanl.gov/abs/hep-ph/0604159}{{\tt hep-ph/0604159}}].

\bibitem{Huang:2015izx}
F.~P. Huang, P.-H. Gu, P.-F. Yin, Z.-H. Yu, and X.~Zhang, {\it {Testing the
  electroweak phase transition and electroweak baryogenesis at the LHC and a
  circular electron-positron collider}},  {\em Phys. Rev.} {\bf D93} (2016),
  no.~10 103515, [\href{http://xxx.lanl.gov/abs/1511.03969}{{\tt 1511.03969}}].

\bibitem{Balazs:2016yvi}
C.~Balazs, G.~White, and J.~Yue, {\it {Effective field theory, electric dipole
  moments and electroweak baryogenesis}},  {\em JHEP} {\bf 03} (2017) 030,
  [\href{http://xxx.lanl.gov/abs/1612.01270}{{\tt 1612.01270}}].

\bibitem{Uesugi:1996rd}
T.~Uesugi, A.~Sugamoto, and A.~Yamaguchi, {\it {Baryogenesis with vector - like
  quark model in charge transport mechanism}},  {\em Phys. Lett.} {\bf B392}
  (1997) 389--394, [\href{http://xxx.lanl.gov/abs/hep-ph/9606302}{{\tt
  hep-ph/9606302}}].

\bibitem{McDonald:1996uz}
J.~McDonald, {\it {CP violation for electroweak baryogenesis from mixing of
  Standard Model and heavy vector quarks}},  {\em Phys. Rev.} {\bf D53} (1996)
  645--654.

\bibitem{Chen:2015uza}
C.-Y. Chen, S.~Dawson, and Y.~Zhang, {\it {Higgs CP Violation from Vectorlike
  Quarks}},  {\em Phys. Rev.} {\bf D92} (2015), no.~7 075026,
  [\href{http://xxx.lanl.gov/abs/1507.07020}{{\tt 1507.07020}}].

\bibitem{Ivanov:2017dad}
I.~P. Ivanov, {\it {Building and testing models with extended Higgs sectors}},
  {\em Prog. Part. Nucl. Phys.} {\bf 95} (2017) 160--208,
  [\href{http://xxx.lanl.gov/abs/1702.03776}{{\tt 1702.03776}}].

\bibitem{Ginzburg:2010wa}
I.~F. Ginzburg, K.~A. Kanishev, M.~Krawczyk, and D.~Sokolowska, {\it {Evolution
  of Universe to the present inert phase}},  {\em Phys. Rev.} {\bf D82} (2010)
  123533, [\href{http://xxx.lanl.gov/abs/1009.4593}{{\tt 1009.4593}}].

\bibitem{Dorsch:2013wja}
G.~C. Dorsch, S.~J. Huber, and J.~M. No, {\it {A strong electroweak phase
  transition in the 2HDM after LHC8}},  {\em JHEP} {\bf 10} (2013) 029,
  [\href{http://xxx.lanl.gov/abs/1305.6610}{{\tt 1305.6610}}].

\bibitem{Basler:2016obg}
P.~Basler, M.~Krause, M.~Muhlleitner, J.~Wittbrodt, and A.~Wlotzka, {\it
  {Strong First Order Electroweak Phase Transition in the CP-Conserving 2HDM
  Revisited}},  {\em JHEP} {\bf 02} (2017) 121,
  [\href{http://xxx.lanl.gov/abs/1612.04086}{{\tt 1612.04086}}].

\bibitem{Basler:2017uxn}
P.~Basler, M.~Mühlleitner, and J.~Wittbrodt, {\it {The CP-Violating 2HDM in
  Light of a Strong First Order Electroweak Phase Transition and Implications
  for Higgs Pair Production}},  \href{http://xxx.lanl.gov/abs/1711.04097}{{\tt
  1711.04097}}.

\bibitem{Ma:2006km}
E.~Ma, {\it {Verifiable radiative seesaw mechanism of neutrino mass and dark
  matter}},  {\em Phys. Rev.} {\bf D73} (2006) 077301,
  [\href{http://xxx.lanl.gov/abs/hep-ph/0601225}{{\tt hep-ph/0601225}}].

\bibitem{Barbieri:2006dq}
R.~Barbieri, L.~J. Hall, and V.~S. Rychkov, {\it {Improved naturalness with a
  heavy Higgs: An Alternative road to LHC physics}},  {\em Phys. Rev.} {\bf
  D74} (2006) 015007, [\href{http://xxx.lanl.gov/abs/hep-ph/0603188}{{\tt
  hep-ph/0603188}}].

\bibitem{Majumdar:2006nt}
D.~Majumdar and A.~Ghosal, {\it {Dark Matter candidate in a Heavy Higgs Model -
  Direct Detection Rates}},  {\em Mod. Phys. Lett.} {\bf A23} (2008)
  2011--2022, [\href{http://xxx.lanl.gov/abs/hep-ph/0607067}{{\tt
  hep-ph/0607067}}].

\bibitem{LopezHonorez:2006gr}
L.~Lopez~Honorez, E.~Nezri, J.~F. Oliver, and M.~H.~G. Tytgat, {\it {The Inert
  Doublet Model: An Archetype for Dark Matter}},  {\em JCAP} {\bf 0702} (2007)
  028, [\href{http://xxx.lanl.gov/abs/hep-ph/0612275}{{\tt hep-ph/0612275}}].

\bibitem{Chowdhury:2011ga}
T.~A. Chowdhury, M.~Nemevsek, G.~Senjanovic, and Y.~Zhang, {\it {Dark Matter as
  the Trigger of Strong Electroweak Phase Transition}},  {\em JCAP} {\bf 1202}
  (2012) 029, [\href{http://xxx.lanl.gov/abs/1110.5334}{{\tt 1110.5334}}].

\bibitem{Borah:2012pu}
D.~Borah and J.~M. Cline, {\it {Inert Doublet Dark Matter with Strong
  Electroweak Phase Transition}},  {\em Phys. Rev.} {\bf D86} (2012) 055001,
  [\href{http://xxx.lanl.gov/abs/1204.4722}{{\tt 1204.4722}}].

\bibitem{Glashow:1976nt}
S.~L. Glashow and S.~Weinberg, {\it {Natural Conservation Laws for Neutral
  Currents}},  {\em Phys. Rev.} {\bf D15} (1977) 1958.

\bibitem{Chivukula:1987py}
R.~S. Chivukula and H.~Georgi, {\it {Composite Technicolor Standard Model}},
  {\em Phys. Lett.} {\bf B188} (1987) 99--104.

\bibitem{Hall:1990ac}
L.~J. Hall and L.~Randall, {\it {Weak scale effective supersymmetry}},  {\em
  Phys. Rev. Lett.} {\bf 65} (1990) 2939--2942.

\bibitem{DAmbrosio:2002vsn}
G.~D'Ambrosio, G.~F. Giudice, G.~Isidori, and A.~Strumia, {\it {Minimal flavor
  violation: An Effective field theory approach}},  {\em Nucl. Phys.} {\bf
  B645} (2002) 155--187, [\href{http://xxx.lanl.gov/abs/hep-ph/0207036}{{\tt
  hep-ph/0207036}}].

\bibitem{Isidori:2012ts}
G.~Isidori and D.~M. Straub, {\it {Minimal Flavour Violation and Beyond}},
  {\em Eur. Phys. J.} {\bf C72} (2012) 2103,
  [\href{http://xxx.lanl.gov/abs/1202.0464}{{\tt 1202.0464}}].

\bibitem{Cline:2011mm}
J.~M. Cline, K.~Kainulainen, and M.~Trott, {\it {Electroweak Baryogenesis in
  Two Higgs Doublet Models and B meson anomalies}},  {\em JHEP} {\bf 11} (2011)
  089, [\href{http://xxx.lanl.gov/abs/1107.3559}{{\tt 1107.3559}}].

\bibitem{Cline2017}
J.~M. Cline, {\it {Is electroweak baryogenesis dead?}},  in {\em {52nd
  Rencontres de Moriond on EW Interactions and Unified Theories (Moriond EW
  2017) La Thuile, Italy, March 18-25, 2017}}, 2017.
\newblock \href{http://xxx.lanl.gov/abs/1704.08911}{{\tt 1704.08911}}.

\bibitem{Camargo-Molina:2016yqm}
J.~E. Camargo-Molina, A.~P. Morais, A.~Ordell, R.~Pasechnik, M.~O. Sampaio, and
  J.~Wessén, {\it {Reviving trinification models through an E6 -extended
  supersymmetric GUT}},  {\em Phys. Rev.} {\bf D95} (2017), no.~7 075031,
  [\href{http://xxx.lanl.gov/abs/1610.03642}{{\tt 1610.03642}}].

\bibitem{Camargo-Molina:2017kxd}
J.~E. Camargo-Molina, A.~P. Morais, A.~Ordell, R.~Pasechnik, and J.~Wessén,
  {\it {Scale hierarchies, symmetry breaking and SM-like fermions in
  $\mathrm{SU}(3)$-family extended SUSY trinification}},
  \href{http://xxx.lanl.gov/abs/1711.05199}{{\tt 1711.05199}}.

\bibitem{Alanne:2016wtx}
T.~Alanne, K.~Kainulainen, K.~Tuominen, and V.~Vaskonen, {\it {Baryogenesis in
  the two doublet and inert singlet extension of the Standard Model}},  {\em
  JCAP} {\bf 1608} (2016), no.~08 057,
  [\href{http://xxx.lanl.gov/abs/1607.03303}{{\tt 1607.03303}}].

\bibitem{Banik:2014cfa}
A.~Dutta~Banik and D.~Majumdar, {\it {Inert doublet dark matter with an
  additional scalar singlet and 125 GeV Higgs boson}},  {\em Eur. Phys. J.}
  {\bf C74} (2014), no.~11 3142, [\href{http://xxx.lanl.gov/abs/1404.5840}{{\tt
  1404.5840}}].

\bibitem{Blinov:2015sna}
N.~Blinov, J.~Kozaczuk, D.~E. Morrissey, and C.~Tamarit, {\it {Electroweak
  Baryogenesis from Exotic Electroweak Symmetry Breaking}},  {\em Phys. Rev.}
  {\bf D92} (2015), no.~3 035012,
  [\href{http://xxx.lanl.gov/abs/1504.05195}{{\tt 1504.05195}}].

\bibitem{Chao:2017vrq}
W.~Chao, H.-K. Guo, and J.~Shu, {\it {Gravitational Wave Signals of Electroweak
  Phase Transition Triggered by Dark Matter}},
  \href{http://xxx.lanl.gov/abs/1702.02698}{{\tt 1702.02698}}.

\bibitem{Camargo-Molina:2016bwm}
J.~E. Camargo-Molina, A.~P. Morais, R.~Pasechnik, and J.~Wess{\'e}n, {\it {On a
  radiative origin of the Standard Model from Trinification}},  {\em JHEP} {\bf
  09} (2016) 129, [\href{http://xxx.lanl.gov/abs/1606.03492}{{\tt
  1606.03492}}].

\bibitem{Blinov:2015vma}
N.~Blinov, S.~Profumo, and T.~Stefaniak, {\it {The Electroweak Phase Transition
  in the Inert Doublet Model}},  {\em JCAP} {\bf 1507} (2015), no.~07 028,
  [\href{http://xxx.lanl.gov/abs/1504.05949}{{\tt 1504.05949}}].

\bibitem{Patrignani:2016xqp}
{\bf Particle Data Group} Collaboration, C.~Patrignani {\em et~al.}, {\it
  {Review of Particle Physics}},  {\em Chin. Phys.} {\bf C40} (2016), no.~10
  100001.

\bibitem{Quiros:1999jp}
M.~Quiros, {\it {Finite temperature field theory and phase transitions}},  in
  {\em {Proceedings, Summer School in High-energy physics and cosmology:
  Trieste, Italy, June 29-July 17, 1998}}, pp.~187--259, 1999.
\newblock \href{http://xxx.lanl.gov/abs/hep-ph/9901312}{{\tt hep-ph/9901312}}.

\bibitem{Curtin:2016urg}
D.~Curtin, P.~Meade, and H.~Ramani, {\it {Thermal Resummation and Phase
  Transitions}},  \href{http://xxx.lanl.gov/abs/1612.00466}{{\tt 1612.00466}}.

\bibitem{Linde1983}
A.~Linde, {\it Decay of the false vacuum at finite temperature},  {\em Nuclear
  Physics B} {\bf 216} (1983), no.~2 421 -- 445.

\bibitem{Dine:1992wr}
M.~Dine, R.~G. Leigh, P.~Y. Huet, A.~D. Linde, and D.~A. Linde, {\it {Towards
  the theory of the electroweak phase transition}},  {\em Phys. Rev.} {\bf D46}
  (1992) 550--571, [\href{http://xxx.lanl.gov/abs/hep-ph/9203203}{{\tt
  hep-ph/9203203}}].

\bibitem{Coleman:1977py}
S.~R. Coleman, {\it {The Fate of the False Vacuum. 1. Semiclassical Theory}},
  {\em Phys. Rev.} {\bf D15} (1977) 2929--2936. [Erratum: Phys.
  Rev.D16,1248(1977)].

\bibitem{Wainwright:2011kj}
C.~L. Wainwright, {\it {CosmoTransitions: Computing Cosmological Phase
  Transition Temperatures and Bubble Profiles with Multiple Fields}},  {\em
  Comput. Phys. Commun.} {\bf 183} (2012) 2006--2013,
  [\href{http://xxx.lanl.gov/abs/1109.4189}{{\tt 1109.4189}}].

\bibitem{Carrington:1991hz}
M.~E. Carrington, {\it {The Effective potential at finite temperature in the
  Standard Model}},  {\em Phys. Rev.} {\bf D45} (1992) 2933--2944.

\bibitem{Camargo-Molina:2016moz}
J.~E. Camargo-Molina, A.~P. Morais, R.~Pasechnik, M.~O.~P. Sampaio, and
  J.~Wessén, {\it {All one-loop scalar vertices in the effective potential
  approach}},  {\em JHEP} {\bf 08} (2016) 073,
  [\href{http://xxx.lanl.gov/abs/1606.07069}{{\tt 1606.07069}}].

\end{thebibliography}\endgroup

\end{document}